\pdfoutput=1

\documentclass[12pt]{article}%
\usepackage[nosort]{cite}
\usepackage{graphicx}
\usepackage{multicol}
\usepackage{amsfonts}
\usepackage{amssymb}
\usepackage{amsmath}
\usepackage{heck}
\usepackage{setspace}
\usepackage{verbatim}
\usepackage{color}
\usepackage{longtable}
\usepackage{float}
\usepackage{epsfig}
\usepackage{epstopdf}

\usepackage{tikz}
\usetikzlibrary{decorations.markings}

\newcommand{\dynkinradius}{.04cm}
\newcommand{\dynkinstep}{.35cm}
\newcommand{\dynkindot}[2]{\fill (\dynkinstep*#1,\dynkinstep*#2) circle (\dynkinradius);}

\newcommand{\dynkinline}[4]{\draw[thin] (\dynkinstep*#1,\dynkinstep*#2) -- (\dynkinstep*#3,\dynkinstep*#4);}

\newenvironment{dynkin}{\begin{tikzpicture}[decoration={markings,mark=at position 0.7 with {\arrow{>}}}]}
{\end{tikzpicture}}

\usepackage[margin=1in]{geometry}%
\usepackage{titletoc}%
\providecommand{\U}[1]{\protect\rule{.1in}{.1in}}
\numberwithin{equation}{section}

\newcommand{\ba}{\begin{eqnarray}}
\newcommand{\ea}{\end{eqnarray}}

\newcommand{\ov}{\overset }
\newcommand{\op}{\oplus }

\begin{document}

\date{February 2015}

\title{Atomic Classification of 6D SCFTs}

\institution{UNC}{\centerline{${}^{1}$Department of Physics, University of North Carolina, Chapel Hill, NC 27599, USA}}

\institution{UCSBmath}{\centerline{${}^{2}$Department of Mathematics, University of California Santa Barbara, CA 93106, USA}}

\institution{UCSBphys}{\centerline{${}^{3}$Department of Physics, University of California Santa Barbara, CA 93106, USA}}

\institution{HARVARD}{\centerline{${}^{4}$Jefferson Physical Laboratory, Harvard University, Cambridge, MA 02138, USA}}

\authors{Jonathan J. Heckman\worksat{\UNC}\footnote{e-mail: {\tt jheckman@email.unc.edu}},
David R. Morrison\worksat{\UCSBmath, \UCSBphys}\footnote{e-mail: {\tt drm@physics.ucsb.edu}},\\[2mm]
Tom Rudelius\worksat{\HARVARD}\footnote{e-mail: {\tt rudelius@physics.harvard.edu}},
and Cumrun Vafa\worksat{\HARVARD}\footnote{e-mail: {\tt vafa@physics.harvard.edu}}}

\abstract{We use F-theory to classify possibly all six-dimensional superconformal field theories (SCFTs).
This involves a two step process:  We first classify all possible tensor branches allowed in F-theory (which
correspond to allowed collections of contractible spheres)
and then classify all possible configurations of seven-branes wrapped over them.
We describe the first step in terms of ``atoms'' joined into ``radicals'' and
``molecules,'' using an analogy from chemistry.
The second step has an
interpretation via quiver-type gauge theories constrained by anomaly cancellation. A very surprising outcome
of our analysis is that all of these tensor branches have the structure of a linear chain of intersecting spheres
with a small amount of possible decoration at the two ends. The resulting
structure of these SCFTs takes the form of a generalized quiver consisting of ADE-type nodes joined by conformal matter.
A collection of highly non-trivial examples involving $E_8$ small instantons probing an ADE singularity is shown to have an F-theory realization.  This yields a classification of homomorphisms from ADE subgroups of $SU(2)$ into $E_8$ in purely geometric terms, matching results obtained in the mathematics literature from an intricate group theory analysis.}

\maketitle

\tableofcontents

\enlargethispage{\baselineskip}

\setcounter{tocdepth}{2}

\newpage

\section{Introduction \label{sec:INTRO}}

Six-dimensional superconformal field theories (SCFTs) occupy a special role in
the study of quantum fields and strings. Six dimensions is the maximal
dimension in which a superconformal field theory can exist \cite{Nahm:1977tg}. However,
constructive evidence that such theories could exist required input from
string theory \cite{Witten:1995ex, Witten:1995zh, Strominger:1995ac,
WittenSmall, Ganor:1996mu,MorrisonVafaII,Seiberg:1996vs, Seiberg:1996qx, Bershadsky:1996nu,
Brunner:1997gf, Blum:1997fw, Aspinwall:1997ye, Intriligator:1997dh, Hanany:1997gh}. Additionally, the constitutive elements of these
theories involve tensionless strings, but are nevertheless still governed by
the rules of local quantum field theory. Finally, the worldvolume theory of
M5-brane probes of geometry are governed by such 6D SCFTs, so determining any
details on the microscopic structure of such theories would constitute a significant
advance in our understanding of M-theory.

Our aim in this paper will be to give an explicit list of all 6D SCFTs. More
precisely, we shall enumerate all possible ways of manufacturing a 6D SCFT
both from a bottom up and top down perspective. An important aspect of our
classification is that the bottom up and top down constraints are more or less isomorphic, so we can freely interchange our terminology, though at the present the top down perspective seems
to have some additional ingredients which have yet to be translated into purely field
theoretic statements. Once this rather small number of extra ingredients are properly
translated into field theoretic terms, we envision that the top down perspective will
be viewed as a tool to organize the classification rather than to impose
extra conditions.\footnote{Analogous progress in narrowing the gap between
the two approaches was made in the context of global F-theory models in
\cite{Kumar:2009us,
Kumar:2009ae,
mapping,
Kumar:2010ru}.}

From a bottom up perspective, the plan will be to pass onto the tensor branch
for any candidate SCFT. In this limit, the resulting low energy effective
field theory is governed by a rather conventional 6D theory. In this 6D
theory, we must demand that all discrete and continuous anomalies can be
cancelled. Additionally, to have an SCFT, we must be able to simultaneously
tune all moduli of the tensor branch to zero.

From a top down perspective, our plan will be to enumerate all possible
compactifications of F-theory which can lead to a 6D SCFT. To generate a 6D
theory with eight real supercharges, we consider all possible F-theory
backgrounds of the form $\mathbb{R}^{5,1} \times X$, where $X$ is a
non-compact elliptically fibered threefold with a non-compact base $B$. To
reach a 6D SCFT, we take curves in the base $B$ and contract them to zero
size. D3-branes wrapped over these curves correspond to strings in the 6D
effective field theory, and shrinking them to zero corresponds to the
tensionless limit of these strings.

The F-theory description also suggests a natural strategy for enumerating 6D
SCFTs. First, we list all non-compact bases $B$ for which there exists an
elliptic fibration $X \rightarrow B$ such that  the curves in the base
can simultaneously contract
to zero size. In field theory terms, this is the condition that we can
simultaneously tune the scalars of all tensor multiplets back to the origin of
the SCFT. Next, we ask what sorts of elliptic fibrations can be supported over
each base, compatible with the condition that we have a non-compact
Calabi-Yau (i.e., what are the allowed types of wrapped F-theory seven-branes). Other possible ingredients, such as T-branes, turn out to already be covered by these considerations and do not seem to be necessary for achieving
a full classification of 6D SCFTs.

As it will form the core of our analysis, let us now describe in more detail
our procedure for building SCFTs. The non-compact bases of relevance to us
will be those in which all compact curves can simultaneously contract to zero
size. This means that each of these curves must be a $\mathbb{P}^{1}$.
Moreover, as found in \cite{Heckman:2013pva}, the condition that we maintain an elliptically fibered Calabi-Yau
implies that pairs of distinct curves can only intersect once, and that the full
intersection pattern must fill out a tree (i.e., there are no closed loops in
the graph). To specify a base geometry, it therefore suffices to list the
self-intersection of each $\mathbb{P}^{1}$, as well as the overall
intersection matrix for these compact curves.

In the context of F-theory, however, we must also demand that any candidate
configuration of such $\mathbb{P}^{1}$'s can serve as a base for an elliptic
Calabi-Yau. An important result from \cite{Morrison:2012np} states that all such
bases are built up from a small number of building blocks known as
non-Higgsable clusters (NHCs), together with ADE configurations of $-2$ curves
and $-1$ curves to join such clusters together. The non-Higgsable clusters consist of
collections of up to three curves where at least one curve has self-intersection between
$-3$ and $-12$. For each cluster, there is a \textit{minimal} singular behavior for
the elliptic fibration, and a corresponding minimal gauge algebra associated
with each curve\footnote{In one of the non-Higgsable clusters, there is a curve with no associated gauge symmetry. The same phenomenon occurs in other cases with non-minimal singularity types.} in the configuration. These clusters and ADE configurations
can be combined by a
pairwise ``gluing'' with a curve of self-intersection $-1$. Iterating this
procedure leads to a large number of possible bases.

Given a consistent base, we can then ask whether we can adjust the \textit{minimal} singularity type to reach
a more singular geometry with the same base. In the effective field theory on
the tensor branch, this corresponds to increasing the rank of the gauge
algebra, and also incorporating additional matter fields. Giving vevs to these
matter fields then induces a flow back to the original minimal configuration.

\begin{figure}[ptb!]
\begin{center}
\includegraphics[trim=20mm 160mm 20mm 60mm,clip,width=120mm]{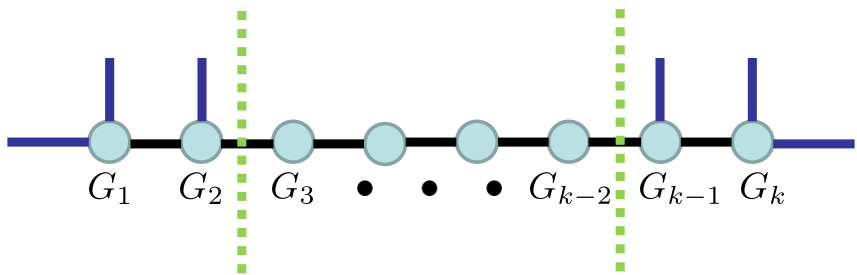}
\end{center}
\caption{The general structure of the base geometry of a 6D SCFT. This geometry is constructed from a number of curves (i.e., circles / nodes) which always have at least a D/E-type gauge symmetry supported over them. These nodes are joined together by ``links'' which are composed of sequences of curves which minimally support no D/E-type gauge symmetry. These links are also SCFTs, so that a base can even have no nodes at all. A striking consequence of our classification is that these bases always have the structure of a single line, with only a small amount of decoration on the two leftmost and rightmost nodes.}%
\label{generalstructure}
\end{figure}

The NHCs and ADE configurations define a list of ``atoms" for generating 6D SCFTs. These atoms join together (by bonding via the $-1$ curves) to form more elaborate radicals and molecules.
In fact, once we start building up such molecules, we can ask whether they can in turn bind
to form additional structures. We indeed find that this is often the case, but that
the resulting structures always take the form of linear chains, with only a small amount of decoration near the ends. For a schematic
depiction of the resulting structure, see figure \ref{generalstructure}.

Along these lines, we give an explicit classification of \textit{all} base
geometries. Quite surprisingly, these base geometries are really just linear chains of curves with a small amount of decoration on
the ends. Moreover, these linear chains have the structure of a generalized quiver in
which certain curves (i.e., ``atoms'') function as the nodes, and other collections (i.e., some of our ``radicals'' and ``molecules'')
serve as links connecting these nodes. An interesting feature of these structures is that the minimal gauge algebra over the nodes
is always a D/E-type algebra, while the links are always composed of curves each one of which has a minimal gauge algebra which is either
empty, or supports a non-simply laced algebra.

On the tensor branch, all of the base geometries can be viewed as generalized quivers. The
nodes of these quivers specify DE-type gauge groups, and the links between
these nodes correspond to the superconformal matter of references \cite{DelZotto:2014hpa, Heckman:2014qba}.\footnote{Note that a
base may contain no nodes at all, i.e. it may be compsed of just links.}
For these DE-type nodes, we also find that there is a partial ordering
constraint on the ranks of these groups. This can be phrased in terms of a
nested sequence of containment relations for $k$ such nodes:%
\begin{equation}
G_{1}\subseteq...\subseteq G_{m}\supseteq...\supseteq G_{k},
\end{equation}
where $G_{m}$ denotes the middle or \textquotedblleft
maximal\textquotedblright\ rank gauge group in the sequence. Similar nested
containment relations have been observed in the context of 6D SCFTs for the
classical groups \cite{Intriligator:1997dh, Hanany:1997gh, Brunner:1997gf} (see also \cite{Gaiotto:2014lca, DelZotto:2014hpa}).

In some cases, the non-D/E-type molecules cannot bind to any other structures. These ``noble molecules'' include some well-known
6D SCFTs such as the D- and E-type $(2,0)$ theories. These theories are realized in F-theory by a configuration of $-2$ curves which intersect according to the corresponding Dynkin diagram. Interestingly, the condition that we get an SCFT means they cannot connect to any other NHCs. Part of our classification also includes cataloguing a list of all such noble molecules.

After whittling away at the possibilities in
this way, we determine all possible links, all possible configurations of nodes, and all possible ways to combine these elements
to form base geometries. Since the combinatorics can become slightly unwieldy,
we collect these data in a set of companion \texttt{Mathematica} files. To
complete the classification of such F-theory geometries, we then turn to a
systematic analysis of ways in which the elliptic fiber can be enhanced. Here,
we find that the options are typically quite limited.
Putting these elements together, we arrive at a classification of all
possible non-compact F-theory backgrounds which can generate a 6D SCFT.

The fact that all of our theories have an essentially linear structure is
rather striking, and is also what is encountered in certain M-theory
constructions of 6D SCFTs. For example, many M-theory realizations of 6D SCFTs
involve M5-branes probing the ADE singularity of the background $\mathbb{R}%
^{5,1}\times\mathbb{R}_{\bot}\times\mathbb{C}^{2}/\Gamma$. On the tensor
branch, the M5-branes separate, and correspond to domain walls of the 7D Super
Yang-Mills theory generated by the ADE singularity. The other way in which
such linear chains occur is via M5-branes probing an ADE singularity near
a Ho\v{r}ava-Witten $E_8$ nine-brane.

In both of these cases, there are additional \textquotedblleft boundary
data\textquotedblright, which lead to additional theories. It
is natural to conjecture that all of these boundary data are
captured by purely \textit{geometric} data of the corresponding F-theory
background. We shall indeed present rather convincing evidence that this is indeed
the case. In particular, we will see that the boundary data of a small instanton configuration in the aforementioned setup, which are known to be in one-to-one correspondence with homomorphisms from $\Gamma$ to $E_8$, are also in one-to-one correspondence with a specific subset of bases and fiber decorations in the F-theory setup.  We verify this correspondence in some highly nontrivial cases where we are able to compare with a detailed study in the mathematics literature of embeddings of finite groups into the Lie group $E_8$ \cite{FREY}.
In other words, these boundary data are actually redundant and are already fully accounted
for by geometric phases of the theory, giving us reason to believe that our classification is complete.

The rest of this paper is organized as follows. First, in section \ref{sec:BOTUP} we
present a brief review of necessary constraints required to reach a 6D
SCFT, from a bottom up perspective. Then, we present the top down, i.e., F-theory
perspective in section \ref{sec:TOPDOWN}. Importantly, nearly all of the F-theory conditions have
analogues in field theory. We then briefly summarize in section \ref{sec:STRATEGY} our strategy for classifying all 6D SCFTs.
After this is in place, we turn in section \ref{sec:BASE} to the first element of our
classification, explicitly determining the structure of all possible base
geometries. Next, in section \ref{sec:FIBERS} we turn to the possible ways to enhance the
fiber type of these geometries. This will constitute a full classification of
possible Calabi-Yau geometries which can support a 6D SCFT. We then present in
section \ref{sec:BOUNDARY} evidence that all of the boundary data for these theories are
actually captured by purely geometric data on the tensor
branch, including a detailed comparison with embeddings of certain finite groups into $E_8$. In section \ref{sec:CONC} we present our conclusions and avenues of future
investigation. Various technical aspects of the classification are deferred to
a set of Appendices, as well as companion \texttt{Mathematica} scripts.

\section{6D\ SCFTs from the Bottom Up \label{sec:BOTUP}}

The strongest evidence for the existence of 6D\ SCFTs come from string
constructions. Nevertheless, the basic elements of these theories can often be phrased in
purely field theoretic terms. This in turn leads to a number of consistency
conditions which must be satisfied for any putative 6D\ SCFT. In this section, we review these bottom up consistency conditions.\footnote{For a review of consistency conditions for 6D supergravity theories, see, e.g., \cite{Taylor:2011wt} and references therein.}

As preliminary comments, we will be dealing with a superconformal theory in
six dimensions. That means operators of our theory must transform in
representations of the $\mathfrak{so}(6,2)$ conformal algebra, and also, that
our theory has $8$ real supercharges $Q$ and $8$ real supercharges $S$ (their
superconformal partners). The spinors assemble to give us a chiral
$\mathcal{N}=(1,0)$ supersymmetry in six dimensions.

All of the theories we shall encounter have a tensor branch. Recall that in a
6D\ theory, the bosonic content of a tensor multiplet consists of a single
real scalar, and a two-form potential with an anti-self-dual three-form field strength.
The rest of the supermultiplet is filled out by fermions. We move onto the tensor branch by activating a vev for
the scalar of this multiplet. On this branch, string-like excitations which couple to the two-form develop
a non-zero tension, which vanishes upon passing back to the origin.

Passing onto the tensor branch, we have a six-dimensional effective field
theory with a UV\ cutoff $\Lambda_{UV}$. Provided we keep the vevs of the
tensor branch scalars below $\Lambda_{UV}$, this description is valid. This
effective field theory may include various gauge groups and matter fields,
but may also include dynamical tensors which only \textquotedblleft come to
life\textquotedblright\ near the conformal fixed point. A necessary condition for anomaly cancellation is that each
simple gauge group factor must come with a corresponding tensor multiplet.
Indeed, the vev of the real scalar controls the value of the gauge coupling.
There can, however, also be tensor multiplets which are not associated with a
gauge theory.

In more detail, the bosonic content of a tensor multiplet consists of a real scalar $s$, and a two-form
potential $B_{\mu \nu}^{-}$, with an anti-self-dual field strength. Including the
(decoupled) gravity multiplet with its two-form potential $B_{\mu \nu}^+$, the two-form
potentials rotate in the vector representation of $SO(1,T)$, and the scalars
$s_{1},...,s_{T}$ provide local coordinates on the coset space
$SO(1,T) / SO(T)$.\footnote{The global topology of the tensor multiplet moduli
space may be quotiented by a further discrete group action.} Quantization of
charge imposes the condition that there is an integral lattice of BPS\ charges for our strings
$\Lambda_{string} \subset\mathbb{R}^{1,T}$. Geometrically, this lattice specifies two-cycles
in the base geometry of an F-theory compactification, so that on a smooth base (i.e., one with no collapsed two-cycles),
we have $H_2(B , \mathbb{Z}) = \Lambda_{string}$.\footnote{For further discussion on the case
with orbifold singularities, see reference \cite{DelZotto:2014fia}.} Further, the dot product is just the intersection
pairing. We will shortly argue that for an SCFT, there are
further restrictions on the form of this matrix.

On the tensor branch of any putative 6D SCFT, there are two constraints which must be
satisfied. First, we must satisfy 6D\ anomaly cancellation for both the
discrete and continuous gauge symmetry factors. Secondly, we must ensure that it is
indeed possible to reach the origin of the tensor branch. We will
shortly see that both of these conditions have clear analogues
in the F-theory construction, and come from demanding consistency of the
elliptic fibration, and the ability to reach a conformal fixed point by
simultaneously contracting curves in the base. We now turn to a more detailed
discussion of each of these constraints.

\subsection{Anomaly Cancellation}

Anomaly cancellation serves as a powerful constraint on the consistency of any
low energy effective field theory. It is particularly
stringent for chiral theories in six dimensions (see, e.g., \cite{Green:1984bx,
Nishino:1985xp, Sagnotti:1992qw, Schwarz:1996, Sadov:1996zm, Kumar:2010ru}). For
continuous anomalies, we must consider box diagrams with four external
insertions of a symmetry current. Cancellation of the anomaly can be arranged
provided the anomaly factorizes into a perfect square. Indeed, in this case,
exchange of a tensor can cancel this box diagram via the Green-Schwarz
mechanism. For discrete anomalies, we must ensure that the matter content of a
corresponding gauge theory appears in appropriate (half-) integer multiples.

\subsubsection{Continuous Anomalies \label{ssec:CONTanom}}

To begin, we ask what constraints are imposed on a 6D\ gauge theory which has
been decoupled from gravity. We can view this as a necessary condition which
must be satisfied for an SCFT on its tensor branch.

Along these lines, we assume that we have a gauge theory consisting of gauge
groups $G_{a}$, and with matter fields transforming in representations of
$G_{a}$. Since we have a gauge group, we can consider the external current
$J_{a}$ associated with such a factor. For a non-abelian gauge theory, anomaly
cancellation means that these external currents must come in pairs, so in the
four-point amplitude, we can restrict attention to two insertions of $J_{a}$
and two insertions of $J_{b}$, where $a$ and $b$ label two gauge group factors
in our list. When $a\neq b$, we shall sometimes refer to this as a
\textquotedblleft mixed anomaly\textquotedblright.

For a representation $R$ of some gauge group $G$, we introduce $Ind_{R}$,
$x_{R}$, and $y_{R}$. In our conventions, these are related to the quadratic
and quartic Casimirs of the group according to:
\begin{equation}
Tr_{R}F^{2}=Ind_{R}trF^{2}\text{ \ \ and \ \ }Tr_{R}F^{4}=x_{R}trF^{4}+
y_{R} (trF^{2})^{2},
\end{equation}
where $tr$ indicates a trace in the defining representation of the
group.\footnote{For $SU(N)$ and $Sp(N)$, this is simply the fundamental
representation. For $SO(5)$ and $SO(6)$ (where there can be an accidental isomorphism with another
classical algebra series) the spinor representations are the defining
representations. For $SO(N)$, $N\geq7$, the fundamental (vector)
representation is the defining representation, but it is normalized to have an
additional factor of 2, so that $trF^{2}=\frac{1}{2}Tr_{f}F^{2}$,
$trF^{4}=\frac{1}{2}Tr_{f}F^{4}$. In other words, $Ind_{f}=2$, $x_{f}=2$,
$y_{f}=0$.}

For four external currents of the same gauge group factor $J_{a}$, the
constraints from anomaly cancellation impose the conditions (see,
e.g., \cite{Bershadsky:1997sb, Kumar:2010ru, Grassi:2011hq, Park:2011ji}):%
\begin{align}
Ind_{Adj_{a}}-\sum_{R}{Ind_{R_{a}}n_{R_{a}}}  &  =6(10-n)\frac{\Omega_{ij}a^i b_a^j}{\Omega_{ij}a^i a^j}\\
y_{Adj}-\sum_{R}{y_{R}n_{R}}  &  =-3(10-n)\frac{\Omega_{ij}b_a^i b_a^j}{\Omega_{ij}a^i a^j}\\
x_{Adj}-\sum_{R}{x_{R}n_{R}}  &  =0,
\end{align}
where $\Omega_{ij}$ is the natural metric on the $n$-dimensional space of antisymmetric tensors, and  $a^i$ and $b_a^i$ come from decomposing the 8-form anomaly $I_8$:%
\begin{equation}
I_8 = \frac{1}{2} \Omega_{ij} X^i X^j\,,~~~~X^i = \frac{1}{2}a^i tr R^2 + 2 b_a^i tr F_a^2.
\end{equation}
Translating from the field theory to the F-theory picture, these conditions become:%
\begin{align}
Ind_{Adj_{a}}-\sum_{R}{Ind_{R_{a}}n_{R_{a}}}    &  =6(\overrightarrow{v}_{K}\cdot
\overrightarrow{v}_{a})\\
y_{Adj}-\sum_{R}{y_{R}n_{R}}  &  =-3(\overrightarrow{v}_{a}\cdot
\overrightarrow{v}_{a})\\
x_{Adj}-\sum_{R}{x_{R}n_{R}} &  =0,
\end{align}
where we have introduced a vector $\overrightarrow{v}_{K} \in \Lambda_{string} \subset \mathbb{R}^{1,T}$, which in the
geometry is identified with the canonical class.\footnote{Indeed, an alternative presentation of the first equation is that the
righthand side is equal to $2 g_{a} - 2 - \overrightarrow{v}_{a} \cdot \overrightarrow{v}_{a}$, where $g_{a}$ is the ``genus'' associated with the
tensor multiplet. For an SCFT, $g_{a}$ will always be zero.} In the above $Adj_{a}$ refers to the adjoint
representation of the gauge group $G_{a}$, and ${n_{R_{a}}}$ refers to the
number of hypermultiplets in a representation $R_{a}$. Additionally, we have
the constraint from mixed anomalies, i.e., where we have two distinct external
currents:%
\begin{equation}
\sum_{R_{a},S_{b}}Ind_{R_{a}}Ind_{S_{b}}n_{R_{a},S_{b}}=\overrightarrow{v}%
_{a}\cdot\overrightarrow{v}_{b},
\end{equation}
where $R_{a}$ and $S_{b}$ refer to representations of $G_{a}$ and $G_{b}$, respectively.

\subsubsection{Discrete Anomalies}

In addition to these continuous anomaly constraints, there can in some cases
also be constraints from discrete anomalies \cite{Bershadsky:1997sb}. Much as in the four-dimensional
case \cite{Witten:1982fp}, these constraints come about from the condition that the
overall phase appearing in the path integral is well-defined. A non-trivial
constraint appears whenever $\pi_{6}(G)$ is non-trivial. It so happens that
this only occurs for the gauge groups $SU(2)$, $SU(3)$ and $G_{2}$, where it
is respectively $%
\mathbb{Z}
_{12}$, $%
\mathbb{Z}
_{6}$ and $%
\mathbb{Z}
_{3}$. Restricting to $SU(2)$ theories with just doublets, $SU(3)$ theories
with matter in the $\mathbf{3}$ and $\mathbf{6}$, and $G_{2}$ theories with
matter in the $\mathbf{7}$, we have the constraints \cite{Bershadsky:1997sb}:
\begin{align}
SU(2) &  :2(n_{\mathbf{2}}-4)\equiv0\text{ }\operatorname{mod}12\\
SU(3) &  :n_{\mathbf{3}}-n_{\mathbf{6}}\equiv0\text{ }\operatorname{mod}6\\
G_{2} &  :n_{\mathbf{7}}-1\equiv0\text{ }\operatorname{mod}3,
\end{align}
where the particular integer we mod out by is dictated by the homotopy group.
Observe that for $SU(2)$, we allow for the possibility that $n_{\mathbf{2}}$
is a half-integer. This can occur because the doublet is a pseudo-real
representation, so we can have a half-hypermultiplet.

\subsection{Motion to the SCFT Point}

In the above, we derived some necessary conditions to make sense of a 6D\ SCFT
on its tensor branch. Now, to really have a 6D\ SCFT, we need to proceed back
to the origin of the tensor branch.
The positive-definiteness of
the metric for the scalars in the tensor
multiplets  imposes the condition that the matrix:%
\begin{equation}
A_{ab}=-\overrightarrow{v}_{a}\cdot\overrightarrow{v}_{b}%
\end{equation}
is positive definite. Note that this condition is specific to requiring the existence of an SCFT decoupled from gravity,
and need not be satisfied in a general 6D theory.

\section{6D\ SCFTs from the Top Down \label{sec:TOPDOWN}}

In this section we turn to the F-theory realization of 6D SCFTs. F-theory provides
a formulation which systematically enumerates possible tensor branches. It is therefore ideally suited
for the purposes of classification.

An F-theory compactification can be defined by starting with M-theory on an
elliptically fibered Calabi-Yau threefold $\widetilde{X}$ over a non-compact
base $B$. To reach the F-theory model, we contract the elliptic fiber to zero
size, arriving at F-theory on the base $B$ with axio-dilaton profile
determined by the fibration $\widetilde{X}\rightarrow B$.
We assume that all fibers of $\widetilde{X}\rightarrow B$ are one-dimensional;
this can always be achieved by blowing up $B$ if necessary \cite{Grassi91}.

Now, to reach a 6D SCFT from such an F-theory model, we must simultaneously contract curves in the base
$B$ to zero size. The reason for this is that D3-branes wrap these curves,
producing effective strings in the 6D effective theory. By shrinking these
curves to zero size, we reach a limit where additional light degrees of
freedom contribute to the theory. To get a 6D SCFT, this contraction must be possible at finite
distance in the Calabi-Yau moduli space. Geometrically, the
condition that we can contract a curve in the base in this way means they are
all $\mathbb{P}^{1}$'s. Labelling these curves as $\Sigma_{i}$, the main
condition for contracting such curves is that the adjacency matrix
\begin{equation}
A_{ab} = - \Sigma_{a} \cdot\Sigma_{b}%
\end{equation}
be positive definite.

As explained in \cite{Heckman:2013pva}, for a 6D SCFT, each such base consists of a connected configuration of curves built from (a) ADE configurations of $-2$ curves, and
(b)
configurations of curves known as ``non-Higgsable clusters'' \cite{Morrison:2012np}, possibly joined by
curves of self-intersection $-1$.  The geometry of elliptic fibrations
then helps to determine
the possible ways that these NHCs and ADE configurations can be joined
together via $-1$ curves.

Now, for each cluster, there is a \textit{minimal} singularity type for the
elliptic fiber. This is found by considering the minimal Weierstrass model:
\begin{equation}
y^{2}=x^{3}+fx+g
\end{equation}
with $f$ and $g$ sections of $\mathcal{O}_{B}(-4K_{B})$ and $\mathcal{O}%
_{B}(-6K_{B})$. The minimal order of vanishing for $f$ and $g$ over each
compact curve dictates the possible appearance of a singular fiber
\cite{Taylor:2011wt}. We list these clusters as a sequence of integers
$n_{1},...,n_{k}$ where $-n_{j}=\Sigma_{j}\cdot\Sigma_{j}$, and the sequence
indicates which curves intersect. The full list of NHCs is:%
\begin{align} \label{eq:onecurve}
\text{One Curve}  &  \text{: }n\text{ \ \ for \ \ }2\leq n\leq8\text{ \ \ or \ \ }n=12\\
\text{Two Curves}  &  \text{: }2,3\text{ \ \ and \ \ }2,2\\
\text{Three Curves}  &  \text{: }2,2,2\text{ \ \ and \ \ }2,2,3\text{ \ \ and
\ \ }2,3,2.
\end{align}
In addition to these NHCs, we also have the ADE graphs of just $-2$ curves.
When the context is clear, we shall sometimes omit the commas between these
integers. For each such NHC, there is a corresponding minimal order of
vanishing for $f$ and $g$ of the Weierstrass model. This order of vanishing
then translates to a gauge algebra, and matter content on the curve.

A single curve of self-intersection $-9$, $-10$ or $-11$ can also occur,
but the associated fibration $\widetilde{X}\rightarrow B$ will have
some two-dimensional fibers; in heterotic language, we say that the
theory with self-intersection $-n$ has $12-n$ small instantons.
To cure this, $12-n$ points along the
curve of self-intersection $-n$ must be blown up to yield a curve
of self-intersection $-12$.  It will be convenient for our combinatorial
analysis to treat curves these self-intersections in parallel with
the other cases, i.e., to replace \eqref{eq:onecurve} with
\begin{equation} \label{eq:onecurvebis}
\text{One Curve}    \text{: }n\text{ \ \ for \ \ }2\leq n\leq12 ,
\end{equation}
and to remember that $12-n$ blowups must be done when $n= 9$, $10$, or $11$.

The corresponding gauge algebra for each cluster is dictated purely by the
self-intersection number in the base. For four curves and above, the base is
already Calabi-Yau, so the fibration can be trivial. For one to three curves,
a gauge algebra, and sometimes matter are also possible. Here is the list of
gauge algebras and matter fields for three curves or less:%
\begin{align}
&
\begin{tabular}
[c]{ccccccccccc}%
Matter & $\{\}$ & $\{\}$ & $\{\}$ & $\{\}$ & $\frac{1}{2}$ $\mathbf{56}$ & $\{\}$ & $\mathbf{I}^{\oplus 3}$ & $\mathbf{I}^{\oplus 2}$ & $\mathbf{I}^{\oplus 1}$ & $\{\}$\\
Algebra & $\mathfrak{su}_{3}$ & $\mathfrak{so}_{8}$ & $\mathfrak{f}_{4}$ &
$\mathfrak{e}_{6}$ & $\mathfrak{e}_{7}$ & $\mathfrak{e}_{7}$ & $\mathfrak{e}%
_{8}$ & $\mathfrak{e}_{8}$ & $\mathfrak{e}_{8}$ & $\mathfrak{e}_{8}$\\
Cluster & $3$ & $4$ & $5$ & $6$ & $7$ & $8$ & $9$  & $10$ & $11$ & $12$
\end{tabular}
\\
&
\begin{tabular}
[c]{cccc}%
Matter & $\frac{1}{2}$ $(\mathbf{2},\mathbf{7}+\mathbf{1})$ & $\frac{1}{2}$
$(\mathbf{2},\mathbf{8,1})\oplus\frac{1}{2}$ $(\mathbf{1}%
,\mathbf{8,2})$ & $\frac{1}{2}$ $(\mathbf{1},\mathbf{2},\mathbf{1}) \oplus \frac{1}{2}$ $(\mathbf{1},\mathbf{2},\mathbf{7})$\\
Algebra & $\mathfrak{su}_{2}\oplus\mathfrak{g}_{2}$ & $\mathfrak{su}_{2}%
\oplus\mathfrak{so}_{7}\oplus\mathfrak{su}_{2}$ & $\{0\} \oplus\mathfrak{sp}_{1}%
\oplus\mathfrak{g}_{2}$\\
Cluster & $2,3$ & $2,3,2$ & $2,2,3$
\end{tabular}
\\
&
\end{align}
where in the above, we have emphasized the difference between $\mathfrak{sp}%
_{1}\simeq\mathfrak{su}_{2}$ and $\mathfrak{su}_{2}$, since the $\mathfrak{sp}%
_{1}$ case arises from monodromy in the elliptic fiber. In the above, the notation $\mathbf{I}^{\oplus k}$
for the $\mathfrak{e}_8$ algebras refers to the presence (in heterotic language) of $k$ small instantons. In F-theory, these arise from
the collision of the $\mathfrak{e}_{8}$ locus with $k$ components of the discriminant locus, each of which supports an $I_1$ fiber.\footnote{Technically,
all such collisions must be blown up to obtain a fibration $\widetilde{X}\rightarrow B$ all of whose fibers are one-dimensional.}  For the clusters with
more than one curve, there is a summand in the algebra corresponding
to each curve, although in the 223 case one of those summands is trivial
(denoted by $\{0\}$ above).  Note that there is still a tensor multiplet
associated to that curve even though there is no gauge algebra.

Now, to put together more general base geometries, we take these NHCs as well as configurations of $-2$ curves and
insert curves of self-intersection $-1$ in between them. To get a CFT, a
number of conditions must be met \cite{Heckman:2013pva}:

\begin{itemize}
\item A $-1$ curve can intersect at most two NHCs. Otherwise, upon blowing down we violate a
condition for normal crossing, and the curves cannot all simultaneously
contract to zero size.

\item For a pair of curves $\Sigma_{L}$ and $\Sigma_{R}$ which intersect the
$-1$ curve, there is a corresponding \textit{minimal} gauge algebra
$\mathfrak{g}_{L}$ and $\mathfrak{g}_{R}$ supported on each curve. A
consistent elliptic model requires that this minimal algebra satisfies the
condition $\mathfrak{g}_{L}\times\mathfrak{g}_{R}\subset\mathfrak{e}_{8}$.
\end{itemize}

Now, as we have already mentioned several times, the minimal gauge algebra on
curve of the base is dictated by its self-intersection \cite{Morrison:2012np}.
In some cases, we can make this fiber more singular, for example, by introducing additional
seven-branes into the system. The main condition we need to check is that
doing this continues to retain a balancing of all brane tensions, or in
geometric terms, that an elliptic fibration satisfying the Calabi-Yau
condition still exists.

Once we enhance the fiber type, the gluing condition used to construct
consistent base geometries must be generalized. Geometrically, the main
condition we need to satisfy is that the order of vanishing for $f$ and $g$ is
such that we can even define a minimal Weierstrass model in the first place.
In practice, this means that the product algebra of two neighbors must fit
inside either an infinite classical series of $\mathfrak{su}$-, $\mathfrak{sp}$- or $\mathfrak{so}$-type, or
must fit inside a subalgebra of $\mathfrak{e}_{8}$.

Enhancing a fiber above the minimal type also means that we both enhance the
gauge algebra on a curve, and also introduce additional matter fields charged
under this algebra. This must be so, because we need to be able to Higgs the theory back down
to the minimal fiber type.

To give an example, consider the case of a single $-5$ curve. This minimally
supports an $\mathfrak{f}_{4}$ algebra. However, we can enhance this to an
$\mathfrak{e}_{6}$ theory with a single hypermultiplet in the $\mathbf{27}$.
Giving a vev to the $\mathbf{27}$ initiates a breaking pattern back to the
minimal gauge algebra.

Now, for all of the curves of self-intersection $-5$ or less, we have an
exceptional algebra, so any enhancement we do must be a subalgebra of
$\mathfrak{e}_{8}$. Indeed, we can in principle enhance the fiber all the way
to an $\mathfrak{e}_{8}$ algebra. For $\mathfrak{e}_{7}$ and its subalgebras,
there is extra matter which determines the corresponding unfolding back down
to the minimal symmetry algebra. For the case of $\mathfrak{e}_{8}$, there is
some number of small instantons (in heterotic language). Dissolving these
small instantons again initiates an unfolding to a lower symmetry algebra.

The case of a $-3$ curve is also rather special since it arises in F-theory
from a non-perturbative bound state of seven-branes with different $(p,q)$ type.
This means it is better thought of as part of the exceptional series.

For the remaining curves, i.e., those of self-intersection $-4$,$~-2$ and $-1$,
it is helpful to first study what algebras \textit{cannot} occur. This is
basically a consequence of the condition that we need to be able to Higgs the
theory down to the minimal singularity type. For a $-1$ curve, we find no
restrictions on $\mathfrak{su}$-type, $\mathfrak{sp}$-type gauge algebras, or exceptional gauge algebras, though high rank $\mathfrak{so}$-type gauge algebras are excluded because any configuration of matter will yield an anomalous gauge theory. For a $-2$ curve, we
find that $\mathfrak{sp}$-type algebras are excluded along with high rank $\mathfrak{so}$-type gauge groups. For a $-4$ curve, we find that both
$\mathfrak{su}$- and $\mathfrak{sp}$-type algebras are excluded, i.e., only $\mathfrak{so}$ and exceptional
algebras are possible.

Let us now examine what sorts of gauge algebra enhancements can in fact occur.
For a $-4$ curve, a further enhancement in the fiber takes us either to a
subalgebra of $\mathfrak{e}_{8}$, or to a higher rank $\mathfrak{so}$-type algebra.
Moreover, to have the option to unfold back down to an $\mathfrak{so}_{8}$
algebra, the available matter content on this enhanced $\mathfrak{so}$-type algebra is
also quite limited. Using the collision rules of \cite{KatzVafa}, we deduce that we
either must unfold from inside $\mathfrak{e}_{8}$ (if the rank of the
$\mathfrak{so}$-type algebra is low enough), or we unfold from inside a high rank
$\mathfrak{so}$-type algebra. In the latter case, we can only get matter in the
fundamental representation. In the former case, we have a few additional
options which were worked out in \cite{BershadskyPLUS,KatzVafa,Grassi:2011hq}, and consist of spinor
representations of the $\mathfrak{so}$-type algebra.

Turning next to a $-2$ curve, we see that we can enhance to an $\mathfrak{su}$-type
algebra, an $\mathfrak{so}$-type algebra, or continue on through the exceptional series.
Now, we can again ask about the matter content which can be charged under this
gauge algebra. We split our analysis up into whether we embed in a higher rank
classical algebra, or an exceptional one. In the case of the classical
algebras, we can embed in a higher rank $\mathfrak{su}$ or $\mathfrak{so}$ algebra. Again following
the collision rules, we learn that we can have matter in the
fundamental and two-index anti-symmetric representation.

In the case of a $-1$ curve, we see that there are no
restrictions on the algebra which can be supported over the curve.
Additionally, this means that the types of matter fields which can also be
supported at points of these curves all follow from unfolding of either
$\mathfrak{e}_{8}$, or a higher rank $\mathfrak{su}$, $\mathfrak{sp}$ or $\mathfrak{so}$-type algebra.

Quite importantly, this analysis also reveals that the geometric content of self-intersection numbers is not ``extra input'' from the top down construction, but is simply a convenient repackaging of data in the 6D effective field theory. For example,
we can either state that we have a curve of self-intersection $-5$ and an algebra $\mathfrak{e}_{6}$, or equivalently, we can state that
we have an $\mathfrak{e}_6$ algebra with a hypermultiplet in the fundamental representation. We will amplify this point
further when we turn in section \ref{sec:FIBERS} to the consistent ways to decorate the base of an F-theory compactification.

\section{Strategy for Classification \label{sec:STRATEGY}}

In the last two sections we observed that the bottom up constraints on
the construction of 6D\ SCFTs have direct avatars in the top down approach via
compactifications of F-theory. Indeed, compared with other top down methods,
the F-theory approach allows for a clean geometric identification of the
tensor branch which directly mirrors the effective field theory construction.
We shall therefore adhere to this approach in what follows.

Now, to classify possible 6D\ SCFTs via F-theory, we shall proceed in the following steps:

\begin{itemize}
\item Step 1:\ Classify all non-compact base geometries

\item Step 2:\ Classify all ways of enhancing the minimal fiber type of these
geometries without inducing further blowups

\end{itemize}

One might think that a third step--classifying all ways of decorating the theory by boundary data such as T-branes--would also be required. In fact, we will present strong evidence that this is unnecessary, namely these boundary data are
already captured by listing all possible elliptic fibrations.

In the remainder of this section we review some of the geometric tools which
will be useful in performing this classification. This will include some of
the salient elements from the classification obtained in \cite{Heckman:2013pva}.

\subsection{Orbifolds and Endpoints}

In reference \cite{Heckman:2013pva} a coarse classification
of 6D\ SCFTs was presented in which every 6D SCFT is labelled by a discrete subgroup of $U(2)$.
One of the central methods from reference \cite{Heckman:2013pva} that we shall heavily exploit in our
classification of bases is the effect that blowing down a $-1$ curve has on
the self-intersection of other curves in the base of an F-theory geometry. As
explained in more detail in \cite{Heckman:2013pva}, for a sequence of three curves with the $-1$ curve in the middle, the
blowdown of this curve shifts the self-intersections as:%
\begin{equation}
x,1,y\rightarrow(x-1),(y-1).
\end{equation}
That is, we shrink down the $-1$ curve to zero size, which changes the geometry of the base.  It may or may not be possible to  perform a
complex structure deformation to move the blown-down point away to general position. But in either case (i.e., whether the blown-down space is a valid
base for F-theory or not), for the
original base to support a 6D\ SCFT, we need the adjacency matrix to be positive
definite. This is equivalent to checking that the adjacency matrix obtained
after blowing down all $-1$ curves is also positive definite.

Let us also note that in a 6D SCFT, a $-1$ curve can never intersect more than two distinct curves. We shall sometimes refer to this
as the ``normal crossing condition'', as this is the geometric condition which would be violated. One can also
see a cruder version of this statement by considering any adjacency matrix where a $-1$ curve acts as a trivalent vertex in a graph. In this case,
one can proceed on a case by case basis through possible ways to attach extra curves compatible with the gluing condition. In
all cases, the adjacency matrix is no longer positive definite. This is a very important restriction, and means that structures such as the $\mathfrak{su}_3 \times \mathfrak{su}_3 \times \mathfrak{su}_3 \subset \mathfrak{e}_6$ ``trifundamentals'' prevalent in 4D $\mathcal{N} = 2$ theories (see, e.g., \cite{Gaiotto:2009we}) cannot arise.

In fact, we can iterate this procedure of successively blowing down the $-1$ curves one after the other. Doing
so, we get a configuration of curves which all have self-intersection $-n < -1$.
It is then enough to check that this final adjacency matrix is positive
definite. We refer to a configuration of curves obtained in this way as an
\textquotedblleft endpoint\textquotedblright: it is a complex surface,
obtained by blowing down an F-theory base, from which $-1$ curves have been eliminated.  In \cite{Heckman:2013pva} all of these endpoints
were classified, where it was found that they all have the structure of
generalized ADE\ Dynkin diagrams where the self-intersections of some of the
curves can be different from $-2$.

Such configurations are all associated with the resolution of orbifold
singularities $\mathbb{C}^{2}/\Gamma$ for $\Gamma$ a discrete subgroup of
$U(2)$. Given this list, we can also perform a minimal set of resolutions so
that the elliptic fiber of the corresponding Calabi-Yau stays in Kodaira-Tate
form over each curve. Said differently, we get the following coarse
classification of 6D\ SCFTs:

\begin{itemize}
\item 1)\ For every 6D\ SCFT, there is a corresponding discrete subgroup
$\Gamma\subset U(2)$.

\item 2) Call this collection of discrete subgroups $\mathfrak{G}_{SCFT}$.
Then, for each $\Gamma\in\mathfrak{G}_{SCFT}$, there is a canonical SCFT
obtained by performing a minimal set of blowups of $\mathbb{C}^2/\Gamma$
to obtain a valid  base for F-theory.
\end{itemize}

This classification is coarse in the sense that more than one 6D\ SCFT could
have the same endpoint, and thus the same $\Gamma$. For example, the trivial
endpoint, i.e., $\Gamma$ isomorphic to the identity covers all the bases
$1,2...2$, with an unlimited number of $-2$ curves.

From this perspective, one potential strategy to refining this classification
would be to see how many extra blowups can be added to the base, and then, to
check what sorts of non-minimal elliptic fibers can be supported over these choices.

Though this is a viable approach, we shall find it more direct to pursue a
somewhat different approach to the classification of bases and fibers. One
consequence of this alternative approach will be that we recover points 1) and 2) with
little additional effort.

\subsection{The ``Chemistry'' of Classification \label{ssec:EXECUTIVE}}

In this subsection, we present a brief summary of the classification.  Details
are spelled out in the following sections. The basic steps of our classification scheme, and the section
where the details can be found are as follows:
\begin{itemize}

\item Step 1: Classify all Base Geometries (section \ref{sec:BASE}).

\item Step 2: Classify all Fiber Enhancements of the Base Geometry (section \ref{sec:FIBERS}).

\end{itemize}
The content of section \ref{sec:BOUNDARY} will be to argue that all of the possible 6D SCFTs, including decorations by boundary data such as T-branes, are already captured by purely \textit{geometric} data in an F-theory compactification. Let us now discuss in further detail each of these steps.

Consider first the structure of the base geometries. Much as in chemistry, all of the 6D SCFTs we will encounter are built up from a small number of building blocks: the non-Higgsable clusters of reference \cite{Morrison:2012np}, together with ADE graphs consisting of $-2$ curves. These building blocks play the role of ``atoms." They can in turn be joined to other atoms by $-1$ curves. To further facilitate our classification scheme, we shall split these building blocks up into those which are of DE-type, and those which are not:
\begin{align} \label{eq:DE}
\text{DE type}  &  \text{: \ \ }4,6,7,8,9,10,11,12\\
\text{non-DE\ type}  &  \text{: \ \ }3,23,232,223,5, \text{ADE graphs}. \label{eq:non-DE}
\end{align}
Our ``DE'' nomenclature references the \textit{minimal} gauge symmetry supported over the curve (if any).
To build a bigger structure we must interpose a $-1$ curve between two NHCs
and/or ADE graphs.

The first step in our classification scheme will be to give an explicit list of all possible ``radicals'' and ``molecules'' which can be formed by combining only non-DE type curves, together with $-1$ curves. We call the DE-type atoms ``nodes'' and call the radicals and molecules ``links'' since they typically connect to one or more DE-type nodes, linking them together. It turns out that the structure of these links is quite limited, so that there is only one link with a quartic vertex, while the rest have at most one trivalent vertex, or are just a single line of curves. We also find examples of ``noble molecules,'' that is, links which can never attach to a DE-type node. An important feature of these links is that the only configurations which can grow to an arbitrary size are the instanton link $1,2...,2$ and the A- and D-type Dynkin diagram configurations of $-2$ curves.

After listing all possible links, we then turn to the ways that they can attach to the nodes. This is where we encounter many families of 6D SCFTs which can sometimes grow to arbitrary size. That being said, these structures are still remarkably constrained. For example, we find that a configuration of DE type nodes forms at most a single line, i.e., there are no tree-like structures at all for linking together such nodes. Moreover, nearly all such DE type nodes attach to only two links. Only at the two leftmost and rightmost nodes can there be three links attached. Finally, we also determine all possible links which can actually attach two such nodes. We find that in general, the links are of ``minimal type'', i.e., they are the ones which would be expected from performing a minimal resolution of colliding singularities in an F-theory compactification. The non-minimal links only attach between the three leftmost or rightmost nodes of such a configuration of curves. For a schematic depiction of the resulting structures, see figure \ref{generalstructure}. We collect a full list of possible links, as well as possible sequences of DE-type nodes in a set of Appendices. The companion \texttt{Mathematica} notebooks allow the reader to further explore our list of theories.

With the classification of bases in hand, we next turn to step 2: the possible ways that we can make the resulting elliptic fibration more singular whilst still retaining the condition that the fiber over each curve remains in Kodaira-Tate form. Here, the options are so limited that it is typically enough to simply list these conditions for each curve individually. Indeed, the vast majority of our bases admit no enhancement at all. The main lesson from this set of examples is that to get an enhancement of the fiber, we typically need to have a sequence of classical gauge groups. The collection of bases which can support such gauge groups is also rather limited, and makes it possible to sort out the generic fiber enhancement.

A complete classification of 6D SCFTs must include the possible ways to supplement a theory by ``boundary data." An important example of such boundary data are T-branes (see, e.g., \cite{Donagi:2003hh, TBRANES, glueI, glueII, FCFT, Chiou:2011js, HVW, D3gen, Heckman:2012jm, Anderson:2013rka, Collinucci:2014qfa, Collinucci:2014taa}). These are non-abelian intersections of seven-branes which have broken gauge / flavor symmetries, with a singular spectral equation. Additional examples of boundary data include M5-branes probing an $E_8$ wall near an ADE singularity in which there are non-trivial boundary conditions for the instanton. Owing to the fact that such data is not captured directly by the complex equations of an F-theory compactification, it is natural to ask whether this extra data must also be included in a full classification scheme.

We find that these data are redundant. That is, upon moving onto the tensor branch of the theory, we will find that every way of supplementing a theory by T-brane / small-instanton boundary data is already accounted for by a purely \textit{geometric} operation where we enhance the singularity type of the elliptic fiber. A rather striking consequence of this perspective is that this physical picture leads us to a beautiful and completely unexpected classification scheme for homomorphisms from discrete subgroups of $SU(2)$ to $E_8$: the boundary data for instantons probing an orbifold singularity are captured by such homomorphisms, and are in turn (as we show) described by the data of an F-theory compactification! In more detail, we present detailed matches between the resulting flavor symmetries on both sides of the correspondence for the ADE discrete subgroups of $SU(2)$.

Putting together these steps, we arrive at a rather complete picture of how to build a 6D SCFT in F-theory. Further since our ``top down'' constraints can often be phrased in purely effective field theory terms, we are led to conjecture that this is the full set of ways to manufacture an SCFT.

We now proceed to the classification of 6D SCFTs.

\section{Classification of Bases \label{sec:BASE}}

We now turn to the first stage of our classification program: We determine an
explicit list of all possible bases for F-theory geometries. In effective
field theory terms, classifying the bases can be viewed as determining all
configurations of tensor multiplets which can support a 6D\ SCFT (compatible
with the conditions of anomaly cancellation and reaching the origin of the
tensor branch of the moduli space). Further, for each such base, there is a
canonically associated theory. In some cases, there is also the possibility of
enhancing the gauge symmetry over some of the curves. We shall turn to this
further refinement in section \ref{sec:FIBERS}.

The big surprise of this section is how limiting the resulting structures turn
out to be: We find that these bases are essentially just linear chains of
curves, with some decorations on the end. The bulk of the combinatorics is
thus reduced to a classification of these decorations, and how to consistently
combine them with possible linear chains.

To tame the combinatorial chemistry of building bases, we shall introduce some
helpful nomenclature (for a brief review see section \ref{sec:STRATEGY}).
Recall that we view the non-Higgsable clusters together with the ADE graphs as the \textquotedblleft
atoms\textquotedblright\ out of which we build an SCFT. It will prove convenient
to further distinguish these atoms according to the minimal gauge algebra
which they support, as spelled out in eqs.~\eqref{eq:DE}-\eqref{eq:non-DE}
above.
We shall often refer to the DE-type curves as \textquotedblleft
nodes\textquotedblright, and to compounds built solely from the non-DE\ type
curves together with $-1$ curves as \textquotedblleft links\textquotedblright.
(Notice that the simplest link is just a single $-1$ curve itself.)
The utility of this
nomenclature is that all of the DE-type curves attach to one another via such
links. Moreover, this distinction will provide us with a systematic way to
blowdown $-1$ curves: on our way to an endpoint, we will often go to an intermediate point involving just
non-DE\ type curves, and only then consider blowing down a DE-type curve.

A priori, a link could be an arbitrarily complicated structure. We find,
however, that this is not the case. To help collect the possibilities, we
shall refer to an \textquotedblleft$n$-link\textquotedblright\ as one in which
there are precisely $n$ curves of self-intersection $-1$ which only attach to
one curve. That is, they are the places where a potential bond to another atom
/ radical / molecule could occur. Here are some examples of $n$-links:%
\begin{equation}
\text{A }4\text{-link}\text{: }1\overset{1}{\underset{1}{5}}1\text{, A
}3\text{-link}\text{: }1\overset{1}{5}1\text{, A }2\text{-link}\text{:
}151\text{, A }1\text{-link}\text{: }15\text{, A }0\text{-link}\text{: }5.
\end{equation}
Of course, in the case of a $0$-link, it attaches to nothing else. Let us also
note that all of the ADE-type configurations of just $-2$ curves are examples
of $0$-links:%
\begin{equation}
\text{More Examples of }0\text{-links: }2...2\text{, \ \ }2\overset{2}{2}%
...2\text{, \ \ }22\overset{2}{2}22\text{, \ \ }22\overset{2}{2}222\text{,
\ \ }22\overset{2}{2}2222.
\end{equation}
In all cases other than the A-series, we cannot attach a $-1$ curve to any of
these $0$-links. This is because blowing down the $-1$ curve successively
eventually inflicts a blowdown on the trivalent vertex. In the case of the
A-series, attaching a single $-1$ curve is allowed and leads to an instanton link.

As we have already mentioned, the $n$-links with $n\geq1$ can often attach to
various nodes. We shall refer to a link as a \textquotedblleft noble
molecule\textquotedblright\ if it can never attach to a node, and we shall
refer to a link as being \textquotedblleft alkali\textquotedblright\ if it can
only potentially attach to precisely one node.\footnote{The terminology is
borrowed (in bowdlerized form) from chemistry, where the noble gases are
chemically inert, and the alkali elements can typically attach to precisely
one other element. We leave a more detailed set of analogies / metaphors to
the reader well-versed in organic chemistry.}

In some cases, a link which could potentially attach to more than one node may
only be affixed to one. For this reason, it is also helpful to reference a
link as being a \textquotedblleft side link\textquotedblright\ if it only
attaches to one node, and to an \textquotedblleft interior
link\textquotedblright\ as one which attaches to at least two nodes. So in
other words, an alkali link is always a side link, but a link which is
interior can also potentially operate as a side link.

To give an example of how to piece together these ingredients, consider a
collection of $-12$ curves to be our nodes. We can join two such nodes
together via the link consisting of eleven curves: $12231513221$. Using this,
we can string together an arbitrarily long repeating pattern of such nodes:%
\begin{equation}
(12)12231513221(12)12231513221(12)....
\end{equation}
This consists of collections of nodes, i.e., the $-12$ curves, and in between
each pair is an $(E_{8},E_{8})$ link: $12231513221$. This and similar
repeating patterns were noted in \cite{Bershadsky:1996nu,Aspinwall:1997ye,Morrison:2012np}. The
minimal links found here are precisely those of the \textquotedblleft6D
conformal matter\textquotedblright\ studied in \cite{DelZotto:2014hpa, Heckman:2014qba}: They originate as
the minimal conformal sector where two $E_{8}$ singularities intersect in F-theory.

It has likely not escaped the reader that the structure of our base looks
quite a bit like a generalization of a quiver. We will soon find that there
are nested containment relations on these algebras, with the largest rank
simple groups residing in the interior of the configuration of curves. To
further reinforce this concept, we shall often omit the self-intersection of a
curve, and will instead simply reference the \textit{minimal} gauge algebra
supported over a node. Observe that just giving the gauge algebra is not
enough to reconstruct the self-intersection of a DE-type curve. For example,
both the $-7$ and $-8$ curves minimally support an $E_{7}$ gauge symmetry,
while the $-9,-10,-11,-12$ curves all support an $E_{8}$ algebra. We shall
therefore introduce the notation of a \textquotedblleft primed
node\textquotedblright:%
\begin{align}
E_{7}^{\prime}  &  :-7\text{ curve}\\
E_{7}  &  :-8\text{ curve}\\
E_{8}^{\prime\prime\prime}  &  :-9\text{ curve}\\
E_{8}^{\prime\prime}  &  :-10\text{ curve}\\
E_{8}^{\prime}  &  :-11\text{ curve}\\
E_{8}  &  :-12\text{ curve.}%
\end{align}

Having introduced some useful terminology, our plan in the remainder of this
section will be to establish a number of lemmas. With these in place, we will
be able to significantly constrain the structure of a base. In order to
systematically classify the bases, we observe that we can consistently blow
down the $-1$ curves to reach an endpoint for a link. Now, upon performing
this sequence of blowdowns, the self-intersection of each node curve will also
change. This change is uniquely fixed once we specify all the links which are
attached to a given node. To see whether we have a consistent base for an
SCFT, we therefore can first blowdown to the endpoints for the links, and only
then consider blowdowns on each of the nodes. As listing all of the
intermediate curves in a link is often unnecessary (being dictated by the
neighboring structure of the node) we shall often write:%
\begin{equation}
g_{L}\overset{s,t}{\oplus}g_{R}%
\end{equation}
to indicate that we have two nodes $g_{L}$ and $g_{R}$ and a link suspended
between them. The superscript by $s$ and $t$ indicates that upon reaching the
endpoint of the link induces a shift in the self-intersection number by $s$ on
$g_{L}$ and $t$ on $g_{R}$. To distinguish the blowdown of links from a full
blowdown of all curves, we shall sometimes use the notation
$\overset{L}{\rightarrow}$ to indicate that we are blowing down just the
interior links. Here is an example of a link blowdown, first in compressed
notation, and then in expanded notation:%
\begin{equation}
E_{8}\overset{5,5}{\oplus}E_{8}\overset{L}{\rightarrow}77
\end{equation}%
\begin{equation}
(12)12231513221(12)\overset{L}{\rightarrow}77.
\end{equation}

Our first task will therefore be to classify the possible links which can
attach to the nodes of our base. A priori, such links could be a linear chain
of non-DE\ type NHCs, or possibly a tree-shape configuration, for example:%
\begin{equation}
1\overset{1}{5}1.
\end{equation}

In a set of Appendices, we give a full list of all possible links, their
endpoints, as well as the number of blowdowns these links induce on
neighboring nodes. \textit{The key point is that the interior structure of a base is quite limited: The
nodes of the base form a single line, and the type of interior link
is minimal except near the very ends of the base.} The combinatorics of classifying bases
is thus reduced to a small amount of decoration on the ends. Since these
options are in turn completely determined by the constraints collected here
(and in the Appendices)\ \textit{we will classify all bases}.

The rest of this section is organized as follows. First, in subsection
\ref{ssec:basequiver}, we give the final result from the classification of
bases. We also give an overview to the various lemmas, establishing where to
find the relevant material. In subsection \ref{ssec:LINKS}, we establish the
main results on the structure of links. First, we give a list of all possible
links, and then determine constraints on where these links can sit. In nearly
all cases, the \textquotedblleft minimal link\textquotedblright\ is the only
available option. We then turn in subsection \ref{ssec:NODES} to constraints
on the locations of nodes in a base. We find a strong partial ordering
constraint, which effectively cuts down the possibilities to a small number of
places where decoration is possible. Finally, in subsection \ref{ssec:ENDS},
we turn to constraints on how to decorate the ends of a base. In a set of
Appendices and in some companion \texttt{Mathematica} programs we collect the
full list of possible interior sequences, as well as all possible links which
can attach to an end.

\subsection{General Structure of a Base \label{ssec:basequiver}}

In this section we provide a brief overview to the results to follow, which
mainly consist of a set of interlocking lemmas which build towards the final
result. The main outcome from this analysis is that the most general base
takes the form of a linear chain of curves:%
\begin{equation}
S_{0,1}\overset{S_{1}}{g_{1}}L_{1,2}\overset{\mathbf{I}^{\oplus s}}{g_{2}%
}...L_{m-1,m}g_{m}L_{m,m+1}...\overset{\mathbf{I}^{\oplus t}}{g_{k-1}%
}L_{k-1,k}\overset{\mathbf{I}^{\oplus u}}{g_{k}}S_{k,k+1}\text{.}
\label{genBaseQuiver}%
\end{equation}
Here, each of the $g_{i}$ refers to a DE-type node, and the $S$'s and $L$'s
refer to the possibility of attaching respectively a side link or an interior
link. Additionally, the notation $\mathbf{I}^{\oplus s}$ refers to attaching
$s$ small intantons to a curve, that is, a sequence of $s$ curves such as
$1,2...,2$. For all of the interior nodes, i.e., $g_{3},...,g_{k-2}$, we find
that no decoration by a side link is possible. That is, they only attach to
two links. Said differently, the only deviation away from a linear chain of
curves occurs on the two leftmost and rightmost nodes of the base. For
example, only these extremal nodes can support a primed gauge group.

Let us note that here we do not distinguish between one long chain, and
shorter small instanton chains with the same number of total curves, i.e., we
identify $\mathbf{I}^{\oplus s}g\mathbf{I}^{\oplus t}$ and $g\mathbf{I}%
^{\oplus(s+t)}$. For now, this is simply a convenient bookkeeping device,
though we should also note that when there is no further decoration of the
fiber, such configurations turn out to flow to the same SCFT point \cite{Heckman:2013pva}.

Additionally, we shall also find that the instanton links attached in the
interior are always limited to at most two curves, i.e., the configuration
$1,2$ or two marked points, each with a single $-1$ curve.
Moreover, at five nodes and above the only option available is zero or
one instanton.

Another outcome from our analysis is that we have a sequence of partially
ordered gauge groups for the nodes:%

\begin{equation}
G_{1}\subseteq G_{2}\subseteq...\subseteq G_{m}\supseteq...\supseteq
G_{k-1}\supseteq G_{k}\text{.}%
\end{equation}
Moreover, the structure of the base is just a linear chain of curves, up to
some possible decoration which can occur at the ends.

Let us note that some bases may be comprised entirely of side links, that is,
there are no DE-type nodes at all. This covers all of the ADE graphs with just
$-2$ curves. Additionally, there are some more exotic tree-like side links. We
collect all of these possibilities in an Appendix.

Finally, here is an overview of the various elements which go into our general
constraints on the structure of base geometries:

\begin{itemize}
\item In subsection \ref{ssec:LINKS} we derive a number of constraints on the
properties of links. We list all interior links, and introduce the notion of a
minimal and non-minimal link. The full list of possible links is collected in
an Appendix.

\item In subsection \ref{ssec:NODES} we turn to the constraints on the nodes
of a base. The major constraint we discover is that a node can join to a
maximum of two other nodes. In particular, this limits the topology of the
base to a line, with only a small amount of decoration by links at the ends.
We also uncover a \textquotedblleft stability condition\textquotedblright\ on
the minimal gauge algebra supported over each node: In a base these nodes obey
a partial ordering condition such that the largest rank algebras appear in the
interior of the base. We also show that the primed nodes $E_{7}^{\prime}$,
$E_{8}^{\prime\prime\prime}$ and $E_{8}^{\prime\prime}$ can only occur on the
two leftmost and rightmost nodes, while the primed node $E_{8}^{\prime}$ can
appear in the middle of a five node base, but otherwise is also constrained to
the two leftmost and rightmost nodes.

\item In subsection \ref{ssec:ENDS}, we turn to the structure of the end
nodes. We find that a non-minimal link can only attach in between the two
leftmost or rightmost nodes, and so there can be at most two such
non-minimal links. Further, only the two leftmost and rightmost nodes can
support a side link. Additionally, we also show that only the leftmost and
rightmost node can support a side link which is not an instanton link, and
that generically (i.e., at five nodes or more) the instanton link is at best a
single $-1$ curve. Finally, the total number of such side links in a base is
at most three.

\item Putting all of these steps together, in subsection \ref{ssec:summary} we
obtain the general claim that all base geometries take the form of line
(\ref{genBaseQuiver}), namely a single linear chain with a small amount of
decoration on the two leftmost and rightmost nodes.

\item The Appendices fill in some remaining details of the classification of
bases. For example, we give an explicit list of all possible sequences of
nodes, and also all possible links which can attach to these nodes.
\end{itemize}

\subsection{Constraints on Links \label{ssec:LINKS}}

In this subsection we determine constraints on the links which can appear in a
base geometry. The first item of business will be to determine all possible
links which could appear. This is readily dealt with through a computer sweep,
and we collect the results in an Appendix. In this Appendix, we also detail
how such links can attach to nodes of a base geometry. The plan in this
subsection will be to further explain various restrictions on how such links
can attach to nodes in a base. We will show in particular that
all of the interior links are $2$-links. This means,
for example, that a tree-shaped $3$-link can attach to a maximum of one node.

In fact, the most common types of links which we shall encounter are the
\textquotedblleft minimal\textquotedblright\ interior links, and an
\textquotedblleft instanton link\textquotedblright. We refer to a link as
minimal if it is completely determined by performing the minimal number of
blowups between a pair of intersecting seven-branes. For example, the minimal
link for $(E_{6},E_{6})$ is $131$. For a full list of these minimal links, see
Appendix A of reference \cite{DelZotto:2014fia}.

\subsubsection{The Linear $2$-Links \label{ssec:twoLinks}}

As a first step in the classification of bases, we first list all linear
$2$-links. These are $2$-links, which can potentially connect to two of our
nodes. It is convenient to organize all such $2$-links according to the number
of $-5$ curves:%
\begin{align}
&  1\text{, }131\text{, }1231\text{, }12321\text{, }12231\text{,}\\
&  151\text{, }15131\text{, }151321\text{, }1513221\text{,}\\
&  1315131\text{, }13151321\text{, }131513221\text{,}\\
&  123151321\text{, }1231513221\text{, }12231513221\text{,}\\
&  1513151\text{, }15123151\text{, }131513151\text{, }1231513151\text{,
}12231513151.
\end{align}
Beyond two $-5$ curves, we cannot produce a consistent linear $2$-link.

Having collected all possible $2$-links which are linear chains, we now turn
to some of their properties. Of the above possibilities, observe that to have an
interior link, the sequence of curves must begin (resp. end) with a pattern other than $1,5$ (resp. $5,1$). The reason is that the gluing
condition does not allow us to pair even a $-4$ with a $-5$ curve. Of the 2-links which are also interior,
we also see that they all have trivial endpoint. Moreover, blowing down the link also
leads to a fixed number of blowdowns on the nodes attached
to it. In many cases, these data actually allow us to uniquely reconstruct the
corresponding link. For example, we can denote a configuration for the
\textquotedblleft long link\textquotedblright\ by the compressed notation:%
\begin{equation}
g_{L}12231513221g_{R}\simeq g_{L}\overset{5,5}{\oplus}g_{R}\text{.}%
\end{equation}
where the notation indicates that five blowdowns are inflicted to the left,
and five to the right.

Now, to obey the gluing rules, there is always a maximal algebra which can be
attached to a given link. There is a strict hierarchy here:%
\begin{equation}
D\subset E_{6}\subset E_{7}\subset E_{8}\text{,}%
\end{equation}
namely, if it is possible to attach an $E_{8}$ node, then the gluing rule also
allows us to attach an $E_{7}$ node. Note, however, that if we can attach an
$E_{7}$ node, there is no guarantee that we can attach an $E_{8}$ node.\ Here,
we do not distinguish between primed and unprimed groups.

Taking this into account, we have the following list of pairings (see also
Appendix B):%
\begin{gather}
D1D\simeq D\overset{1,1}{\oplus}D\label{DDpair}\\
E_{6}131E_{6}\simeq E_{6}\overset{2,2}{\oplus}E_{6}\\
E_{7}12321E_{6}\simeq E_{7}\overset{3,3}{\oplus}E_{6}\\
E_{7}12321E_{7}\simeq E_{7}\overset{3,3}{\oplus}E_{7}\\
E_{7}1231D\simeq E_{7}\overset{3,2}{\oplus}D\\
E_{8}12231D\simeq E_{8}\overset{4,2}{\oplus}D\\
E_{6}1315131E_{6}\simeq E_{6}\overset{3,3}{\bigcirc}E_{6}\\
E_{6}13151321E_{7}\simeq E_{6}\overset{3,4}{\oplus}E_{7}\\
E_{6}131513221E_{8}\simeq E_{6}\overset{3,5}{\oplus}E_{8}\\
E_{7}123151321E_{7}\simeq E_{7}\overset{4,4}{\oplus}E_{7}\\
E_{7}1231513221E_{8}\simeq E_{7}\overset{4,5}{\oplus}E_{8}\\
E_{8}12231513221E_{8}\simeq E_{8}\overset{5,5}{\oplus}E_{8}. \label{E8E8pair}%
\end{gather}
Again, here we do not need to distinguish between primed and unprimed groups.
The remaining cases of $2$-links\ which are linear chains cannot be joined
consistently to two nodes. Hence, these correspond to at best either a side
link, or a noble molecule.

\subsubsection{All Interior Links are Linear $2$-Links \label{ssec:ALLINT}}

In fact, it is possible to show that all interior links are actually
$2$-links, and moreover, they are exactly linear chains. This means in
particular that the more exotic types of tree-like links encountered
previously can only attach to at most one node.

To see this, suppose that we have two nodes $g$ and $g^{\prime}$ which are
joined by one of these more exotic links. Now, this link must also contain a
sublink which is just a $2$-link comprised of a linear chain. On the other
hand, we have already seen that all the interior $2$-links which are linear
chains have trivial endpoint. That means in particular that we simply cannot
add anything else to these links, without violating the normal crossing
condition (remember, a $-1$ curve cannot attach to three distinct curves). So,
this means that any tree-like $2$-link, or any $n$-link with $n>2$ cannot
attach to two nodes.

One corollary of this result is that we cannot join three nodes with any such
link. Another corollary of this result is that the tree-shaped links can only
attach to a maximum of one node.

\subsubsection{Minimal and Non-Minimal Interior Links \label{ssec:NODEblow}}

In preparation for later, here we collect some properties of the minimal and
non-minimal links. Recall that we refer to an interior link as \textquotedblleft
minimal\textquotedblright\ if all of the blowups between the two nodes are
forced, and \textquotedblleft non-minimal\textquotedblright\ otherwise. For
example, a non-minimal $(D,D)$ link is $1,3,1$, with the minimal link being a
single $-1$ curve. An important feature of this structure is that each minimal
link leads to a fixed number of blowdowns on a neighboring node. Running over
the list of possible links, we see that the minimal number of blowdowns from
attaching via a minimal link is:%
\begin{equation}%
\begin{tabular}
[c]{|c|c|c|c|c|}\hline
& $D$ & $E_{6}$ & $E_{7}$ & $E_{8}$\\\hline
blowdowns from a minimal link & $1$ & $2$ & $3$ & $4$ or $5$\\\hline
\end{tabular}
\ \ .
\end{equation}
We note that some minimal links can induce more blowdowns, for example, the
link between $(D,E_{8})$. Indeed, for the $E_{8}$ case, $4$ blowdowns only
occurs when attaching to a $D$-type node. Otherwise, the minimal number of
blowdowns is $5$. A non-minimal link always induces \textit{at least} one more
blowdown:%
\begin{equation}%
\begin{tabular}
[c]{|c|c|c|c|c|}\hline
& $D$ & $E_{6}$ & $E_{7}$ & $E_{8}$\\\hline
blowdowns from a non-minimal link & $\geq2$ & $\geq3$ & $\geq4$ & $5$\\\hline
\end{tabular}
\ \ ,
\end{equation}
that is, a non-minimal link leads to more blowdowns. Observe that non-minimal
links are necessarily rather sparse since too many will lead to an
inconsistent base.

To demonstrate the utility of this notion, we shall now show that after
blowing down all links, the self-intersection of an interior node (i.e., one
that attaches to two or more nodes) is either $-1$, $-2$, $-3$ or $-4$.
Moreover, the latter two cases can only occur for an $E_{8}$ node which has a
minimal link joining to a $D$-node.

We establish this result simply by considering the minimal interior link which
connect any two nodes. For a $D$-node, we have at least two blowdowns on a
$-4$ curve. For an $E_{6}$ node, we have at least four blowdowns on a curve of
self-intersection $-6$. After these blowdowns, we are left with at best a $-2$
curve. For an $E_{7}$ node, we have at least six blowdowns on a curve of
self-intersection either $-7$ or $-8$. For an $E_{8}$ node we can potentially
attach to a link such as $\overset{4,2}{\oplus}$. So in this case, blowing
down a link can lead to a curve of self-intersection $-1$, $-2$, $-3$ or $-4$.
For example, we get a $-4$ curve from a subconfiguration such as $D\oplus
E_{8}\oplus D$, and we get a $-3$ curve from a subconfiguration such as
$D\oplus E_{8}\oplus E_{8}$.

\subsubsection{Number of Links on a Node\label{ssec:NONINST}}

Let us now show that a node can only attach to at most three non-instanton
links, i.e., links that are not of the form $1,2...,2$. We establish this by
showing on a case by case basis that for four non-instanton links, we always
generate an inconsistent blowdown of the base. For the instanton links, we have a choice of how
we partition up the small instantons into specific marked points. For $k$ small instantons,
this amounts to a choice of $k$-box Young tableau. The fact that a tableau labels a different
theory is explained further in reference \cite{BackToTheFuture}.

Consider first a $D$-type node. If we attach four $-1$ curves, we have:%
\begin{equation}
1\overset{1}{\underset{1}{4}}1\text{,}%
\end{equation}
which does not blowdown consistently. Next, consider an $E_{6}$-type node. If
we attach four non-instanton links, we at least have:%
\begin{equation}
31\overset{3}{\underset{3}{\overset{1}{\underset{1}{6}}}}13\rightarrow
2\overset{2}{\underset{2}{2}}2.
\end{equation}
This comes from the fact that any non-instanton link attached to a $-6$ curve must at the very least induce one blowdown and blow down to a $-2$ curve (see Appendix \ref{linkappendix} for the list of links that can attach to a $-6$ curve, the number of blowdowns they induce, and the configuration after blowdown).
But the adjacency matrix for this configuration of $-2$ curves is not positive
definite. So, this also is not allowed. Next, consider an $E_{7}$-type node.
If we attach four non-instanton links, we at least have:%
\begin{equation}
321\overset{3}{\underset{3}{\overset{2}{\underset{2}{\overset{1}{\underset{1}{8}%
}}}}}123,
\end{equation}
since any non-instanton link attached to a $-8$ curve must at the very least induce two blowdowns (see Appendix \ref{linkappendix}).  This generates eight blowdowns on the $-8$ curve, again a contradiction.
Finally, consider an $E_{8}$-type node. If we attach four non-instanton links,
we at least have:%
\begin{equation}
3221\overset{3}{\overset{2}{\underset{3}{\underset{2}{\overset{2}{\underset{2}{\overset{1}{\underset{1}{(12)}%
}}}}}}}1223.
\end{equation}
From this, we generate twelve blowdowns, again a contradiction. Based on this, we
conclude that any of our nodes can attach to a maximum of three non-instanton links.

\subsection{Constraints on Nodes \label{ssec:NODES}}

An important aspect of links, i.e. molecules built of soley non-DE type curves is that the only
infinite series are the instanton links $1,2...2$, and the $A$- and $D$-series
of $-2$ curves. All of the other links have a size which is bounded above. To
build more general structures, we must combine these links with nodes.

The main consistency condition we will be applying repeatedly is that the
adjacency matrix for the configuration of curves is positive definite. A
necessary condition is that in any connected subconfiguration, the resulting
adjacency matrix must also be positive definite. In more geometric terms, we
need to be able to consistently blowdown all the $-1$ curves.

To constrain the structure of nodes in the base, we will use the general
procedure introduced in subsection \ref{ssec:NODEblow}: We will first blowdown
all the links to their endpoints, and we will then analyze the resulting
structure of the graph, and in particular the self-intersection of the DE-type
curves (i.e., the nodes).

The strongest constraint comes from the fact that blowing down the links
attached to an interior node (namely one which attaches to at least two other
nodes) usually leaves us with a $-1$ or $-2$ curve, with the case of a $-3$ or
$-4$ curve restricted to special circumstances where an $E_{8}$-node links to
at least one D-type node. Systematically applying this condition, we shall
derive a number of constraints on the ways to string together multiple nodes.

First, we shall establish that a node can join to at most two other nodes.
Combined with the result of subsection \ref{ssec:ALLINT} that an interior link
can only attach to two nodes, we will demonstrate in subsection
\ref{ssec:TRIVALENT} that a configuration of nodes is always a line. That is,
the data about the nodes is completely captured by specifying a sequence of
the form $G_{1},...,G_{k}$ for a configuration with $k$ nodes.

The remaining items will be to determine all possible sequences of nodes, and
moreover, what sorts of links can attach to such nodes. We will establish that
the minimal gauge group supported on a node obeys a strict partial ordering
constraint:%
\begin{equation}
G_{1}\subseteq...\subseteq G_{m}\supseteq...\supseteq G_{k}\text{,}%
\end{equation}
namely the biggest gauge symmetries happen in the interior of a base.

Finally, we will establish that the primed nodes $E_{7}^{\prime}$,
$E_{8}^{\prime\prime\prime}$ and $E_{8}^{\prime\prime}$ can only occur on the
two leftmost or rightmost curves, while the $E_{8}^{\prime}$ node can occur in
the middle of a five node base. Otherwise, it too is also constrained to the
two leftmost and rightmost nodes.

The resulting structure for all such base geometries will be of the form:%
\begin{equation}
S_{0,1}\overset{S_{1}}{\underset{...}{g_{1}}}L_{1,2}\overset{S_{2}%
}{\underset{...}{g_{2}}}...\overset{S_{m}}{\underset{...}{g_{m}}%
}...\overset{S_{k-1}}{\underset{...}{g_{k-1}}}L_{k-1,k}\overset{S_{k}%
}{\underset{...}{g_{k}}}S_{k,k+1},
\end{equation}
where the $L$'s refer to interior links, and the $S$'s refer to side links. We shall further cut down the structure of possible base geometries in
subsection \ref{ssec:ENDS}.

\subsubsection{No Trivalent Nodes \label{ssec:TRIVALENT}}

To constrain the possible structures of a base, we now show that the general
topology of a base is essentially just a linear chain, with some possible
tree-like structure only near the ends. In other words, we now eliminate the possibility of
trivalent vertices in a base. Consider an interior node, that is, one which is attached to at least two
interior links, and suppose it attaches to a third node to form a configuration such as:
\begin{equation}
\text{This cannot happen: \ \ }g_{L}\oplus\overset{\overset{g_{U}}{\oplus
}}{g_{\text{mid}}}\oplus g_{R},
\end{equation}
To arrive at this conclusion, consider the possible gauge groups which could
be supported on $g_{\text{mid}}$. We observe that the minimal number of
blowdowns on an $E_{6}$, $E_{7}$ or $E_{8}$ node would be $6$, $9$, or $12$,
respectively. This means the middle curve will not be contractible at the end
of these reductions on the links. That leaves us with the case of a $D$-type node.

For a $D$-type node, we have at least three blowdowns on a $-4$ curve. This
leaves us with a $-1$ curve at the middle of a trivalent vertex:%
\begin{equation}
g_{L}\oplus\overset{\overset{g_{U}}{\oplus}}{g_{\text{mid}}}\oplus
g_{R}\overset{L}{\rightarrow}\widetilde{g_{L}}\overset{\widetilde{g_{U}}%
}{1}\widetilde{g_{R}}.
\end{equation}
However, this sort of configuration violates the normal crossing condition for
a $-1$ curve:\ too many curves are attached to it.

This leaves us to contend with bases where each node attaches to at most two other nodes. It can also potentially
attach to some side links, but this part of the base cannot extend to form a
new direction. As a consequence, we can fully specify the
connectivity of nodes in a base just by listing a sequence of the form
$g_{1},...,g_{k}$.

\subsubsection{Partial Ordering on Nodes \label{ssec:PARTIAL}}

The next restriction we claim is that in the interior of a base, the ordering
of the nodes is not arbitrary. The main idea is that if we introduce the
partial ordering of nodes:%

\begin{equation}
D\subset E_{6}\subset E_{7}\subset E_{8}\text{,}%
\end{equation}
for the corresponding gauge group / algebra, then for a pattern such as:%
\begin{equation}
G_{L}\oplus G_{\text{mid}}\oplus G_{R},
\end{equation}
we cannot have $G_{L}\supsetneq G_{\text{mid}}\varsubsetneq G_{R}$.

To see why, let us return to the list of blowdowns inflicted in a given
pairing. For a $D/E_{6}/E_{7}$-type node paired with anything higher, the
number of blowdowns inflicted is respectively $2$, $3$, and $4$. Now, if this
happens on two sides, the number of blowdowns inflicted is respectively $4$, $6$, and $8$.  But the original self-intersections of the three curves was respectively $4$, $6$, and $8$, as well, and hence at the end of the blowdown process we would find in all three cases a curve with self-intersection number $0$, a contradiction.

Putting these considerations together, we deduce the following structure for a
general base. First, it suffices to list a sequence of nodes with
corresponding gauge group:%
\begin{equation}
G_{1},...,G_{k}.
\end{equation}
Second, the entries of this sequence satisfy the partial ordering constraint:%
\begin{equation}
G_{1}\subseteq G_{2}\subseteq...\subseteq G_{m}\supseteq...\supseteq
G_{k-1}\supseteq G_{k}\text{.}%
\end{equation}
This in turn means that the biggest rank gauge group factors will occur in the interior.

Having listed such a partial ordering of the nodes, we can now decorate either
by a choice of interior links joining two such nodes, or by side links which
only attach to one such node. The full structure of the base thus take the form:%
\begin{equation}
S_{0,1}\overset{S_{1}}{g_{1}}L_{1,2}...L_{m-1,m}\overset{S_{m}}{g_{m}%
}L_{m,m+1}...L_{k-1,k}\overset{S_{k}}{g_{k}}S_{k,k+1}\text{,}%
\end{equation}
in the obvious schematic notation. To avoid overloading the notation, we have
suppressed the possible presence of additional side links attached to each
node (which we will soon exclude anyway).

\subsubsection{Number of Interior Primed Nodes \label{ssec:PRIMED}}

In this subsection we turn to further constraints on admissible sequences of
base nodes. We claim that a base can support at most two interior primed
nodes, and a maximum of three total primed nodes. Moreover, we can support a
maximum of two interior primed nodes of any kind, no interior $E_{8}%
^{\prime\prime\prime}$ nodes, an interior $E_{8}^{\prime\prime}$ node at four
nodes or less, at most one $E_{7}^{\prime}$ node, and at most two interior
$E_{8}^{\prime}$ nodes.

Our first claim is that there is at most one interior $E_{7}^{\prime}$ node.
To see this, suppose to the contrary, i.e., there are at least two interior
$E_{7}^{\prime}$ nodes:%
\begin{equation}
...\oplus E_{7}^{\prime}\oplus...\oplus E_{7}^{\prime}\oplus...
\label{E7primeCONST}%
\end{equation}
Now, from our partial ordering constraint, all of the nodes in between these
two $E_{7}^{\prime}$'s need to be either $E_{7}$ or $E_{8}$ nodes (which might
be primed). Observe, however, that in this case blowing down the links turns
the $E_{7}^{\prime}$ and $E_{7}^{\prime}$ into $-1$ curves (each have six
blowdowns), and all of the curves in between these two nodes turn into $-1$ or
$-2$ curves. In the best case, that leads to a sequence of curves $1,2...2,1$,
a contradiction. We conclude we can have at most one interior $E_{7}^{\prime}$ node.

Similar considerations apply for the $E_{8}^{\prime\prime\prime}$ and
$E_{8}^{\prime\prime}$ nodes. For example, in the case of $E_{8}^{\prime
\prime}$, having two such nodes means that the general structure of the
interior is:%
\begin{equation}
...D\oplus E_{8}^{\prime\prime}\oplus...\oplus E_{8}^{\prime\prime
}\oplus D...,
\end{equation}
where the $D$-type nodes could be on the edge of the configuration of base curves.
Now, this means we again induce nine blowdowns, and a similar argument used near line (\ref{E7primeCONST})
rules out this possibility.

Finally, we come to the case of interior $E_{8}^{\prime}$ nodes. In this case,
the constraint is somewhat milder, i.e., we can have at most two interior
$E_{8}^{\prime}$ nodes. This can happen for example provided we attach to
$D$-type nodes:%
\begin{equation}
...\oplus D\oplus E_{8}^{\prime}\oplus...\oplus E_{8}^{\prime}\oplus
D\oplus....
\end{equation}
However, a quite similar argument to that use near line (\ref{E7primeCONST})
reveals that we can only have at most two such nodes.

In the above argument, the main idea we used was that the partial ordering
constraint would tend to force enough blowdowns to leave the primed node as a
$-1$ curve. That means in particular that the\textit{ total} number of
interior primed nodes is at most two. The only case where we can have two
interior primed nodes is where at least one is an $E_{8}^{\prime}$ node.

\paragraph{Three or Less Primed Nodes}

We can also show that there are at most three primed nodes in any base. To establish this,
observe that to even have four primed nodes, we need at least one of the
interior nodes to be an $E_{8}^{\prime}$ node. Since we need both end nodes to
be primed, we conclude that the full sequence (via the partial ordering
constraint) consists of $E$-type nodes. That in turn means that each of our
interior primed nodes blows down to a $-1$ curve, with (in the best case
situation) $-2$ curves in between these two interior nodes. This again
generates a contradiction.

We can, however, have three primed nodes in a base. For example a consistent
base and its endpoint is:%

\begin{equation}
E_{7}^{\prime}\oplus E_{7}^{\prime}\oplus E_{7}^{\prime}%
\overset{L}{\rightarrow}4,1,4\rightarrow3,3.
\end{equation}

\paragraph{No Interior $E_{8}^{\prime\prime\prime}$ Nodes}

In fact, we have a much tighter constraint on some primed nodes. We claim that
an $E_{8}^{\prime\prime\prime}$ node never resides in the interior of a base.
To see this, consider the three possible interior subsequences and their
blowdowns of the interior links:%
\begin{align}
&  E\oplus E_{8}^{\prime\prime\prime}\oplus E\overset{L}{\rightarrow}%
g_{L},(-1),g_{R}\\
&  D\oplus E_{8}^{\prime\prime\prime}\oplus E\overset{L}{\rightarrow
}2,0,\widetilde{g}\\
&  D\oplus E_{8}^{\prime\prime\prime}\oplus D\overset{L}{\rightarrow}2,1,2,
\end{align}
that is, in the first case a total of \textit{ten} blowdowns are induced on a
$-9$ curve, so we immediately get a contradiction. In the other cases, we also
derive a contradiction.

\paragraph{Interior $E_{8}^{\prime\prime}$ Only for at Most Four Nodes}

We can also see that an interior $E_{8}^{\prime\prime}$ node can only be
supported for four or fewer nodes in the base. Along these lines, consider the
possible four node subsequences and their blowdowns of the interior links:%
\begin{align}
&  E\oplus E_{8}^{\prime\prime}\oplus E\oplus E\overset{L}{\rightarrow
}\widetilde{g_{i-1}},0,2,\widetilde{g_{i+2}}\\
&  D\oplus E_{8}^{\prime\prime}\oplus E\oplus E\overset{L}{\rightarrow
}2,1,2,\widetilde{g_{i+2}}\\
&  E\oplus E_{8}^{\prime\prime}\oplus E\oplus D\overset{L}{\rightarrow
}\widetilde{g_{i-1}},0,3,2\\
&  D\oplus E_{8}^{\prime\prime}\oplus E\oplus D\overset{L}{\rightarrow
}2,1,3,2\\
&  E\oplus E_{8}^{\prime\prime}\oplus D\oplus D\overset{L}{\rightarrow
}\widetilde{g_{i-1}},1,1,3\\
&  D\oplus E_{8}^{\prime\prime}\oplus D\oplus D\overset{L}{\rightarrow
}2,2,1,3.
\end{align}
By inspection, we generate a contradiction at four nodes in all cases but the
sequence $D\oplus E_{8}^{\prime\prime}\oplus E\oplus D$. So we conclude that
the unique four node sequence with an $E_{8}^{\prime\prime}$ node is:%
\begin{equation}
\text{Unique Four Node Sequence with }E_{8}^{\prime\prime}\text{: \ \ }D\oplus
E_{8}^{\prime\prime}\oplus E\oplus D.
\end{equation}
We also see that adding another node always leads to an inconsistent endpoint.
So, we conclude that an interior $E_{8}^{\prime\prime}$ node can only occur at four nodes or less.

\subsection{Decoration Only Near the Ends \label{ssec:ENDS}}

Our analysis so far has constrained the global structure of a base to take the
form of a single line of nodes with possible decorations by side links and
non-minimal links. In this subsection we pare down these possibilities further.

The central result of this subsection will be that any non-minimal decoration,
be it by a non-minimal interior link or any sort of side link is restricted to
the two leftmost or rightmost nodes of a base. Moreover, we will also
establish that a side link which is not of the form $1,2...2$ can only attach
to the leftmost or rightmost node. We will also show that the total number of
such non-instanton side links is limited to three. Thus, the general structure
of a base geometry will be:%
\begin{equation}
S_{0,1}\overset{S_{1}}{g_{1}}L_{1,2}\overset{\mathbf{I}^{\oplus s}}{g_{2}%
}...L_{m-1,m}g_{m}L_{m,m+1}...\overset{\mathbf{I}^{\oplus t}}{g_{k-1}%
}L_{k-1,k}\overset{\mathbf{I}^{\oplus u}}{g_{k}}S_{k,k+1}\text{,}%
\end{equation}
as in line (\ref{genBaseQuiver}).

\subsubsection{Non-Minimal Interior Link Only at the End
\label{ssec:minINTlinks}}

When we specify a pair of nodes in a base, there is a minimal number of
blowdowns which will be inflicted by the corresponding interior link. In some
special cases, we can attempt to switch out this interior link for another
non-minimal one. For example, the minimal link between two $E_{7}$ nodes is
$12321$, but we can also entertain the possibility of a link such as
$12231513221$.

Our central claim in this subsection is that such non-minimal interior links
can only occur near the end of a base. More precisely, we show that a
non-minimal link can attach only to the two leftmost or rightmost nodes. So,
only the leftmost or rightmost interior link can be non-minimal.

To establish this, we consider a base with at least four nodes, i.e.,
$g_{1}...g_{k}$ for $k\geq4$. We proceed by assuming we have a non-minimal
link between $g_{2}$ and $g_{3}$. This will suffice in determining a potential contradiction.

First, we observe that from our sweep over all possible interior links, any
interior node which is not $E_{8}$ which attaches via a non-minimal link will
--upon blowing down the links-- become a $-1$ curve. Additionally, for an
$E_{8}$-type node which attaches via a non-minimal link, we either have a $-1$
or $-2$ curve. Finally, we observe that if we have a non-minimal link which connects an interior node
to an $E_{8}$ node, the links which induce the fewest number of blowdowns are:
\begin{align}
D\overset{3,5}{\oplus}E_{8}  &  \simeq D131513221E_{8}\\
E_{6}\overset{4,5}{\oplus}E_{8}  &  \simeq E_{6}1231513221E_{8}\\
E_{7}\overset{5,5}{\oplus}E_{8}  &  \simeq E_{7}12231513221E_{8}.
\end{align}

Now, we next observe that we can never have a non-minimal link to $E_{8}$ in
the interior. The reason is that if a non $E_{8}$-type node attaches to
$E_{8}$, we always induce too many blowdowns. Indeed, for a $D$-type node, we
inflict at least three blowdowns, and any other interior link gives one more -- a
contradiction--. Similarly, for $E_{6}$, we induce at least four blowdowns, and two
more always occur for any interior link, again a contradiction. Finally, for
an $E_{7}$ node, we have five blowdowns from this non-minimal link, but we
always have at least three blowdowns from an interior link.

So, we conclude that our non-minimal link cannot involve an interior $E_{8}$
node on either side. On the other hand, any other non-minimal link converts the
attached nodes to $-1$ curves. So, the only other option is two $-1$ curves
which touch, again a contradiction.

We therefore conclude that a non-minimal link can only occur on the ends of a
base. That is to say, in a $k$ node base, the only place to have a non-minimal
link is between $g_{1}$ and $g_{2}$, or between $g_{k-1}$ and $g_{k}$.

\subsubsection{Locations of Primed Nodes \label{ssec:PRIMEend}}

We can also deduce that primed nodes cannot sit too far into the interior of a
graph. We have already excluded an $E_{8}^{\prime\prime\prime}$ from ever
sitting in the interior, and we have also already established that an
$E_{8}^{\prime\prime}$ node can only be in the interior for a four or three node base,
so it is enough to restrict our attention to the $E_{7}^{\prime}$ and
$E_{8}^{\prime}$. We show that in a base, only the two leftmost and rightmost
nodes can support an $E_{7}^{\prime}$ node. We also show that for a base
with five nodes, an $E_{8}^{\prime}$ node can reside on the middle (i.e., the
third) node. However, at six curves and above, an $E_{8}^{\prime}$ can only
reside on the two leftmost or rightmost nodes.

Consider first the case of an $E_{7}^{\prime}$ node. Suppose our primed node
$g_{i}$ sits \textquotedblleft deep in the interior\textquotedblright, that
is, we have $2<i<k-2$:%
\begin{equation}
g_{i-2}\oplus g_{i-1}\oplus g_{i}\oplus g_{i+1}\oplus g_{i+2}.
\end{equation}
Blowing down the interior links, we learn that $g_{i}$ is at best a $-1$
curve. This happens when neither $g_{i-1}$ nor $g_{i+1}$ is an $E_{8}$-type
node. But this means that they will convert to (at best) $-2$ curves, so we
have the subconfiguration $2,1,2$, a contradiction.

Consider next the location of an interior $E_{8}^{\prime}$ node. As we have
already mentioned, there is a distinction here between the case of a five node
base, and that of six nodes and more. We claim that an $E_{8}^{\prime}$ node
can reside in the middle of a five node base, but for six nodes and above, an
$E_{8}^{\prime}$ node can only reside on the two leftmost or rightmost nodes
(which is the same as the other primed nodes).

To establish this result, it is convenient to first consider a general
sequence of five nodes. We split up our analysis according to whether the
neighboring nodes are of $D$- or $E$-type. This leaves us with six
possibilities to analyze. Blowing down the links in these cases leads (in the
best case)\ to the configurations:%
\begin{align}
&  E\oplus E\oplus E_{8}^{\prime}\oplus E\oplus E\overset{L}{\rightarrow
}\widetilde{g_{i-2}},2,1,2,\widetilde{g_{i+2}}\\
&  D\oplus E\oplus E_{8}^{\prime}\oplus E\oplus E\overset{L}{\rightarrow
}2,3,1,2,\widetilde{g_{i+2}}\\
&  D\oplus E\oplus E_{8}^{\prime}\oplus E\oplus D\overset{L}{\rightarrow
}2,3,1,3,2\\
&  D\oplus D\oplus E_{8}^{\prime}\oplus E\oplus E\overset{L}{\rightarrow
}3,1,2,2,\widetilde{g_{i+2}}\\
&  D\oplus D\oplus E_{8}^{\prime}\oplus E\oplus D\overset{L}{\rightarrow
}3,1,2,3,2\\
&  D\oplus D\oplus E_{8}^{\prime}\oplus D\oplus D\overset{L}{\rightarrow
}3,1,3,1,3,
\end{align}
Here, we have left the $E$-type nodes as being general, though in the end
those which support an $E_{8}$ node yield the best case scenario (because then
we get a $-3$ curve).

By inspection of the above list, we see that most of the above cases blowdown
to an inconsistent endpoint. Thus, an $E_{8}^{\prime}$ node can only be
supported for a five node base in a limited number of ways:%
\begin{align}
&  D\oplus E\oplus E_{8}^{\prime}\oplus E\oplus E\\
&  D\oplus E\oplus E_{8}^{\prime}\oplus E\oplus D\\
&  D\oplus D\oplus E_{8}^{\prime}\oplus E\oplus D.
\end{align}

Now suppose that we attempt to add one more node either to the left or the
right in such a sequence. We show that in all cases, we do not get a
consistent endpoint.

Consider first the sequence $D\oplus E\oplus E_{8}^{\prime}\oplus E\oplus E$.
There are three distinct ways for us to add another node to this sequence, and
upon blowing down the links, we reach:%
\begin{align}
&  D\oplus D\oplus E\oplus E_{8}^{\prime}\oplus E\oplus
E\overset{L}{\rightarrow}3,1,3,1,2,\widetilde{g}\\
&  D\oplus E\oplus E_{8}^{\prime}\oplus E\oplus E\oplus
E\overset{L}{\rightarrow}2,3,1,2,2,\widetilde{g}\\
&  D\oplus E\oplus E_{8}^{\prime}\oplus E\oplus E\oplus
D\overset{L}{\rightarrow}2,3,1,2,2,2.
\end{align}
which in all cases generates an inconsistent endpoint.

Consider next the sequence $D\oplus E\oplus E_{8}^{\prime}\oplus E\oplus D$.
Here, the partial ordering constraint only allows us to attach a $D$-type node
on the ends. So, it is enough to consider a single sequence, which upon
blowdown of the links leads to:%
\begin{equation}
D\oplus E\oplus E_{8}^{\prime}\oplus E\oplus D\oplus D\overset{L}{\rightarrow
}2,3,1,3,1,2,
\end{equation}
which again leads to an inconsistent endpoint.

Finally, consider the sequence $D\oplus D\oplus E_{8}^{\prime}\oplus E\oplus
D$. In this case, we can either append a $D$-type node on the left, or on the
right\ (by the partial ordering constraint). So, we consider the blowdowns of
the links for the two possible sequences:%
\begin{align}
&  D\oplus D\oplus E_{8}^{\prime}\oplus E\oplus D\oplus
D\overset{L}{\rightarrow}3,1,2,3,1,3\\
&  D\oplus D\oplus D\oplus E_{8}^{\prime}\oplus E\oplus
D\overset{L}{\rightarrow}3,2,1,2,3,3
\end{align}
which always leads us to an inconsistent endpoint. This establishes the claim
that at six nodes and above, an $E_{8}^{\prime}$ node can only occur for the
two leftmost or rightmost nodes.

\subsubsection{Constraints on Side Links \label{ssec:SIDElinks}}

To further pare down the possible structures, we now argue that only the two
leftmost or rightmost nodes can support any sort of side link. Moreover, we
will establish that only the leftmost and rightmost nodes can support a
non-instanton link. We also find that in nearly all cases, an instanton side
link can only be supported on the two leftmost and rightmost
nodes.\footnote{As a brief comment, we recall that a node in the base can also
refer to a primed node, and in the case of the $E_{8}$ series, these nodes
automatically come with some nunber of small instantons attached. In fact,
returning to subsection \ref{ssec:PRIMED}, we recall that in nearly all cases,
a primed node only exists on the two leftmost and rightmost nodes. The only
exception to this is the five node base, where an $E_{8}^{\prime}$ node can
reside in the middle.} The only exception is the five node base, which can
support an $E_{8}^{\prime}$ node at the middle.

The general structure we shall be considering is a candidate base of the form:%
\begin{equation}
S_{0,1}\overset{S_{1}}{g_{1}}L_{1,2}...L_{m-1,m}\overset{S_{m}}{g_{m}%
}L_{m,m+1}...L_{k-1,k}\overset{S_{k}}{g_{k}}S_{k,k+1}\text{.}%
\end{equation}
Suppose we now perform a blowdown of all of the interior links $L_{i,i+1}$ for
$i=1,...,k-1$. These links all have trivial endpoint, leaving us with a
sequence of the form:%
\begin{equation}
S_{0,1}\overset{S_{1}}{\widetilde{g_{1}}}...\overset{S_{m}}{\widetilde{g_{m}}%
}...\overset{S_{k}}{\widetilde{g_{k}}}S_{k,k+1}\text{,}%
\end{equation}
where for all $\widetilde{g_{2}},...,\widetilde{g_{k-1}}$ in the interior, we
have a curve of self-intersection $-1,-2,-3,-4$. The latter two cases only
occur at an interface between a $D$-type node and an $E_{8}$-type node.

We shall first establish that a non-instanton side link can only occur on the
leftmost and rightmost node. By assumption, a non-instanton side link is not
of the form $12...2$. That means it must terminate with something other than a
$-2$, so it is of the general form $1....x$ for some $x\neq2$. Now, if we do
not have an $E_{8}$-type node (primed or not), then blowing down the interior
links leaves us at best with a $-2$ curve. One more blowdown leaves us with a
$-1$ curve. However, since this curve occurs in the interior of a graph, it
now touches three other curves, violating the normal crossing condition.

Next, suppose we have an $E_{8}$-type node. In this case, the minimal
structure for a side link attached to such a node has the form $1223$, that
is, it induces at least three blowdowns. On the other hand, the minimal number
of blowdowns induced by an interior link is four (when it interfaces with a
$D$-type node). That means we have minimally eleven blowdowns, leaving us with
a $-1$ curve (if it is an unprimed $E_{8}$ node). But again, we now see that
normal crossing will be violated since our $-1$ curve touches three other
curves. We therefore conclude that our node cannot have two interior links,
i.e., it must reside at the end.

Next, let us show that aside from the case of a five node base, an instanton
side link can only occur on the two leftmost or rightmost nodes. The case of a
five node base is special, since as we have already seen it can support an
$E_{8}^{\prime}$ node in the middle (see subsection \ref{ssec:PRIMED}). So,
suppose first that we are dealing with an $E_{8}$-type node. Then, since we
have already seen that a primed node can only exist on the two leftmost or
rightmost nodes, the claim follows.

Next, consider the case of any node other than an $E_{8}$-type node
(primed or otherwise). Then, after blowing down the interior links, we have a
curve which has self-intersection $-1$ or $-2$. We can only attach a side link
if we have a $-2$ curve after this first stage of blowdowns. In fact, we can
immediately see that the only side link available to us is a single
$-1~$\ curve. Observe that in a sequence $\widetilde{g_{i-1}}\widetilde{g_{i}%
}\widetilde{g_{i+1}}$, we need at least one of these curves to have
self-intersection $-3$ or $-4$. If this is not the case, the further blowdown
induced by the side link on $g_{i}$ would generate a contradiction, for
example $2\overset{1}{2}2$, which is inconsistent. To get a curve of
self-intersection $-3$ or $-4$, we therefore need either $g_{i-1}$ or
$g_{i+1}$ to refer to an $E_{8}$-type node, so without loss of generality take
it to be $g_{i-1}$. We now step through the possible nodes. For $g_{i}$ an
$E_{7}$-type node, we get four blowdowns from the link with the $E_{8}$,
another three from the link to the right, and one more from our small
instanton. This is a total of eight blowdowns, a contradiction. For $g_{i}$ an
$E_{6}$-type node, we get three blowdowns from the link with the $E_{8}$ node,
another two from the link to the right, and one more from the small instanton.
This is a total of six blowdowns, a contradiction. Finally, for $g_{i}$ a
$D$-type node, we get two blowdowns from the link with the $E_{8}$, another
one from the link to the right, and one more from the small instanton. This is
a total of four blowdowns, a contradiction.

We therefore conclude that only the two leftmost and rightmost nodes can
support a small instanton link at all. In the case of the end nodes $g_{1}$
and $g_{m}$, side links can be attached, and a priori, more than one can be
consistently added on. The reason is that fewer blowdowns are inflicted on the sides.

\paragraph{Bounds on an Interior Instanton Link}

In fact, a small extension of the above argument reveals that in nearly all
cases, the total number of small instantons for a side link is at most one.
The only exception to this is the $E_{8}^{\prime\prime\prime}$ and
$E_{8}^{\prime\prime}$ nodes. However, we have already seen that there are no
interior $E_{8}^{\prime\prime\prime}$ nodes, while the interior $E_{8}%
^{\prime\prime}$ nodes only occur in a base with four or fewer nodes. This
means that at five or more nodes in a base, decoration by a \textquotedblleft
small instanton link\textquotedblright\ in the interior involves adding on at
most one $-1$ curve.

\subsubsection{Maximum of Two Side Links on an End Node
\label{ssec:MAXsidelinks}}

Having cut down the possible ways that side links can attach to a quiver, we
now turn to further restrictions on the side links which can attach to the
nodes $g_{1}$ and $g_{k}$. As throughout this section, we assume that $k>1$.
Again, it is helpful to split up the types of side links into one long
instanton link $1,2...2$, and up to two non-instanton links $\alpha_{1}$ and
$\beta_{1}$ which can attach to $g_{1}$. Similar conventions hold for $g_{m}$.
Our main result from this subsection is that the maximum number of
non-instanton side links is three, and moreover, only $E_{6}$ and $E_{8}$ can
tolerate more than one non-instanton side link.

To establish this, we will first determine the number of non-instanton side
links which can attach to a given node. Then, we shall determine the global
structure of how these links can attach together.

To begin, suppose we have a D-type node. We claim that it can attach to a
maximum of one non-instanton side link. Indeed, suppose to the contrary. Then,
we will have a structure of the form:%
\begin{equation}
\overset{x}{\underset{y}{\underset{...}{\overset{...}{\overset{1}{\underset{1}{4}%
}}}}}\oplus g_{2}.
\end{equation}
Now, upon blowing down the $-1$ curves adjacent to the $-4$ curve, we reach a
$-1$ curve which attaches to three curves, a violation of the normal crossing condition.

This means we can only attach side links to a $D$-type end node via:%
\begin{equation}
\underbrace{1}\overset{\alpha}{D}\oplus g_{2},
\end{equation}
where the notation $\underbrace{1}$ indicates either adding or omitting this
curve, and $\alpha$ is a non-instanton side link.

Next, consider the case of a $-6$ curve. Again, suppose we have two
non-instanton side links. Then, we will at least have the structure:%
\begin{equation}
\underset{k}{\underbrace{2...21}}%
\underset{3}{\overset{3}{\overset{1}{\underset{1}{6}}}}131g_{2}.
\end{equation}
We claim that in this configuration, we must have $k=0$, that is, there is no
instanton link in this case. To see this, observe that if such an instanton
link is present, we induce at least five blowdowns on the $-6$ curve, and the
resulting $-1$ curve will touch three curves, violating the normal crossing condition.

As a consequence, we can have at most two side links attached to this node.
They can be of the form:%
\begin{equation}
\underset{k}{\underbrace{2...21}}\overset{\alpha}{E_{6}}\oplus g_{2}\text{
\ \ or \ \ }\overset{\alpha}{\underset{\beta}{E_{6}}}\oplus g_{2},
\end{equation}
that is, at most two side links can attach to the $-6$ curve.

Next, consider the case of an $E_{7}$ node. For simplicity, we assume this is
given by a $-8$ curve since the case of a $-7$ curve has even further
restrictions. Again, we ask whether we can attach two side links to this
configuration. A non-instanton side link attached to an $E_{7}$ node will have
the form $123...$ or $1223...$. That means at least two blowdowns will be
induced. Further, for an interior link, at least three blowdowns will be
induced. Counting up, we see that if we have two side links and an interior
link, we have already reached seven blowdowns. Moreover, the resulting curve
will have self-intersection $-1$, and will be attached to three curves. This
violates the normal crossing condition, so this cannot occur. This leaves us
with the possibility:
\begin{equation}
\underset{k}{\underbrace{2...21}}\overset{\alpha}{E_{7}}\oplus g_{2},
\end{equation}
i.e., at most one non-instanton side link.

Finally, consider the case of an $E_{8}$ node. Again, for simplicity, we
assume this is given by a $-12$ curve. In this case, a non-instanton side like
will necessarily have the form $1223...$. \ The resulting configuration of
curves has the form:%
\begin{equation}
\underset{k\geq0}{\underbrace{2...21}}%
\overset{3}{\overset{2}{\underset{3}{\underset{2}{\overset{2}{\underset{2}{\overset{1}{\underset{1}{(12)}%
}}}}}}}12231...g_{2}.
\end{equation}
Now, we claim that at most two side links can be tolerated. Indeed, with two
non-instanton side links, we already induce ten blowdowns. Adding one more
yields a $-1$ curve which violates normal crossing. That leaves us with two
options:%
\begin{equation}
\underset{k\geq0}{\underbrace{2...21}}%
\overset{3}{\overset{2}{\overset{2}{\overset{1}{E_{8}}}}}\oplus g_{2}\text{
\ \ or \ \ }%
\overset{3}{\overset{2}{\underset{3}{\underset{2}{\overset{2}{\underset{2}{\overset{1}{\underset{1}{E_{8}%
}}}}}}}}\oplus g_{2}.
\end{equation}
Summarizing then, we see that we can tolerate at most two non-instanton side
links on an end node.

\paragraph{At Most Three Non-Instanton side links}

The next item of our analysis will be to show that at most three non-instanton
side links can be included at all. To see how this comes about, we suppose to
the contrary. The only two cases which have this structure are possible
combinations of an $E_{8}$ and $E_{6}$ node, namely one for each side. Along
these lines, consider first a pair of $E_{8}$ nodes on the ends:%
\begin{equation}
\overset{3}{\overset{2}{\underset{3}{\underset{2}{\overset{2}{\underset{2}{\overset{1}{\underset{1}{E_{8}%
}}}}}}}}\oplus...\oplus
\overset{3}{\overset{2}{\underset{3}{\underset{2}{\overset{2}{\underset{2}{\overset{1}{\underset{1}{E_{8}%
}}}}}}}}.
\end{equation}
Then, we also know from the partial ordering constraint that since only
$E_{8}$ nodes can appear in the middle, we must have an interior link of the
form $12231513221$. This means five blowdowns are induced from the interior
link, and another six are induced from the non-instanton side links. This
would leave us with a $-1$ curve touching three curves, violating normal
crossing. As this cannot occur, we can only tolerate two non-instanton side
links in this case:%
\begin{equation}
2...21\overset{\alpha_{1}}{E_{8}}\oplus...\oplus\overset{\alpha_{m}}{E_{8}%
}12...2.
\end{equation}
Next, consider an $E_{8}$ and an $E_{6}$ node, each with two non-instanton
side links. The case with a minimal number of blowdowns from the non-instanton
side links is:%
\begin{equation}
\overset{3}{\overset{2}{\underset{3}{\underset{2}{\overset{2}{\underset{2}{\overset{1}{\underset{1}{E_{8}%
}}}}}}}}\oplus...\oplus\overset{3}{\underset{3}{\overset{1}{\underset{1}{E_{6}%
}}}}.
\end{equation}
Again, we can only have one non-instanton side link attach to the $E_{8}$
node, since the interior link will contain the contribution $1223151...$ at
least. For the $E_{6}$ end node, the interior link could potentially be $131$.
This means there would only be four blowdowns. So, in other words, we could
have a topology of the form:%
\begin{equation}
2...21\overset{\alpha_{1}}{E_{8}}\oplus...\oplus\underset{\beta_{m}%
}{\overset{\alpha_{m}}{E_{6}}}.
\end{equation}
Finally, we come to the case of an $E_{6}$ node at each end. In this case, the
putative structure for the base is:%
\begin{equation}
\overset{3}{\underset{3}{\overset{1}{\underset{1}{E_{6}}}}}\oplus
...\oplus\overset{3}{\underset{3}{\overset{1}{\underset{1}{E_{6}}}}}.
\end{equation}
Now, in the interior of the quiver, we must have $E_{7}$ or $E_{8}$ nodes.
Since there are no D-type nodes available, blowing down all of the interior
links will convert the interior nodes to $-1$ or $-2$ curves. The end nodes
will become $-2$ curves. Since the end nodes are attached to three nodes, we
see that we cannot tolerate a $-1$ curve in the interior. That means that upon
blowdown of all links, we reach the configuration:%
\begin{equation}
\overset{2}{\underset{2}{2}}2...2\overset{2}{\underset{2}{2}},
\end{equation}
which does not have a positive definite adjacency matrix. That means at most
one end can tolerate two non-instanton side links, leaving us with:%
\begin{equation}
\overset{\alpha_{1}}{\underset{\beta_{1}}{E_{6}}}\oplus...\oplus
\overset{\alpha_{m}}{E_{6}}12...2.
\end{equation}

\subsection{General Structure of a Base \label{ssec:summary}}

Assembling each of these smaller results, we now show how to piece them
together to constrain the general form of a base. Throughout, we restrict to
the case of $k>1$ base nodes.

\begin{itemize}
\item 1) Because there are no trivalent interior links, and because no node
can link to three other nodes, the nodes form a linear chain. We call this
sequence of nodes $g_{1},...,g_{k}$ in the obvious notation. These two results
follow from subsections \ref{ssec:TRIVALENT} and \ref{ssec:ALLINT}.

\item 2)\ Moreover, the groups on the nodes satisfy a partial ordering
condition $G_{1}\subseteq G_{2}\subseteq...\subseteq G_{m}\supseteq
...\supseteq G_{k-1}\supseteq G_{k}$. Additionally, aside from the five node
base, the only locations where a primed node can occur are the two leftmost
and two rightmost nodes. An $E_{8}^{\prime}$ node can reside in the middle of
a five node base. Furthermore, an interior $E_{8}^{\prime\prime\prime}$ node
never appears, and an interior $E_{8}^{\prime\prime}$ node can only occur at
four nodes or less. The partial ordering result follows from subsection
\ref{ssec:PARTIAL} and the conditions on primed nodes follows from subsections
\ref{ssec:PRIMED} and \ref{ssec:PRIMEend}.

\item 3)\ For the deep interior nodes, i.e., for $g_{i}$ with $2<i<k-1$, we
have also seen that no side links can be attached. Moreover, no non-instanton
side link can attach for $1<i<k-1$, i.e., nowhere in the interior. The only
places to attach a side link are the two leftmost nodes and the two right most
nodes (with one caveat: at five nodes, there is a single option to have an
$E_{8}^{\prime}$ node, which is really an $E_{8}$ node attached to one $-1$
curve). Moreover, the only sort of side link which can attach to $g_{2}$ and
$g_{k-1}$ is an instanton side link (with one caveat: at fives nodes and
above, these interior instanton links are at best a single $-1$ curve). These
results follows from subsections \ref{ssec:SIDElinks} and \ref{ssec:NONINST}.

\item 4)\ All interior links, i.e., $L_{i,i+1}$ for $1<i<k-1$, are minimal,
i.e., the only non-minimal links we can support are $L_{1,2}$ and $L_{k-1,k}$.
This result follows from subsection \ref{ssec:minINTlinks}.

\item 5) For the end nodes, i.e., for $g_{1}$ and $g_{k}$, at most two side
links can attach. Moreover, the total number of non-instanton side links which
can attach to the full graph is three. This result follows from subsection
\ref{ssec:MAXsidelinks}. Futher, the number of instantons on the interior
nodes (i.e., $g_{2}$ and $g_{k-1}$) is zero or one when $k>5$.

\item 6) Putting together Items 1), 2), 3), 4) and 5), we learn that the
general structure of a base is given as in line (\ref{genBaseQuiver}):%
\begin{equation}
S_{0,1}\overset{S_{1}}{g_{1}}L_{1,2}\overset{\mathbf{I}^{\oplus s}}{g_{2}%
}...L_{m-1,m}g_{m}L_{m,m+1}...\overset{\mathbf{I}^{\oplus t}}{g_{k-1}%
}L_{k-1,k}\overset{\mathbf{I}^{\oplus u}}{g_{k}}S_{k,k+1}\text{,}%
\end{equation}
where we have also used the fact that a general base can have at most three
side links (i.e., when $k>1$). Here, the only decoration, either by a choice of
side link, or by adding a primed node, occurs on the two leftmost or rightmost
nodes. Furthermore, the only place where a non-minimal interior link can occur
is on the three leftmost or rightmost nodes.\ Finally, for five nodes and
above, the value of $s$ and $t$ is at most one.

\item Special Cases) Finally, there are a few special cases at five nodes or
less. With the same notation indicated previously, this is their structure:%
\begin{align}
\text{Zero Nodes}  & \text{: \ \ \ \ \ \ \ \ \ \ \ }n\text{-link}\\
\text{One Node}  & \text{: \ \ \ \ \ \ \ \ \ \ \ }S_{0,1}\overset{S_{1}%
}{\underset{\mathbf{I}^{\oplus u}}{g_{1}}}S_{1,2}\text{ \ \ \ \ }u\leq11\\
\text{Two Nodes}  & \text{: \ \ \ \ \ \ \ \ \ \ \ }S_{0,1}\overset{S_{1}%
}{g_{1}}L_{1,2}\overset{\mathbf{I}^{\oplus u}}{g_{2}}S_{2,3}\text{ \ \ \ \ }u\leq6\\
\text{Three Nodes}  & \text{: \ \ \ \ \ \ \ \ \ \ \ }S_{0,1}\overset{S_{1}%
}{g_{1}}L_{1,2}\overset{\mathbf{I}^{\oplus s}}{g_{2}}L_{2,3}%
\overset{\mathbf{I}^{\oplus u}}{g_{3}}S_{3,4}\text{ \ \ \ \ }s\leq2, u \leq 6\\
\text{Four Nodes}  & \text{: \ \ \ \ \ \ \ \ \ \ \ }S_{0,1}\overset{S_{1}%
}{g_{1}}L_{1,2}\overset{\mathbf{I}^{\oplus s}}{g_{2}}L_{2,3}%
\overset{\mathbf{I}^{\oplus t}}{g_{3}}L_{3,4}\overset{\mathbf{I}^{\oplus
u}}{g_{4}}S_{4,5}\text{ \ \ \ \ }s,t\leq2, u \leq 6\\
\text{Five Nodes}  & \text{: \ \ \ \ \ \ \ \ \ \ \ }S_{0,1}\overset{S_{1}%
}{g_{1}}L_{1,2}\overset{\mathbf{I}^{\oplus s}}{g_{2}}L_{2,3}%
\overset{\mathbf{I}^{\oplus r}}{g_{3}}L_{3,4}\overset{\mathbf{I}^{\oplus
t}}{g_{4}}L_{4,5}\overset{\mathbf{I}^{\oplus u}}{g_{5}}S_{5,6}\text{
\ \ \ \ }s,t,r\leq1, u \leq 6\\
\text{Six Nodes}  & \text{: \ \ \ \ \ \ \ \ \ \ \ }S_{0,1}\overset{S_{1}%
}{g_{1}}L_{1,2}\overset{\mathbf{I}^{\oplus s}}{g_{2}}L_{2,3}g_{3}L_{3,4}%
g_{4}L_{4,5}\overset{\mathbf{I}^{\oplus t}}{g_{5}}L_{5,6}\overset{\mathbf{I}%
^{\oplus u}}{g_{6}}S_{6,7}\text{ \ \ \ \ \ \ \ }s,t\leq1, u \leq 6.
\end{align}
At six nodes and above, the generic pattern begins. Further, to have an
instanton link on $g_{3}$ at five nodes requires this node to be a $-12$ curve.
\end{itemize}

To round out our analysis, in a set of Appendices, we catalogue the full list
of possible sequences using just unprimed nodes (for the sake of brevity in
the exposition), as well as possible side links which can join on to one base
node. In a companion \texttt{Mathematica} program, we also provide an
interface for the user to explore the full list of bases, including the case
of primed nodes.

This completes the classification of bases.

\section{Enhancing Gauge Groups / Adding Extra Matter \label{sec:FIBERS}}

In the previous section we determined the full list of base geometries which
can support a 6D\ SCFT. In field theoretic terms, this amounts to specifying
all possible configurations of tensor multiplets which could a priori be compatible with the
simultaneous conditions of anomaly cancellation, and the requirement that the
origin of the tensor branch defines a conformal fixed point. For each such configuration, we have also
identified a canonical 6D SCFT, namely the one dictated by the minimal singular behavior of the elliptic
fiber of the F-theory compactification.

Now, given one such tensor branch structure, i.e., one such base geometry, we can
ask how many \textit{different} types of 6D\ SCFTs can be supported over the \textit{same} base.
In field theory terms, we ask whether we can retain the same
configuration of tensor multiplets whilst supplementing the minimal gauge group and matter content.
In geometric terms, this corresponds to making the elliptic fiber more singular over some of the curves
in the base. In string theoretic terms, we are wrapping additional
seven-branes over curves in the base geometry.

Just as in our classification of base geometries, we will find that most of the F-theory constraints have
field theory formulations, modulo a few cases which would be interesting to understand further.
In field theory terms, we will need to demand that 6D anomaly
cancellation is still respected. Further, adding extra matter means that there
is a Higgs mechanism available which takes us down to the minimal base geometry.
The plan will first be to study possible fiber enhancements for isolated non-Higgsable clusters, and to then
study possible enhancements for the various links in our base quivers. The
main upshot of our analysis is that in a generic base quiver with exceptional
groups, it is typically not possible to enhance the fiber of any of the
curves. Rather, the vast majority of fiber enhancements of a geometry only occur on
those nodes which support classical groups.

\subsection{Matter for a Single Simple Factor \label{ssec:MATTER}}

In this subsection we determine constraints on matter fields charged under a
gauge group. First of all, to have a consistent anomaly cancellation
mechanism, we need each non-abelian simple gauge group factor to come with a
corresponding tensor multiplet. Geometrically, this just means each gauge
group is associated with a seven-brane wrapping a $\mathbb{P}^{1}$ in the base of the
geometry. Now, this gauge group may also have matter fields which transform in
representations of the gauge group. Our plan will be to use 6D anomaly
cancellation as a tool to understand what sorts of matter can contribute to
this theory. In F-theory terms, these matter fields are associated with
special points on the $\mathbb{P}^{1}$ where the elliptic fibration becomes
more singular.

To begin, let us recall the general constraints for 6D anomaly cancellation
reviewed in subsection \ref{ssec:CONTanom}, now stated in geometric terms
adapted to the existence of a CFT:%
\begin{align}
Ind_{Adj_{a}}-\sum_{R}{Ind_{R_{a}}n_{R_{a}}}  &  =6(K\cdot\Sigma_{a})=-12-6\Sigma_{a}%
\cdot\Sigma_{a}\\
y_{Adj}-\sum_{R}{y_{R}n_{R}}  &  =-3(\Sigma_{a}\cdot\Sigma_{a})\\
x_{Adj}-\sum_{R}{x_{R}n_{R}}   &  =0\\
\sum_{R_{a},S_{b}}Ind_{R_{a}}Ind_{S_{b}}n_{R_{a},S_{b}}  &  =\Sigma_{a}%
\cdot\Sigma_{b},
\end{align}
where in the first line we have used the fact that the genus formula relates
$\Sigma\cdot(\Sigma+K)=-2$ since all of our theories are supported over
$\mathbb{P}^{1}$'s (i.e., genus zero curves). Indeed, the lattice of vectors
$\Lambda\subset\mathbb{R}^{T,1}$ is, in geometric terms just $H_{2}%
^{\text{cpct}}(B,%
\mathbb{Z}
)$ the lattice of compact two-cycles in our base geometry.

The resulting constraints from these conditions have been worked out in the
literature, for example in \cite{Bershadsky:1996nu}, modulo a few omissions. Along
these lines, we first determine the constraints on the matter content for a
gauge group factor just from anomalies where all four external currents are
the same. We then turn to the additional constraints imposed by anomalies with
two different sets of external currents (i.e., the \textquotedblleft
mixed\textquotedblright\ anomalies).

To begin, we ask what sorts of matter fields can be supported with a single
tensor multiplet, i.e., over a single $\mathbb{P}^{1}$. Recall that for a $-1$
curve, there are no restrictions on the gauge group which can be supported.
For $-2$ curves, we cannot support an $Sp$ gauge theory, and for $-4$ curves,
we cannot support an $Sp$ or $SU$ gauge theory. This essentially follows from
the condition that an NHC\ supports a minimal gauge group, and moreover, a
further enhancement must be allowed to Higgs down to the minimal gauge group.

As a point of notation, let $-n=\Sigma\cdot\Sigma$ denote the
self-intersection of a curve. Also, let $d_{s}$ denote the dimension of the
spinor representation, $d_{s}=2^{k-1}$ for $N=2k$, and $d_{s}=2^{k}$ for
$N=2k+1$. Note that in the cases of $SO(11)$, $SO(12)$, and $SO(13)$, the
resulting $n_{s}$ is sometimes a half-integer. This simply means we are
dealing with a half hypermultiplet (as can happen since the representation is
pseudo-real). The case of enhancement all the way to an $E_{8}$ gauge theory
is the one case which is rather difficult to treat in purely field theoretic
terms. This corresponds to a theory with $n_{\text{inst}}$ some number of
small instantons. Motion on the instanton moduli space translates to a
breaking pattern for the theory, allowing it to descend back down to a
non-Higgsable cluster. For completeness, we also include this possibility in
what follows.

\newpage

Aside from the case of the $E_8$ theories with small instantons, anomaly cancellation tells us all possible single curve theories:

\begin{itemize}
\item $n=1$:

\begin{itemize}
\item $\mathfrak{su}(N)$, $N \geq2$, $n_{f} = N+8$, $n_{\Lambda^{2}}=1$.

\item $\mathfrak{su}(6)$, $n_{f} = 15$, $n_{\Lambda^{3}}= 1/2$.

\item $\mathfrak{sp}(N)$, $N \geq1$, $n_{f} = 8 + 2N$.

\item $\mathfrak{so}(N)$, $N = 5,...,12$, $n_{f} = N-5$, $n_{s} = 48/d_{s}$.

\item $\mathfrak{g}_{2}$, $n_{f} =7 $.

\item $\mathfrak{f}_{4} $, $n_{f} = 4$.

\item $\mathfrak{e}_{6}$, $n_{f} = 5$.

\item $\mathfrak{e}_{7}$, $n_{f}=7/2$

\item $\mathfrak{e}_{8}$, $n_{\text{inst}}=11$
\end{itemize}

\item $n=2$:

\begin{itemize}
\item $\mathfrak{su}(N)$, $N\geq2$, $n_{f}=2N$.

\item $\mathfrak{so}(N)$, $N=7,...,13$, $n_{f}=N-6$, $n_{s}=32/d_{s}$.

\item $\mathfrak{g}_{2}$, $n_{f}=4$.

\item $\mathfrak{f}_{4}$, $n_{f}=3$.

\item $\mathfrak{e}_{6}$, $n_{f}=4$.

\item $\mathfrak{e}_{7}$, $n_{f}=3$

\item $\mathfrak{e}_{8}$, $n_{\text{inst}}=10$
\end{itemize}

\item $n=3$

\begin{itemize}
\item $\mathfrak{su}(3)$, $n_{f}=0$

\item $\mathfrak{so}(N)$, $N=7,...,12$, $n_{f}=N-7$, $n_{s}=16/d_{s}$.

\item $\mathfrak{g}_{2}$, $n_{f}=1$.

\item $\mathfrak{f}_{4}$, $n_{f}=2$.

\item $\mathfrak{e}_{6}$, $n_{f}=3$.

\item $\mathfrak{e}_{7}$, $n_{f}=5/2$

\item $\mathfrak{e}_{8}$, $n_{\text{inst}}=9$
\end{itemize}

\item $n=4$

\begin{itemize}
\item $\mathfrak{so}(N)$, $N\geq8$, $n_{f}=N-8$.

\item $\mathfrak{f}_{4}$, $n_{f}=1$.

\item $\mathfrak{e}_{6}$, $n_{f}=2$.

\item $\mathfrak{e}_{7}$, $n_{f}=2$

\item $\mathfrak{e}_{8}$, $n_{\text{inst}}=8$
\end{itemize}

\item $n=5$

\begin{itemize}
\item $\mathfrak{f}_{4}$, $n_{f}=0$

\item $\mathfrak{e}_{6}$, $n_{f}=1$.

\item $\mathfrak{e}_{7}$, $n_{f}=3/2$

\item $\mathfrak{e}_{8}$, $n_{\text{inst}}=7$
\end{itemize}

\item $n=6$

\begin{itemize}
\item $\mathfrak{e}_{6}$, $n_{f}=0$

\item $\mathfrak{e}_{7}$, $n_{f}=1$

\item $\mathfrak{e}_{8}$, $n_{\text{inst}}=6$
\end{itemize}

\item $n=7$

\begin{itemize}
\item $\mathfrak{e}_{7}$, $n_{f}=1/2$

\item $\mathfrak{e}_{8}$, $n_{\text{inst}}=5$
\end{itemize}

\item $n=8$

\begin{itemize}
\item $\mathfrak{e}_{7}$, $n_{f}=0$

\item $\mathfrak{e}_{8}$, $n_{\text{inst}}=4$
\end{itemize}

\item $n=9$

\begin{itemize}
\item $\mathfrak{e}_{8}$, $n_{\text{inst}}=3$
\end{itemize}

\item $n=10$

\begin{itemize}
\item $\mathfrak{e}_{8}$, $n_{\text{inst}}=2$
\end{itemize}

\item $n=11$

\begin{itemize}
\item $\mathfrak{e}_{8}$, $n_{\text{inst}}=1$
\end{itemize}

\item $n=12$

\begin{itemize}
\item $\mathfrak{e}_{8}$, $n_{\text{inst}}=0$
\end{itemize}
\end{itemize}

The above list can also be viewed as a complete classification of theories
with a single tensor multiplet and gauge algebra.  We will discuss theories with unpaired tensors (i.e. theories with a tensor multiplet but no gauge algebra) in section \ref{ssec:Unpaired}.

\subsection{Matter for Multiple Simple Factors}

In the previous subsection, we focused exclusively on constraints coming from
a single simple factor.\ In our classification of base geometries, we have
also seen that many F-theory geometries support a quiver-like structure for
the resulting 6D\ SCFT. In this subsection we turn to the constraints imposed
by mixed anomalies, i.e., when the two symmetry currents are distinct.

First of all, given a collection of simple gauge algebra factors $\mathfrak{g}_{1}%
,...,\mathfrak{g}_{k}$, we can ask how many gauge groups a matter field could be charged
under. For example, experience from lower-dimensional SCFTs (see, e.g.,
\cite{Gaiotto:2009we}) points to the possibility of matter in tri-fundamental
representations. In the case of 6D\ SCFTs, this cannot occur. The reason is
already apparent from the structure of admissible F-theory base geometries. In
that context, we have pairwise intersections of curves. Any potential triple
intersection of curves in the base is non-generic, and can be moved to general
position through a high order complex structure deformation. In field theory
terms, these deformations to move intersection points to general position
correspond to irrelevant deformations. For this reason, such triple
intersections do not lead to any new 6D\ SCFTs.

As a consequence, the matter fields of our
system will be charged under at most two simple non-abelian gauge algebra
factors. Geometrically, this means it is enough for us to focus attention on
pairwise intersections of curves. Additionally, we also know that each
intersection number is at most one. That means the condition to cancel mixed
anomalies is simply:%
\begin{equation}
\sum_{R_{a},S_{b}}Ind_{R_{a}}Ind_{S_{b}}n_{R_{a},S_{b}}=\overrightarrow{v}%
_{a}\cdot\overrightarrow{v}_{b}=\Sigma_{a}\cdot\Sigma_{b}=1,
\label{mixed}
\end{equation}
where $n_{R_{a},S_{b}}$ is the number of hypermultiplets in the mixed
representation $(R_{a},S_{b})$ of $\mathfrak{g}_{a}\oplus \mathfrak{g}_{b}$. For all of the
representations mentioned above, $Ind_{R}\geq1$, which means that
$n_{R_{a},S_{b}}=1$ for precisely one $R_{a}$ and $S_{b}$. This puts strong
restrictions on the gauge algebras which are allowed on consecutive nodes. First
off, there must be some representation $R_{a}$ of $\mathfrak{g}_{a}$ and $S_{b}$ of
$\mathfrak{g}_{b}$ such that $Ind_{R_{a}}Ind_{S_{b}}=1$. The only representations with
index less than or equal to 1 are the fundamental representations of the
special unitary Lie algebras and the symplectic Lie algebras, so one of these must always
pair up in the mixed representation between any two adjacent curves carrying
gauge algebras. The fundamental representations of the special orthogonal Lie algebras
$\mathfrak{so}(N),N\geq7$ and the exceptional group $\mathfrak{g}_{2}$ all have index 2, so these
can only pair with half-fundamentals of the symplectic algebras (which all have
index $\frac{1}{2}$). The spinor reps of $\mathfrak{so}(7)$ and $\mathfrak{so}(8)$ and the
$\Lambda^{2}$ rep of $\mathfrak{su}(4)$ also have index $2$, so these may also pair up
with half-fundamentals of the symplectic algebras. However, no other
representations have indices $\leq2$, so they are unable to satisfy
(\ref{mixed}) whenever a gauge algebra appears on a neighboring node.

In particular, exceptional gauge algebras $\mathfrak{f}_{4}$, $\mathfrak{e}_{6}$, $\mathfrak{e}_{7}$, or $\mathfrak{e}_{8}$
have no representations with index $\leq2$. Whenever these gauge algebras
appear, their neighbors must be empty. Note that any curve with self
intersection $-5$ or below must hold one of these gauge groups, and so all of
its neighbors must be empty $-1$ curves.

We may thus classify the matter stretching between adjacent curves simply by
these two representations. In particular, the following representations may be
paired between adjacent curves:

\begin{itemize}
\item $\mathfrak{g}_{a} = \mathfrak{su}(N_{a})$, $\mathfrak{g}_{b} = \mathfrak{su}(N_{b})$, $R_{a} = f_{a}$, $R_{b} =
f_{b}$.

\item $\mathfrak{g}_{a} = \mathfrak{su}(N_{a})$, $\mathfrak{g}_{b} = \mathfrak{sp}(N_{b})$, $R_{a} = f_{a}$, $R_{b} =
f_{b}$.

\item $\mathfrak{g}_{a} = \mathfrak{sp}(N_{a})$, $\mathfrak{g}_{b} = \mathfrak{sp}(N_{b})$, $R_{a} = f_{a}$, $R_{b} =
f_{b}$.

\item $\mathfrak{g}_{a} = \mathfrak{sp}(N_{a})$, $\mathfrak{g}_{b} = \mathfrak{so}(N_{b})$, $R_{a} = \frac{1}{2} f_{a}$,
$R_{b} = f_{b}$.

\item $\mathfrak{g}_{a} = \mathfrak{sp}(N_{a})$, $\mathfrak{g}_{b} = \mathfrak{so}(N_{b})$, $N_{b} = 7, 8$, $R_{a} =
\frac{1}{2} f_{a}$, $R_{b} = s_{b}$ or $c_{b}$.

\item $\mathfrak{g}_{a} = \mathfrak{sp}(N_{a})$, $\mathfrak{g}_{b} = \mathfrak{su}(4)$, $R_{a} = \frac{1}{2} f_{a}$,
$R_{b} = \Lambda^{2}_{b}$.

\item $\mathfrak{g}_{a} = \mathfrak{sp}(N_{a})$, $\mathfrak{g}_{b}= \mathfrak{g}_{2}$, $R_{a} = \frac{1}{2} f_{a}$, $R_{b}
= f_{b}$.
\end{itemize}

The rules of subsection \ref{ssec:MATTER} make it clear how one can decorate a curve without
any neighbors. But once we start including neighbors, the allowed
representations and algebras are restricted by the mixed anomaly condition. In
order to satisfy equation (\ref{mixed}), there will be some minimal number of
hypermultiplets on each gauge algebra. If this number is greater than the number
required for gauge anomaly cancellation, the configuration is not allowed. For
instance, consider a configuration consisting of two adjacent curves with
self-intersection $-2$ carrying gauge algebras $\mathfrak{su}(2)$ and $\mathfrak{su}(5)$,
respectively. Gauge anomaly cancellation puts exactly $4$ fundamentals on the
$\mathfrak{su}(2)$ node. But, mixed anomaly cancellation requires the presence of a
bifundamental hypermultiplet $(\textbf{2},\textbf{5})$. This means there must be at least 5
$\mathfrak{su}(2)$ fundamentals which are rotated into each other under the $\mathfrak{su}(5)$
action, contradicting the gauge anomaly cancellation condition. We conclude
that this is not an acceptable theory.

On the other hand, suppose the $\mathfrak{su}(5)$ gauge group is replaced with an
$\mathfrak{su}(4)$. Then, the mixed representation will be $(\bf{2},\bf{4})$. There are 4
fundamentals on the $\mathfrak{su}(2)$ and 8 on the $\mathfrak{su}(4)$ of which 4 and 2 pair up,
respectively. Thus, one is left with an $\mathfrak{su}(2) \times \mathfrak{su}(4)$ quiver gauge theory with
6 extra fundamental hypermultiplets transforming under the $\mathfrak{su}(4)$ gauge symmetry.

This provides a systematic way to determine if a particular gauge group
content is allowed on a specified configuration of curves: list the necessary
matter content on each curve $\Sigma_{a}$ to satisfy the gauge anomaly constraints
for that gauge algebra, $\mathfrak{g}_{a}$. List the mixed representations $(R_{a},
R_{b_{i}})$ necessary to satisfy the mixed anomaly constraints between $\Sigma_{a}$
and its neighbors, $\Sigma_{b_{i}}$. If the number of hypermultiplets of $R_{a}$
required to satisfy these mixed anomaly constraints--given either by
$R_{b_{i}}$ or $\frac{1}{2}R_{b_{i}}$--is greater than the number of hypermultiplets of $R_{a}$ required
for anomaly cancellation, then this configuration is not allowed. This
procedure is iterated for all curves $\Sigma_{a}$, and if it passes all of them,
then the configuration is allowed.

In most cases, the allowed gauge algebras can be determined from the following
abbreviated list of rules, derived from the aforementioned constraints:

\begin{itemize}
\item Any $\mathfrak{so}(N), N \geq7$ appearing on a curve of self-intersection $-3$ or
greater can only have $\mathfrak{sp}(N^{\prime})$ algebras living on the adjacent curves.
In these cases, a half-fundamental of $\mathfrak{sp}(M)$ pairs with a fundamental of
$\mathfrak{so}(N)$, for $N \geq9$, or a spinor, in the special cases of $\mathfrak{so}(7)$ and
$\mathfrak{so}(8)$.

\item Any $-4$ curve with a single neighbor with $\mathfrak{sp}(M)$ gauge algebra must have
a gauge algebra $\mathfrak{so}(N), N\geq M + 8$. Any $-4$ curve with a neighbor on both the
left and right, gauge algebras $\mathfrak{sp}(M_{L}), \mathfrak{sp}(M_{R})$ must have a gauge algebra
$\mathfrak{so}(N), N \geq M_{L} + M_{R} + 8$. Any $-4$ curve with three neighbors forming
a T shape of gauge algebras $\mathfrak{sp}(M_{1}), \mathfrak{sp}(M_{2}),\mathfrak{sp}(M_{3})$ must have gauge
algebra $\mathfrak{so}(N), N \geq M_{1} + M_{2} + M_{3} + 8$. In these cases, a
half-fundamental of $\mathfrak{sp}(M_{i})$ pairs with the fundamental of $\mathfrak{so}(N)$.

\item Conversely, any $-1$ curve of gauge algebra $\mathfrak{sp}(M)$ with neighbors
carrying gauge algebra $\mathfrak{so}(N_{L}), \mathfrak{so}(N_{R})$ must satisfy $M \geq\frac{1}%
{4}(N_{L} + N_{R} + \delta_{N_{L},7} + \delta_{N_{R},7} -16)$. Here, $N_{L}$
or $N_{R}$ are set to zero if the $-1$ curve does not have a neighbor on the
left or right, respectively, and the Kronecker delta arises whenever the
spinor rep $\mathbf{8}$ of $\mathfrak{so}(7)$ is used rather than the fundamental, as
must be the case whenever the curve has self-intersection $-3$ or lower.

\item In a string of $-2$ curves, an $\mathfrak{su}(N)$ algebra between adjacent
$\mathfrak{su}(N_{L})$ and $\mathfrak{su}(N_{R})$ must satisfy the convexity condition $N \geq
\frac{1}{2}(N_{L} + N_{R})$, with $N_{L}$, $N_{R}$ set to zero if there is no
neighbor to the left or right, respectively.  A $-2$ curve carrying $\mathfrak{su}(N)$ with three $-2$ curve neighbors carrying $\mathfrak{su}(N_{L})$, $\mathfrak{su}(N_{R})$, $\mathfrak{su}(N_{T})$ must satisfy $N \geq
\frac{1}{2}(N_{L} + N_{R} + N_T)$

\item A $-2$ curve adjacent to a $-3$ curve must have $\mathfrak{su}(2)$ gauge algebra, and
the total dimensionality of the reps on the left and right of the $-2$ curve
can be no more than $8$. In particular, this means that the other curve
adjacent to this $-2$ curve cannot carry a gauge algebra.

\item A $-3$ curve with two $-2$ neighbors carrying gauge algebra $SU(2)$ can
only have gauge algebra $\mathfrak{so}(7)$. A $-3$ curve with one $-2$ curve
neighbor carrying gauge algebra $\mathfrak{su}(2)$ can have gauge algebra $\mathfrak{so}(7)$
or $\mathfrak{g}_{2}$.

\item $\mathfrak{su}(3)$ can only appear on a $-3$ curve if that curve has no neighbors
with gauge groups.

\item $\mathfrak{f}_{4}$, $\mathfrak{e}_{6}$, $\mathfrak{e}_{7}$, and $\mathfrak{e}_{8}$ do not permit their neighbors to
have gauge groups.

\end{itemize}
In addition, one also often needs the ``gauging condition," elaborated upon in subsection \ref{ssec:Unpaired}.
\begin{itemize}

\item The sum of the gauge algebras on curves touching a single $-1$ curve must be smaller than $\mathfrak{e}_8$.

\end{itemize}

\subsubsection{The Top Down View}

Occasionally, the constraints imposed by anomaly cancellation for continuous gauge symmetries are
insufficient--particularly for geometries with small numbers of curves. Indeed, some putatively self-consistent
field theory realizations of a tensor branch theory appear to have an obstruction to an embedding in
F-theory. The close correspondence between F-theory and field theory found elsewhere leads us to strongly suspect that there is
a pathology in these field theories.

To illustrate these general points, observe that anomaly cancellation considerations alone do not exclude the possibility of a $-3$, $-2$, or $-1$ curve holding gauge algebra $\mathfrak{so}(8)$ touching a $-2$ curve with gauge algebra $\mathfrak{su}(2)$, nor does it rule out the possibility of three $-2$ curves holding $\mathfrak{so}_7$, $\mathfrak{su}_2$, and no gauge algebra, respectively (provided the mixed representation of $\mathfrak{so}_7 \oplus \mathfrak{su}_2$ is $\frac{1}{2}(\mathbf{7},\mathbf{2})$).  Nonetheless, an investigation of F-theory fiber types reveals that these cases are not allowed.  Furthermore, one must also take into account the fact that the collision point of a $-2$ curve with type $II$ fiber with a curve with gauge algebra $\mathfrak{su}(2)$ holds some matter of the $\mathfrak{su}(2)$.  (There is also an ``extra'' tensor associated to the type $II$ curve, as will be discussed in more detail in the next section.)  This is particularly relevant to the cluster $3,2,2$, where we might have na\"ively attempted to enhance from a $\mathfrak{g}_2$ algebra on the $-3$ curve to either $\mathfrak{so}_7$  or $\mathfrak{so}_8$. In
Appendix~\ref{appendixF} we analyze this and other possibilities, and find that an enhancement to $\mathfrak{so}_8$ can never occur without also
introducing further blowups in the base. In the language of $(p,q)$ seven-branes this is because the $-2$ curve is wrapped by a
$A^3 C$ bound state (in the terminology of references
\cite{Gaberdiel:1997ud, DeWolfe:1998bi, DeWolfe:1998zf}\footnote{See also \cite{Grassi:2013kha,Grassi:2014sda,Grassi:2014zxa,arXiv:1410.6817} for more recent work using this approach.}), so roughly speaking the $C$ factor leads to an orientifold plane. The collision with the orientifold plane from the
$\mathfrak{so}_8$ factor yields a contradiction.

Though we leave a full analysis of these field theories for future work, it is helpful to use the lack of an F-theory realization
as a guide to the location of potential pathologies for these field theories. In F-theory, the basic issue centers around an $\mathfrak{sp}_1$
gauge theory which is realized by a non-split type $IV$ fiber. The fact that this is distinct from a type $I_2$ fiber, or a non-split type $I_3$ fiber suggests that the global structure of the gauge group and flavor symmetries may be different. For example, when we decorate a $-2$ curve with
a non-split type $IV$ fiber, we have eight half hypermultiplets of $\mathfrak{sp}_1$. Whenever this fiber type meets a $\mathfrak{so}$ algebra, we must have matter in a spinor representation of this algebra. This additional structure immediately excludes the case of a sequence of three $-2$ curves which respectively support gauge algebras $\mathfrak{so}_7$, $\mathfrak{sp}(1)$ and a type $II$ fiber. Since there is already a single half hypermultiplet of $\mathfrak{sp}_1$ trapped at the collision of the type $II$ and $\mathfrak{sp}_1$, this would in turn mean that the remaining matter for the $\mathfrak{sp}_1$ factor are half hypermultiplets in
the $(\mathbf{7} , \mathbf{2})$ of $\mathfrak{so}_7 \times \mathfrak{su}_2$, rather
than a spinor of $\mathfrak{so}_7$--a contradiction.

By a similar token, we can also consider an isolated $-2$ curve with non-split type $IV$ fiber. The flavor symmetry algebra is $\mathfrak{so}_8$. What seems to be suggested by the F-theory realization of this theory is that the flavor group is $Spin(8) / \mathbb{Z}_2$ instead of $SO(8)$ or $Spin(8)$. Indeed, the case of $SO(8)$ is problematic since it contains no spinor representations. Further, we have already observed difficulty in gauging this algebra. Anomaly cancellation requires a triality invariant spectrum consisting of an equal number of $8_s$'s, $8_c$'s and $8_v$'s. This is broken by working with $Spin(8) / \mathbb{Z}_2$ since one of these spinors is projected out. This would provide a potential explanation for why the $\mathfrak{so}_8$ flavor symmetry algebra cannot be consistently gauged in the above example.

\subsection{Unpaired Tensors \label{ssec:Unpaired}}

Although each gauge theory must come with a corresponding tensor
multiplet (which controls the value of the gauge coupling) the converse need
not hold. Indeed, the $(2,0)$ theories have no non-abelian gauge group on
their tensor branch, but have many tensor multiplets. Another example is the
$(1,0)$ E-string theory coming from a single $-1$ curve. In this subsection we
examine in detail the theories which have such unpaired tensors.

Let $\Sigma$ be one of the curves which is contracted in order to produce
a given SCFT, and let us compare that given SCFT with the theory
$\mathcal{T}_\Sigma$ obtained from
contracting $\Sigma$ alone.  On the tensor branch of the original theory,
the left and right neighbors of $\Sigma$ will be associated to gauge algebras
$\mathfrak{g}_{L}$ and $\mathfrak{g}_{R}$ (either of which may be absent),
and in contracting $\Sigma$ alone, the couplings of those gauge fields will
go to $0$, leaving us with a subgroup $G_L\times G_R$ of the global
symmetry group of $\mathcal{T}_\Sigma$.

If $\Sigma$ has an associated gauge algebra, then field theory can be
used to predict the global symmetry group of $\mathcal{T}_\Sigma$:  the
matter content can be determined from anomaly cancellation, and field
theory then predicts the global symmetry group of that matter
representation.\footnote{A detailed verification that the constraints
from F-theory are compatible with this field theory prediction will be
made in \cite{BMM}.}
However, if $\Sigma$ corresponds to an unpaired tensor, we must use
other methods.  Note that any such $\Sigma$ has $\Sigma^2=-1$ or
$\Sigma^2=-2$.

In the case of $\Sigma^2=-1$, the theory $\mathcal{T}_\Sigma$ has
an $E_8$ global symmetry at the conformal point,
which means that the gauge algebras on the
left and right $\mathfrak{g}_{L},\mathfrak{g}_{R}$ of the $-1$ curve must satisfy $\mathfrak{g}_{L}\oplus
\mathfrak{g}_{R}\subset \mathfrak{e}_{8}$.  Ordinarily, the commutant of $\mathfrak{g}_{L}\oplus
\mathfrak{g}_{R}\subset \mathfrak{e}_{8}$ will be the global symmetry associated to $\Sigma$.  However, F-theory at times imposes tighter restrictions than we would expect from a field theory analysis.  Consider the configuration:
\begin{equation*}
{\overset{\mathfrak{e_{7}%
}}{8}}\,\, \underset{[SU(2)]?}{1} \,\,2\,\,\overset{\mathfrak{su_{2}}%
}{2}\,\,\overset{\mathfrak{g_{2}}}{3}
\end{equation*}
The $[SU(2)]?$ indicates that we might initially expect an $SU(2)$ flavor symmetry could live on the curve of self-intersection $-1$.  However, upon realizing this configuration in F-theory, we see that
the 2-2-3 non-Higgsable clusters
must include a Kodaira type $II$ fiber on the corresponding curves with
no gauge symmetry, leading to the refined description:
\begin{equation*}
{\overset{\mathfrak{e_{7}%
}}{8}}\,\,\underset{[SU(2)]?}{1}\,\,\overset{II}2\,\,\overset{\mathfrak{su_{2}}%
}{2}\,\,\overset{\mathfrak{g_{2}}}{3}
\end{equation*}
Now the $-1$ curve meets the $\mathfrak{e}_7$ brane, one of
the type
$II$ branes, and the global $[SU(2)]$ brane.  The discriminant must vanish
$12$ times along the $-1$ curve; however, it vanishes to order $9$ at
the intersection with the $\mathfrak{e}_7$ brane, it vanishes to order $2$
at the type $II$ brane, and it must vanish to order at least $2$ at any hypothesized
global $[SU(2)]$ brane.  Thus, the total order of vanishing is at least
$9+2+2=13$, which is a contradiction.  In other words, some restrictions
from F-theory beyond a purely field-theory analysis show that this model
does not have any $SU(2)$ flavor symmetry.

If $\Sigma^2=-2$, we find that the theory $\mathcal{T}_\Sigma$
at the conformal point is the tensor product
of the $A_1$ (2,0) theory with $N$ free (uncharged) hypermultiplets,
with the possible values of $N$ being $0$, $1$, $2$, or $4$.  The
precise value is determined by the details of the F-theory
compactification as described in Appendix~\ref{appendixF}.  When $N>0$,
the observed global symmetry from F-theory is always $SU(2)$
which acts nontrivially on the hypermultiplets but has trivial
action on the (2,0) sector of the theory.
(The hypermultiplets transform in $N/2$ copies of the fundamental
representation of $SU(2)$.)
For $N=1$, this suggests that the single hypermultplet is real, which
would give a global symmetry of $Sp(1)\cong SU(2)$.  For $N=2$,
this suggests that the hypermultiplets are complex, which would give
a global symmetry of $SU(2)$.  One possible explanation of the case
$N=4$ is that the hypermultiplets are pseudo-real, which would give
a global symmetry of $SO(4)=SU(2)_L\times SU(2)_R$, of which F-theory
only realizes one of the $SU(2)$'s.  We leave a detailed investigation
of this point to future work.

The case $N=0$ corresponds to Kodaira type $I_0$ along $\Sigma$,
the case $N=2$ corresponds to Kodaira type $I_1$, but both the
cases $N=1$ and $N=4$ correspond to Kodaira type $II$.  When the $SU(2)$
global symmetry is gauged, the localization
of the matter is different depending on whether the $\mathfrak{su}(2)$
gauge symmetry is realized on a curve of Kodaira type $III$ or
Kodaira type $IV$.
The total number of hypermultiplets on the curve with gauge symmetry $\mathfrak{su}_2$ is fixed by anomaly cancellation, so the fact that some of these hypermultiplets localize on the intersection with the empty $-2$ curve has significant implications for mixed anomaly considerations.

Let us give some examples of the interaction of these unpaired tensors with the remainder of the theory. As a first example,
consider the configuration of curves:
$$(12),1,2,2,3,1,5,1,3,2,2,1,(12)$$
i.e., the conformal matter between two $E_{8}$ factors. Anomaly cancellation
requires the $-2$ curves adjacent to $-3$ curves to carry the gauge algebra
$\mathfrak{sp}_{1}$, and it requires the other $-2$ curves to be empty. Furthermore, since seven of the eight half-fundamentals on the $-2$ curves with gauge symmetry $\mathfrak{su}_2$ transform under the mixed representation $\frac{1}{2}(\textbf{2},\textbf{7})$ there is only a single half-fundamental left.  As a result, the empty
$-2$ curves must have global symmetry $SU(2)$ acting on a single half-fundamental (which implies that the Kodaira type is $II$).
Furthermore, the $-1$ curves adjacent to these $-2$ curves must also be empty,
since we have already accounted for all of the hypermultiplets acted upon by the flavor symmetry of the $-2$ curve theory.

Consider also the $-1$ curves adjacent to the $-5$ curve. Anomaly cancellation
requires the $-1$ curve theory to not carry a gauge group, and in F-theory terms the fiber is of type II.
But then, the product of the gauge algebra on the $-5$ curve and the product of the gauge
algebra on the $-3$ curves must be a subset of $E_{8}$. Anomaly considerations
require the $-5$ curve to hold gauge algebra $\mathfrak{f}_{4}$, $\mathfrak{e}_{6}$, or $\mathfrak{e}_{7}$ and
the $-3$ curve to hold gauge algebra $\mathfrak{g}_{2}$. But we see
now that not all possible pairs are allowed: $\{\mathfrak{e}_{6},\mathfrak{e}_{7}\}\oplus
\{\mathfrak{g}_{2}\}$ are not subsets of $\mathfrak{e}_{8}$, so they are not allowed.  This leaves only the
possibility $\mathfrak{f}_{4}\oplus \mathfrak{g}_{2}$, so the $-5$ curve must carry gauge algebra
$\mathfrak{f}_{4}$.  This recovers the
general structure already predicted by the F-theory geometry (see appendix C
in \cite{Heckman:2013pva}).

For another example, consider a chain of three $-2$ curves, the first of which has no gauge group, the second of which has gauge group $\mathfrak{su}_2$, and the third of which has yet to be constrained:
$$
2 \,\, \overset{\mathfrak{su_2}}2 \,\, \overset{?}2
$$
If studied in isolation, the $-2$ curve without a gauge symmetry must have a global symmetry $SU(2)$, which is then gauged by the $-2$ curve with $\mathfrak{su}_2$ gauge symmetry.  There can be either one, two, or four half-fundamentals of $\mathfrak{su}_2$ living on the intersection of these two curves.  However, there are only eight half-fundamentals allowed on the $-2$ curve with gauge algebra $\mathfrak{su}_2$, so the other $-2$ curve cannot hold a gauge algebra that is too big.  If one were to try to place a $\mathfrak{so}_8$ gauge algebra on it, for instance, one would need at least 9 half-fundamentals on the $\mathfrak{su}_2$ $-2$ curve--8 for the mixed $\frac{1}{2}(\textbf{2},\textbf{8})$ representation and one for the intersection with the empty $-2$ curve.  This clearly violates the anomaly cancellation condition on the $-2$ curve, and so we conclude that the only possible gauge symmetries on the third $-2$ curve are $\mathfrak{su}_2$, $\mathfrak{su}_3$, $\mathfrak{g}_2$, or no gauge symmetry.  One might also have expected $\mathfrak{so}_7$ as a possibility from anomaly considerations alone, but as discussed in the last subsection, this does not occur in F-theory.

Finally, note that in the conformal limit of a shrinking $-2$ curve,
there is an additional $SO(5)$ global symmetry of the (2,0) theory (the $R$-symmetry),
which does not act on the decoupled hypermultiplets.  This
same symmetry is present for any of the ADE (2,0) theories.

\subsection{Decorating a Base}

Having derived a number of consistency conditions on possible ways to enhance
a fiber, we now return to our original task of how to enhance the fibers of a
given base. It is helpful to break up our analysis into possible enhancements
on the nodes of a base, and the possible enhancements on the links.

The first point is that in a base, we generically cannot enhance any of
the nodes or links in the interior. First of all, the exceptional interior
nodes of a base generically do not support any enhancement in the fiber
type. The reason is that in subsection \ref{ssec:minINTlinks}, we already
argued that a node could only be joined by minimal interior links. The
exceptions to this rule occur at the two leftmost and rightmost nodes of the
base quiver. By the same token, we cannot enhance the fiber type for our
minimal interior links of a base quiver.

The main class of base geometries where a fiber enhancement is possible are
those where we exclude the possibility of an exceptional base node.
This occurs for bases comprised solely of $-1$, $-2$ and $-4$ curves. In fact,
these are the cases which can also be realized in perturbative IIB\ string
theory, and so we will refer to them as \textquotedblleft semi-classical
bases\textquotedblright -- they are not completely classical because we shall allow
for matter in spinor representations--. We also classify all possible ways to
enhance these base geometries. This covers all perturbative type II\ string
theory realizations of 6D\ SCFTs. The remaining item in our classification
will then be to study whether such classical configurations can in fact
\textquotedblleft rejoin\textquotedblright\ to a configuration that contains
exceptional base nodes. This occurs whenever one of the ends reduces
back to the minimal fiber type. Then, it can successfully rejoin the more
general base quiver.

\subsubsection{Decorating the $(2,0)$ Theories}

One class of geometries where we can consider adding additional seven-branes are the $(2,0)$ theories. One straightforward construction involves adding
stacks of D7-branes, i.e., $I_N$ type fibers, which introduces
perturbative $\mathfrak{su}(N)$ gauge symmetry.
For illustrative purposes, we focus on the case where we just have $\mathfrak{su}_N$ factors. Anomaly cancellation imposes the condition that we have only matter fields transforming in the fundamental representation of each $\mathfrak{su}$ factor. So, for an $\mathfrak{su}(N)$ gauge algebra, we have $2N$ total flavors on each node.

To satisfy the condition of anomaly cancellation on a given stack of seven-branes, it will often be necessary to introduce additional flavor symmetries. Labelling the nodes as $i=1,...,r$ for the Dynkin diagram with $r$ compact $-2$ curves, we can introduce $F_i$ flavors. Introducing the adjacency matrix $A_{ij}$ for each Dynkin diagram, we get the constraint:
\begin{equation} \label{weightroot}
\underset{j}{\sum} A_{ij} N_{j} = F_{i}.
\end{equation}
In this basis, the inverse of the adjacency matrix has positive entries valued in the rational numbers.\footnote{Each cofactor also defines a positive definite adjacency matrix, so its determinant is also positive. The claim then follows.} This means that we can solve the linear equations to find values of the $N_j$. The main constraint is that we need to have a solution over the integers. One simple class of solutions is obtained by multiplying each flavor symmetry factor by $\det A$, though of course there are others.

Note also that for the $E_8$ Dynkin diagram, \textit{any} choice of flavor symmetry will generate a consistent solution since the intersection form is unimodular (i.e., $\det A = 1$). At this point it should be clear that we have significantly constrained the possible structure of such decorations. We will meet a few additional constraints, for example, that the ranks of the gauge groups must increase as we proceed towards the interior. This echoes the constraint found on the structure of the base geometries.

Turning the discussion around, we can ask whether a given configuration of $N_{i}$'s can lead to a consistent theory, namely by introducing an appropriate flavor symmetry. For this to be the case, we need to show that there exists a collection of $F_{i}$'s which are all non-negative. There is an interesting bit of geometric structure here: All of the consistent 6D theories obtained in this way form a cone, that is, we have a collection of vectors such that any positive linear combination of them over the natural numbers gives us yet another element of the cone. Indeed, suppose we have two vectors of solutions $\overrightarrow{N}$ and $\overrightarrow{N}^{\prime}$, with corresponding flavor vectors $\overrightarrow{F}$ and $\overrightarrow{F}^{\prime}$, respectively. Then, we also observe that $\overrightarrow{N} + \overrightarrow{N}^{\prime}$ also leads to a consistent theory with flavor vector $\overrightarrow{F} + \overrightarrow{F}^{\prime}$. This is the condition for us to form a cone. Furthermore, observe that equation (\ref{weightroot}) can be interpreted as saying that each vector $\overrightarrow{N}$ is a positive vector in the weight lattice, and $\overrightarrow{F}$ is a positive vector in the root lattice for the corresponding Lie algebra. To complete the classification, it is enough to observe that each $F_{i}$ just needs to be non-negative. So a necessary and sufficient condition is:
\begin{equation}\label{convexityADE}
\underset{j}{\sum} A_{ij} N_{j} \geq 0.
\end{equation}

\subsubsection{The Semi-Classical Configurations \label{ssec:SEMIclass}}

In this section, we give some further examples, which we shall refer to as ``semi-classical''. These
are configurations where all of the maximally Higgsed gauge groups are classical, but where we also entertain the possibility of
spinors for the $\mathfrak{so}$ factors (upon further enhancement). First, we detail the
case of the classical configurations, namely those which do not have spinor representations or fibers of type $II$. We then turn to examples with spinor representations.

\begin{itemize}
\item The ADE configurations of $-2$ curves. The main condition is the convexity condition detailed in line
(\ref{convexityADE}). It is possible also for none of the curves to have a gauge group.
But as soon as any of the curves is given a gauge
group, all of the other $-2$ curves automatically get one as well with the exception that the outer $-2$ curves can have fiber type $I_1$.  Nonetheless, we may still use the convexity condition (\ref{convexityADE}) as long as we consider the $-2$ curves with fiber type $I_1$ to have $N_j =1$, i.e. morally we consider the empty $-2$ curves to have gauge symmetry $\mathfrak{su}(1)$.  With this caveat, the convexity condition is necessary and sufficient to determine the classical fiber enhancements of ADE configurations of $-2$ curves.

DE configurations of $-2$ curves obey the same rules, with the addition of a single extra rule regarding the trivalent vertex $2\overset{2}22$.  Here, if the gauge symmetries are of the form $\mathfrak{su}(N_L),\overset{\mathfrak{su}(N_T)}{\mathfrak{su}(N_M)},\mathfrak{su}(N_R)$, we have the condition $2 N_M \geq N_L + N_R + N_T$.  Once again, the case of an empty $-2$ curve is handled by setting $N_j=1$.

\begin{figure}[ptb!]
\begin{center}
\includegraphics[trim=5mm 25mm 10mm 10mm,clip,width=140mm]{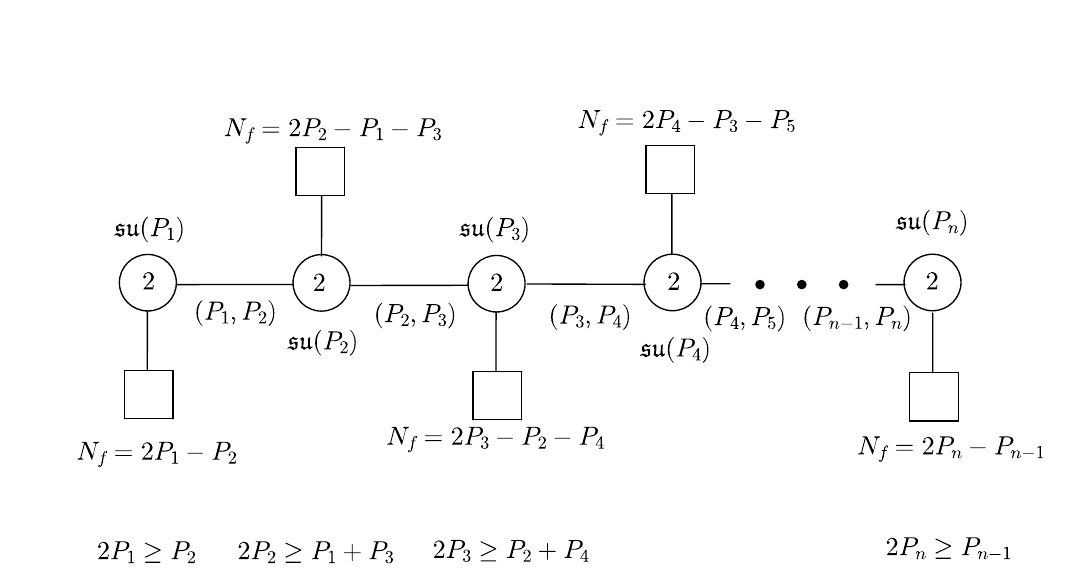}
\end{center}
\caption{Quiver Diagram for the 2...2 configurations.}%
\end{figure}

\item For 1222...2, the $-2$ curves each carry gauge algebra $\mathfrak{su}(P_{i})$,
satisfying the convexity constraint $2P_{i}\geq P_{i-1}+P_{i+1}$. If the gauge
algebra on the $-1$ curve is $\mathfrak{sp}(M^{\prime})$, we get the constraints
$4M^{\prime}\geq 2 P_{1}-16$, $2P_{1}\geq2M^{\prime}+P_{2}$. If the gauge algebra
on the $-1$ curve is $\mathfrak{su}(P^{\prime})$, then we get the constraints $P^{\prime
}+8+\delta_{P^{\prime},3}+\delta_{P^{\prime},6}\geq P_{1}$, $2P_{1}\geq
P^{\prime}+P_{2}$. Here, the Kronecker deltas arise because the gauge algebras
$\mathfrak{su}(3)$ and $\mathfrak{su}(6)$ each may have an extra fundamental hypermultiplet as a part of
their matter content on a $-1$ curve.  The $-1$ curve can also be empty
provided $P_{1}\leq9$, so that the gauge group on the adjacent $-2$ curve is a
subgroup of $\mathfrak{e}_{8}$.  As in the case of configurations with only $-2$ curves, we may think of an empty $-2$ curve as morally having gauge symmetry $\mathfrak{su}(1)$ and applying the usual convexity conditions to it.

\begin{figure}[ptb!]
\begin{center}
\includegraphics[trim=10mm 18mm 10mm 20mm,clip,width=140mm]{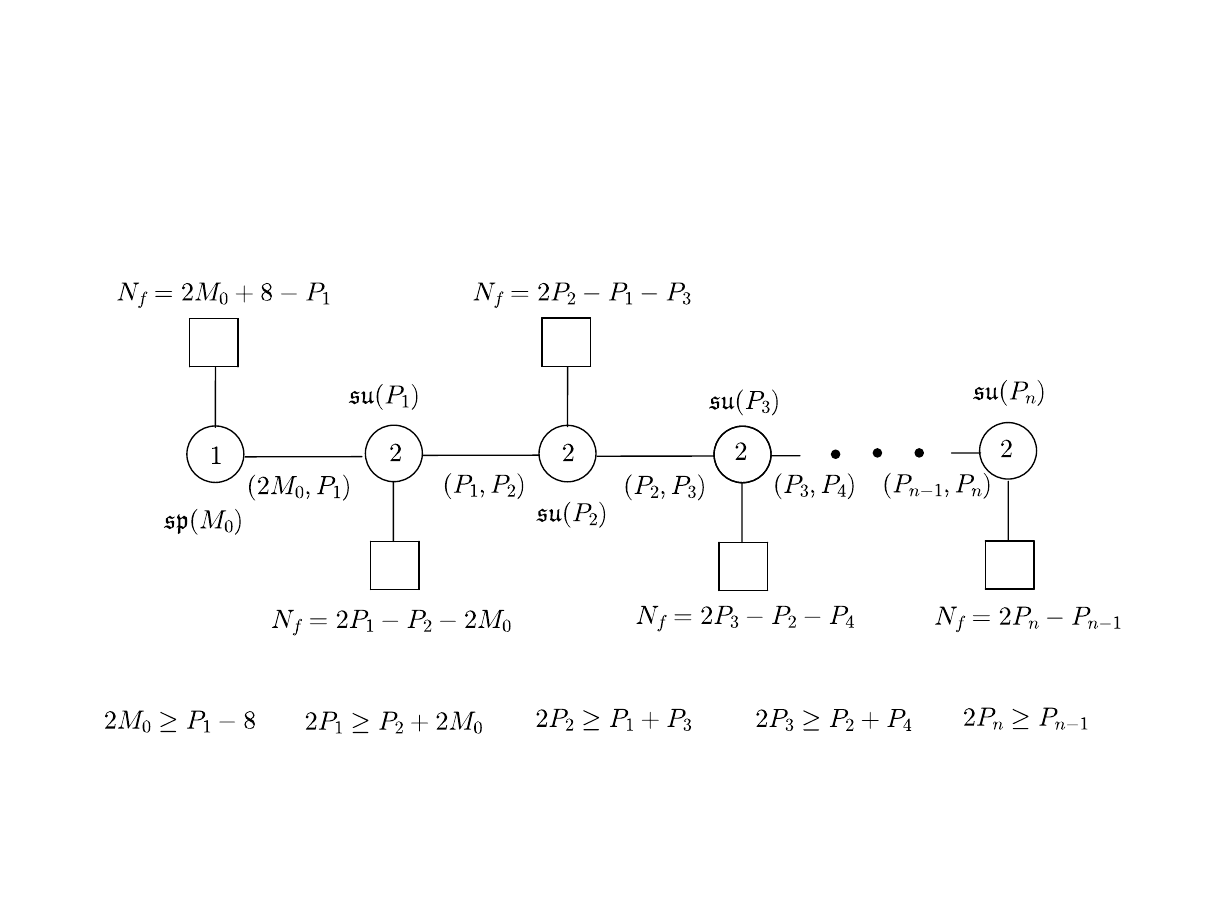}
\end{center}
\caption{Quiver Diagram for the 12...2 configurations.}%
\end{figure}

\item The $-4$ curves of the blowups (1)4141...414(1) must have gauge algebra
$\mathfrak{so}(N_{i}),N_{i}\geq8$. The $-1$ curves must have gauge algebra $\mathfrak{sp}(M_{i})$.
These must satisfy $4M_{i}\geq N_{i}+N_{i+1}-16$, $N_{i}\geq M_{i}+M_{i-1}+8$.
In particular, this imposes convexity constraints, $2N_{i}\geq N_{i+1}%
+N_{i-1}$, $2M_{i}\geq M_{i+1}+M_{i-1}$.

The inner $-1$ curves can be empty provided all of the $-4$ curves have gauge
algebra $\mathfrak{so}(8)$. Any outer $-1$ curves just need their adjacent $-4$ curve to
have $\mathfrak{so}(N)$, $N\leq16$.

\begin{figure}[ptb!]
\begin{center}
\includegraphics[trim=5mm 20mm 10mm 14mm,clip,width=140mm]{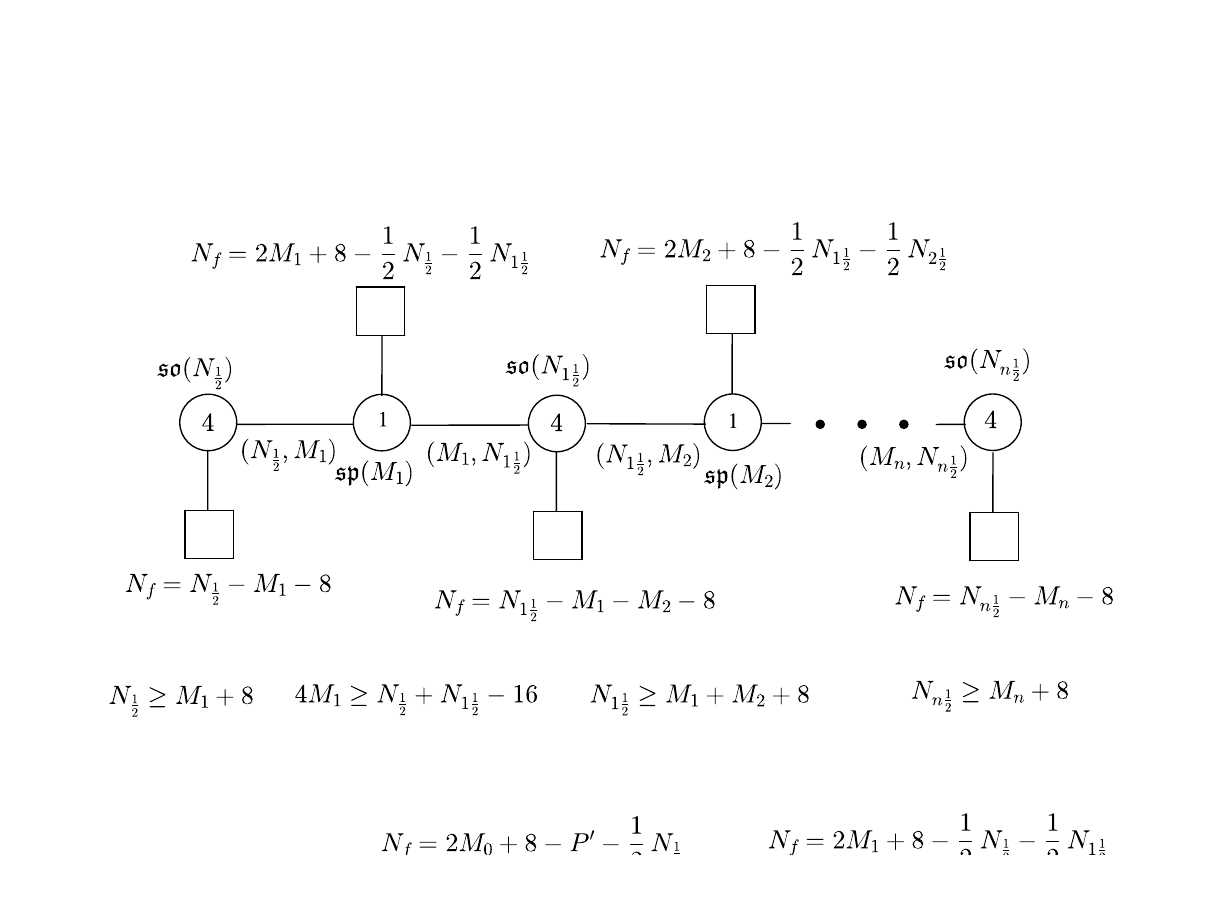}
\end{center}
\caption{Quiver Diagram for the 41...14 configurations.}%
\end{figure}

\item For 214...14 the $-4$ curves must have gauge algebra $\mathfrak{so}(N_{i}),N_{1}%
\geq8$. The $-1$ curves must have gauge algebra $\mathfrak{sp}(M_{i})$. The $-2$ has gauge
algebra $\mathfrak{su}(P^{\prime})$, with $2P^{\prime}\geq 2 M_{1}$, $4M_{1}\geq
N_{1}+2P^{\prime}-16$.

The $-1$ curve adjacent to the $-2$ curve can be empty provided $\mathfrak{so}(N_{1}%
)\oplus \mathfrak{su}(P^{\prime})$ is contained $\mathfrak{e}_{8}$.
The $-2$ curve can be empty provided the adjacent $-1$ curve is also empty.

\begin{figure}[ptb!]
\begin{center}
\includegraphics[trim=5mm 22mm 10mm 15mm,clip,width=140mm]{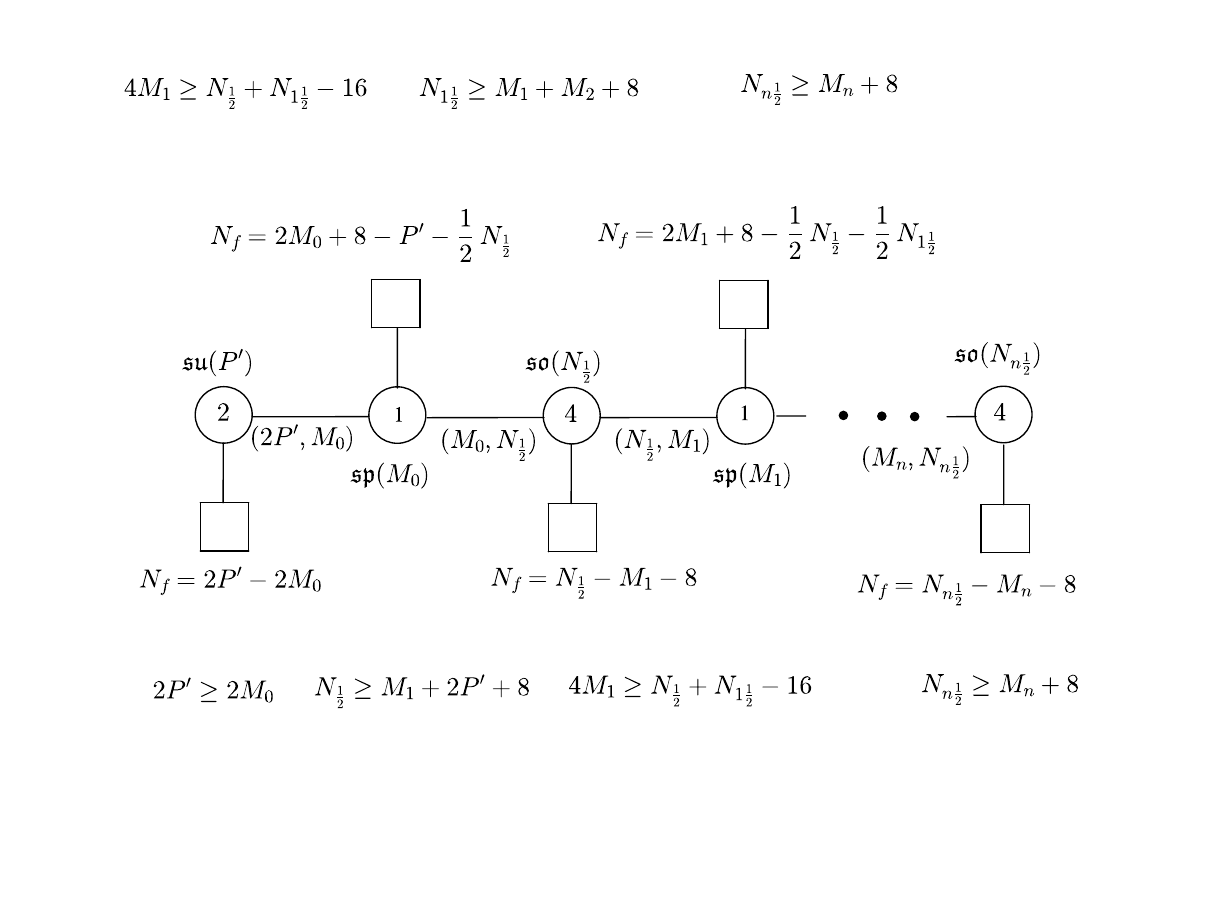}
\end{center}
\caption{Quiver Diagram for the 2141...14 configurations.}%
\end{figure}

\item For (4)141...41$\overset{1}{4}$1(4), the $-4$ curves must have gauge
algebra $\mathfrak{so}(N_{i}),N_{i}\geq8$.  The $-1$ curves must have gauge algebra
$\mathfrak{sp}(M_{i})$. These must satisfy $4M_{i}\geq N_{i}+N_{i-1}-16$, $N_{i}\geq
M_{i}+M_{i+1}+8$. The extra $-1$ curve must have gauge algebra $\mathfrak{sp}(M^{\prime})$,
with $4M^{\prime}\geq N_{k}-16$, where the $-4$ curve touching this extra $-1$
has gauge algebra $\mathfrak{so}(N_{k})$. Further, $N_{k}\geq M_{k}+M_{k+1}+M^{\prime}+8$.

The inner $-1$ curves can be empty provided their adjacent $-4$ curves have
gauge group $\mathfrak{so}(8)$.  The outer $-1$ curves (including the one at the top of
the T) just need their adjacent $-4$ curve to have $\mathfrak{so}(N)$, $N\leq
16$.

\begin{figure}[ptb!]
\begin{center}
\includegraphics[trim=5mm 7mm 5mm 8mm,clip,width=140mm]{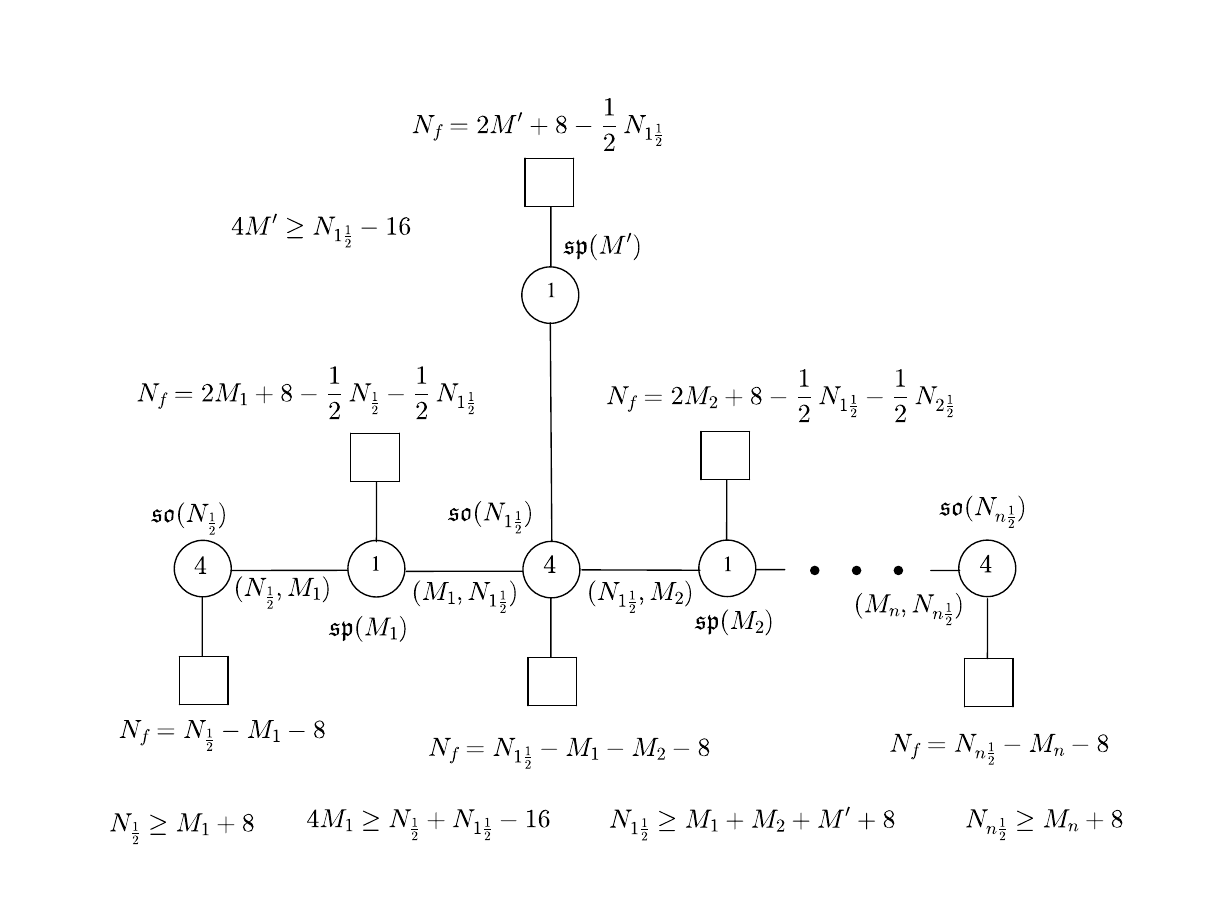}
\end{center}
\caption{Quiver Diagram for the (4)141...41$\protect\overset{1}{4}$1(4)...
configurations.}%
\end{figure}

\end{itemize}

If one allows spinor representations to live on the fibers above the classical bases, the possibilities are only slightly more complicated.

\begin{itemize}
\item
Configurations with alternating $-4$ and $-1$ curves cannot be enhanced at all except in the case that there is only a single $-4$ curve.  In such a configuration, the gauge symmetry on the $-4$ curve may be enhanced to $\mathfrak{f}_4$, $\mathfrak{e}_6$, or $\mathfrak{e}_7$ with the appropriate matter, and in all such cases the adjacent $-1$ curves must be empty.
\item
The configurations with only $-2$ curves also admit a limited number of possible enhancements.  In particular, a $-2$ curve may hold gauge symmetry $\mathfrak{g}_2$, $\mathfrak{so}(7)$, or $\mathfrak{so}(8)$, but in the $\mathfrak{so}(8)$ case, there can only be a single $-2$ curve.  In the case of a curve with $\mathfrak{so}(7)$ or $\mathfrak{g}_2$, any adjacent $-2$ curves must hold $\mathfrak{su}(2)$ gauge symmetry.  Further, these $-2$ curves with $\mathfrak{su}(2)$ gauge symmetry cannot touch any other $-2$ curves in the case of $\mathfrak{so}(7)$, and they can only touch empty curves in the case of $\mathfrak{g}_2$ (these will in fact have fibers of Kodaira type II).

Empty $-2$ curves obey a convexity condition similar to that for curves with gauge algebras.  In particular, any $-2$ curve without gauge symmetry can only touch another $-2$ curve if it has gauge symmetry $\mathfrak{su}(2)$ or smaller.  If it touches one curve with gauge symmetry $\mathfrak{su}(2)$, it cannot touch any other $-2$ curve.  As a result, we cannot have chains of more than one empty $-2$ curve followed by a curve with gauge symmetry $\mathfrak{su}(2)$.  This is not so different from the fact that chains of more than one $-2$ curve with gauge symmetry $\mathfrak{su}(M)$ cannot be followed by a curve with gauge symmetry $\mathfrak{su}(N)$ if $N > M$, which is a consequence of the convexity condition described previously.

As a result, any configuration of $-2$ curves can either take a classical form or a non-classical form.   Non-classical configurations include a $\mathfrak{g}_2$ or $\mathfrak{so}(7)$ symmetry touching nothing but $\mathfrak{su}(2)$ curves.  If the symmetry is $\mathfrak{g}_2$, the $-2$ curves holding $\mathfrak{su}(2)$ gauge symmetry can touch empty $-2$ curves, which will have Kodaira type $II$.  Two examples of such configurations are:
$$
2 \,\, \overset{\mathfrak{su_2}}2 \,\, \underset{[Sp(2)]}{\overset{\mathfrak{g_2}}2} \,\, \overset{\mathfrak{su_2}}2 \,\, 2
$$
$$
{[Sp(3) \times Sp(1)]} \,\, {\overset{\mathfrak{so_7}}2} \,\, \overset{\mathfrak{su_2}}2
$$
Here and henceforth, flavor symmetries are shown in square brackets.  Often, there will be additional abelian flavor symmetries, but such global $U(1)$ symmetries will typically be anomalous. Finally, a product of flavor groups is simply shorthand for the fact that the gauge theory may
contain matter fields in different irreducible representations.

In addition, non-classical configurations also include configurations which from the F-theory perspective contain empty curves with type $II$ fibers rather than type $I_1$ fibers.  When such a curve intersects a $-2$ curve with gauge symmetry $\mathfrak{su}(2)$ of Kodaira type $IV$, there will only be a half-fundamental of the $\mathfrak{su}(2)$ gauge symmetry at the intersection rather than a full fundamental.  As a result, this $\mathfrak{su}(2)$ curve can actually intersect two empty $-2$ curves along with another $-2$ curve holding $\mathfrak{su}_3$.  This introduces new configurations only in the case of D configurations of $-2$ curves.  The following sequences arise in this way:

$$
2 \,\, {\overset{2}{\overset{\mathfrak{su}_2}{2}}} \,\, \overset{\mathfrak{su}_3}{2} \,\, ... \,\, \overset{\mathfrak{su}_3}{2}
$$
\begin{equation}
2 \,\, {\overset{2}{\overset{\mathfrak{su}_2}{2}}} \,\, \overset{\mathfrak{su}_3}{2} \,\, ...\,\, \overset{\mathfrak{su}_3}{2} \,\, \overset{\mathfrak{su}_2}{2}
\label{threebonuscases}
\end{equation}
$$
2 \,\, {\overset{2}{\overset{\mathfrak{su}_2}{2}}} \,\, \overset{\mathfrak{su}_3}{2} \,\, ...\,\, \overset{\mathfrak{su}_3}{2} \,\, \overset{\mathfrak{su}_2}{2} \,\, 2
$$
The empty $-2$ curves here all have fiber type $II$ in the F-theory picture.  These configurations are not classical configurations, for if one were to attempt to realize them with $I_1$ fibers on the empty $-2$ curves, one would violate the convexity conditions on the $-2$ curve with three neighbors.  Note that it is not possible to achieve $\mathfrak{su}_n$ for $n>3$--this is due to the fact that the $II$, $III$, $IV$ sequence of Kodaira fiber types has a maximal gauge group of $\mathfrak{su}_3$.

Of course, it is also possible to duplicate the gauge symmetries of many of the classical configurations using the non-classical fiber types.  For instance, one could achieve the configuration

$$
2 \,\, \overset{\mathfrak{su}_2}{2} \,\, \overset{\mathfrak{su}_3}{2}
$$
using either classical fiber types $I_1$, $I_2$, $I_3$ or non-classical fiber types $II$, $IV^{ns}$, $IV^s$.  However, we do not ordinarily expect these to give rise to distinct field theories.  This is confirmed in large part by our correspondence between certain F-theory configurations and homomorphisms $\mathbb{Z}_N \rightarrow E_8$ in subsection \ref{ssec:ASERIES}.  As a result, we will usually specify only the gauge algebra rather than the specific fiber type.  An exception to this is the case of a single empty $-2$ curve discussed in subsection \ref{ssec:Unpaired}, which gives rise to four different theories depending on the global symmetry and matter content, which in turn depends on the fiber type of this $-2$ curve.  To differentiate it from an empty $-2$ curve with trivial global symmetry, we will often use the symbol $\overset{\mathfrak{su}_1}2$ to indicate a $-2$ curve with global symmetry $SU(2)$.

The only remaining non-classical DE configurations involve a $\mathfrak{g_2}$ or $\mathfrak{so_7}$ gauge symmetry on the central $-2$ curve of the trivalent vertex $2\overset{2}{2}2$.  The three surrounding $-2$ curves here must each hold $\mathfrak{su}_2$ gauge symmetry.  These may touch empty $-2$ curves only in the $\mathfrak{g_2}$ case.

\item
Configurations consisting of chain of $-2$ curves with a single $-1$ curve attached to the end will have similar pockets of classical or non-classical configurations.  The only novelty here are the non-classical configurations involving the $-1$ curves.  Suppose first that our configuration has at least two $-2$ curves.  If we enhance the $-2$ curve touching the $-1$ curve to have gauge symmetry $\mathfrak{g}_2$ or $\mathfrak{so}(7)$, then the adjacent $-1$ curve may have gauge symmetry $\mathfrak{sp}(1)$, $\mathfrak{sp}(2)$, or $\mathfrak{sp}(3)$, or it may be empty.  The other $-2$ curve attached to this $-2$ curve must have gauge symmetry $\mathfrak{su}(2)$.  If the first $-2$ curve has gauge symmetry $\mathfrak{so}(7)$, this $-2$ curve with gauge symmetry $\mathfrak{su}(2)$ cannot touch another $-2$ curve, but if the first has $\mathfrak{g}_2$ symmetry, then the $\mathfrak{su}(2)$ curve may touch a curve with no gauge symmetry.  Conversely, if the $-1$ curve carries an enhanced $\mathfrak{g}_2$, then the adjacent $-2$ curve must hold an $\mathfrak{su}(2)$ gauge symmetry, and the $-2$ curve touching that one must be empty.

In the special case of the base 12, there are even more possibilities.  Here, the gauge symmetry on the $-2$ curve can also be $\mathfrak{so}(N), N=9,10,11,12,13$.  The $-1$ curve may hold gauge algebra $\mathfrak{sp}(M)$ with $M \leq N-6$, or it may be empty.  Additionally, the $-2$ curve may hold gauge symmetry $\mathfrak{f}_4$, $\mathfrak{e}_6$, or $\mathfrak{e}_7$, provided the $-1$ curve is empty.  The $-1$ curve may hold $\mathfrak{so}_7$ gauge symmetry if the $-2$ curve holds $\mathfrak{su}_2$.  Finally, the $-1$ curve may hold gauge algebra $\mathfrak{so}(7)$ if the adjacent $-2$ curve has gauge group $\mathfrak{su}(2)$.  The $-2$ curve of the base 12 may be empty if the $-1$ curve holds $\mathfrak{sp}_1$.

\item
The only remaining classical bases are of the form 21414....  Once again, a non-classical enhancement of the $-4$ gauge symmetry is only possible if there is a single $-4$ curve, and in this case enhancement to $\mathfrak{g}=\mathfrak{f}_4$, $\mathfrak{e}_6$, or $\mathfrak{e}_7$ is possible.  The adjacent $-1$ curve must then be empty, and the $-2$ curve may hold any gauge algebra provided the gauging condition on the $-1$ curve is satisfied.  This means in particular that if $\mathfrak{g}=\mathfrak{f}_4$, the $-2$ curve may hold $\mathfrak{su}(2)$, $\mathfrak{su}(3)$, or $\mathfrak{g}_2$, or it may remain empty.  If $\mathfrak{g}=\mathfrak{e}_6$, the $-2$ curve may hold $\mathfrak{su}(2)$ or $\mathfrak{su}(3)$, or it may remain empty.  If $\mathfrak{g}=\mathfrak{e}_7$, the $-2$ curve may hold $\mathfrak{su}(2)$ or remain empty.  For any number of $-4$ curves, we may enhance the gauge symmetry on the $-2$ curve to $\mathfrak{g}_2$ or $\mathfrak{so}(N)$, $N=7,...,13$.  The $-1$ curve must in such a case hold an $\mathfrak{sp}(M)$ gauge algebra or be empty.  The allowed values for $M$ are determined by anomaly cancellation, and are shown in Table \ref{214table} as a function of the gauge algebra on the $-2$ curve and the adjacent $-4$ curve.  The remaining curves in the diagram, all of self-intersection $-1$ or $-4$, have gauge symmetries determined by the usual convexity conditions observed in the classical case.  The $-2$ curve can also be empty if the $-1$ curve holds $\mathfrak{sp}_1$.

\begin{table}
\begin{center}
\begin{tabular}{|c|c|c|} \hline
$-2$ & $-1$ & $-4$ \\ \hline \hline
$\emptyset$ & $\mathfrak{sp}(1), 4 \geq 1+N-16$  & $\mathfrak{so}(N),N=9,10,...$ \\ \hline
$\mathfrak{g}_2$ & $\mathfrak{sp}(M), 4 M \geq 7+N-16, M \leq 4, N-8$  & $\mathfrak{so}(N),N=9,10,...$ \\ \hline
$\mathfrak{g}_2$ & $\emptyset$ & $SO(N),N=8,9$ \\ \hline
$\mathfrak{so}(7)$ & $\mathfrak{sp}(M), 4 M \geq 8+N-16, M \leq 4, N-8$  & $\mathfrak{so}(N),N=9,10,...$ \\ \hline
$\mathfrak{so}(7)$ & $\mathfrak{sp}(1), 4  \geq 7+N-16$  & $\mathfrak{so}(N),N=9,10,...$ \\ \hline
$\mathfrak{so}(7)$ & $\emptyset$ & $SO(N),N=8,9$ \\ \hline
$\mathfrak{so}(8)$ & $\mathfrak{sp}(M), 4 M \geq 8+N-16, M \leq 2, N-8$  & $\mathfrak{so}(N),N=9,10,...$ \\ \hline
$\mathfrak{so}(8)$ & $\emptyset$ & $SO(8)$ \\ \hline
$\mathfrak{so}(9)$ & $\mathfrak{sp}(M), 4 M \geq 9+N-16, M \leq 3, N-8$  & $\mathfrak{so}(N),N=9,10,...$ \\ \hline
$\mathfrak{so}(10)$ & $\mathfrak{sp}(M), 4 M \geq 10+N-16, M \leq 4, N-8$  & $\mathfrak{so}(N),N=9,10,...$ \\ \hline
$\mathfrak{so}(11)$ & $\mathfrak{sp}(M), 4 M \geq 11+N-16, M \leq 5, N-8$ & $\mathfrak{so}(N),N=9,10,...$ \\ \hline
$\mathfrak{so}(12)$ & $\mathfrak{sp}(M), 4 M \geq 12+N-16, M \leq 6, N-8$  & $\mathfrak{so}(N),N=9,10,...$ \\ \hline
$\mathfrak{so}(13)$ & $\mathfrak{sp}(M), 4 M \geq 13+N-16, M \leq 7, N-8$  & $\mathfrak{so}(N),N=9,10,...$ \\ \hline
\end{tabular}
\end{center}
\caption{The gauge symmetries permitted on the leftmost $-2$, $-1$, and $-4$ curve in the configuration 21414.... There are additional restrictions on all curves except the $-2$ curve and the adjacent $-1$ curve, which are precisely the convexity conditions on the sequence 141414... obtained in the classical case.}
\label{214table}
\end{table}

\end{itemize}

\subsubsection{Fiber Enhancements of Non-classical Bases}

Since a generic base consists of DE-type nodes connected by interior links and with extra side links attached, we may divide our classification of fiber enhancements into three categories: enhancements of DE-type nodes, enhancements of interior links, and enhancements of side links/noble molecules.  Due to the large number of side links/noble molecules, the enhancements of the side links/noble molecules are left to a \texttt{Mathematica} notebook included with our arXiv submission.  We first discuss the enhancements of the DE-type nodes.

\begin{itemize}

\item
A $-7$ or $-8$ curve minimally carries $\mathfrak{e}_7$ gauge symmetry.  However, this may be enhanced to $\mathfrak{e}_8$ as long as the necessary gauging condition on the adjacent $-1$ curve is satisfied.  In particular, this means that enhancement may occur provided the only links attached to the $-7$ or $-8$ curve take the form $1223...$ or $12...2$, such that both the $-1$ curve and the adjacent $-2$ curve are devoid of a gauge group.  This enhancement to $\mathfrak{e}_8$ requires the addition of small instantons, so that in fact the self-intersection of the $-7$ or $-8$ curve will actually be decreased to $-12$.

\item
Similarly, a $-6$ curve minimally carries $\mathfrak{e}_6$ gauge symmetry, but it may be enhanced to either $\mathfrak{e}_7$ or $\mathfrak{e}_8$ provided the necessary gauging conditions on the adjacent $-1$ curves are satisfied.  The requirements for $\mathfrak{e}_8$ enhancement are the same as for a $-7$ or $-8$ curve, and once again small instantons must be added to lower the intersection number to $-12$.  Additionally, the gauge symmetry may be enhanced to $\mathfrak{e}_7$ as long as the only links attached to the $-7$ or $-8$ curve take the form $1223...$, $123...$ or $12...2$.  In these cases, the $-1$ curve must be empty, and the adjacent $-2$ curve cannot hold any gauge symmetry larger than $\mathfrak{su}(2)$.

\item
The enhancements of a $-4$ curve have already been discussed to a large extent in the description of classical bases.  The gauge symmetry living on the $-4$ curve can be enhanced to $\mathfrak{e}_8$ or $\mathfrak{e}_7$ under the same conditions as it can for a $-6$ curve.  It can also be enhanced to $\mathfrak{e}_6$ provided the adjacent links are of the form $1223...$, $123...$, $12...2$, $13$, or $131...$.  In such situations, the $-1$ curve attached to the $-4$ curve must be empty, and the next curve cannot hold any gauge symmetry but $\mathfrak{su}(2)$ or $\mathfrak{su}(3)$.  Similarly, the gauge symmetry can be enhanced to $\mathfrak{f}_4$ provided the adjacent links are of the form $1223...$, $123...$, $12...2$, or $13...$.  In these cases, the $-1$ curve attached to the $-4$ curve must be empty, and the next curve cannot hold any gauge symmetry but $\mathfrak{su}(2)$, $\mathfrak{su}(3)$, or $\mathfrak{g}_2$.

The gauge symmetry of a $-4$ curve may be enhanced to $\mathfrak{so}(N)$ in accordance with the classical convexity conditions and the gauging conditions.  To be more precise, suppose there is a $-1$ curve stretched between a $-4$ curve of gauge symmetry $\mathfrak{so}(N)$ and another curve of intersection $-n$, supporting gauge symmetry $\mathfrak{g}$.  Then, the $-1$ curve may be empty provided $\mathfrak{g} \oplus \mathfrak{so}(N) \subset \mathfrak{e}_8$.  Alternatively, the $-1$ curve may hold gauge symmetry $\mathfrak{sp}(M)$ subject to the condition $4 M \geq N + N_\mathfrak{g}-16$, where $N_\mathfrak{g}$ is the size of the representation which mixes with the half-fundamental of $\mathfrak{sp}(M)$ to satisfy the mixed anomaly condition between these two gauge algebras.  This should be viewed as a straightforward generalization of the classical convexity condition for the base 41414..., in which case $\mathfrak{g} = \mathfrak{so}(N')$ and $N_\mathfrak{g} = N'$.

\end{itemize}

We now describe each of the allowed fiber enhancements for the interior links between any two given DE-type nodes.

\begin{itemize}

\item The interior link $\overset{5,5}\oplus = 12231513221$ carries no gauge symmetry on the leftmost $-1$ curve, no gauge symmetry on the next $-2$ curve (with a type II fiber), $\mathfrak{sp}(1)$ on the next $-2$ curve, $\mathfrak{g}_2$ on the $-3$ curve, nothing on the next $-1$ curve, $\mathfrak{f}_4$ on the $-5$ curve, and the same on the right side.  There are no allowed fiber enhancements, regardless of which of the DE-type nodes are attached to the two ends.

\item The interior link $\overset{4,4}\oplus = 123151321$ carries no gauge symmetry on the left $-1$ curve, $\mathfrak{su}(2)$ on the next $-2$ curve, $\mathfrak{g}_2$ on the $-3$ curve, nothing on the next $-1$ curve, $\mathfrak{f}_4$ on the $-5$ curve, and similarly for the right side.  There are no allowed fiber enhancements, regardless of which of the DE-type nodes are attached to the two ends.

\item The interior link $\overset{3,3}\bigcirc = 1315131$ admits different options depending on which DE-type nodes are attached to the sides.  We present the story for the left hand side, and the story on the right hand side is just the mirror image.  If an $E_6$-type $-6$ curve is attached to the far left of the link, there is no gauge symmetry on the left $-1$ curve, $\mathfrak{su}(3)$ on the adjacent $-3$ curve, nothing on the next $-1$ curve, and either $\mathfrak{f}_4$ or $\mathfrak{e}_6$ on the $-5$ curve.  If, on the other hand, a D-type $-4$ curve is attached to the left hand side or there is no node attached, then there is an additional possibility: the $\mathfrak{su}(3)$ on the $-3$ curve may be enhanced to $\mathfrak{g}_2$ as long as the $-5$ curve is not enhanced to $\mathfrak{e}_6$.  A further enhancement of the $-1$ curve on the far left to $\mathfrak{sp}(1)$ is also possible, and in this case the adjacent $-3$ curve must again support $\mathfrak{g}_2$ gauge symmetry.  Furthermore, as long as neither of the $-3$ curves are enhanced to $\mathfrak{g}_2$, the $-5$ curve may be enhanced to $\mathfrak{e}_6$.

\item The left half of the interior link $\overset{4,5}\oplus = 1231513221$ is the same as the left half of $\overset{4,4}\oplus$, and the right half is the same as the right half of the interior link $\overset{5,5}\oplus$.  Once again, there are no allowed fiber enhancements, regardless of which of the DE-type nodes are attached to the two ends.

\item The left hand side of the link $\overset{3,5}\oplus = 131513221$ follows the same story as $\overset{3,3}\bigcirc$, while the right hand side is fixed to be the same as for $\overset{5,5}\oplus$.  The middle $-5$ curve cannot be enhanced to $\mathfrak{e}_6$.

\item The interior link $\overset{3,4}\oplus = 13151321$ admits the same options as $\overset{3,5}\oplus$ on the left half, and the right half is fixed to be the same as in the $\overset{4,4}\oplus$ case.  The middle $-5$ curve cannot be enhanced to $\mathfrak{e}_6$.

\item The interior link $\overset{4,2}\oplus = 12231$ always carries no gauge symmetry on the left $-1$ curve, no gauge symmetry on the next $-2$ curve (with a type II fiber), and $\mathfrak{su}(2)$ on the next $-2$ curve.  The $-3$ curve carries $\mathfrak{g}_2$.  The far right $-1$ curve must be empty.

\item The interior link $\overset{3,3}\oplus = 12321$ always carries no gauge symmetry on the $-1$ curves and $\mathfrak{su}(2)$ on the $-2$ curves.  The $-3$ curve carries $\mathfrak{so}(7)$.

\item The interior link $\overset{3,2}\oplus = 1231$ always carries no gauge symmetry on the left $-1$ curve and $\mathfrak{su}(2)$ on the $-2$ curve.  The $-3$ curve minimally carries $\mathfrak{g}_2$, but it may be enhanced to $\mathfrak{so}(7)$.  The far right $-1$ curve must be empty if the $-3$ curve is holding $\mathfrak{g}_2$, but if the $-3$ curve is enhanced to $\mathfrak{so}(7)$, this $-1$ curve can support an $\mathfrak{sp}(1)$ gauge symmetry.

\item The interior link $\overset{2,2}\oplus = 131$ minimally carries just a $\mathfrak{su}(3)$ on the $-3$ curve.  If an $\mathfrak{e}_6$-type $-6$ curve is attached at either the left or the right, no enhancements are allowed, and the $-1$ curves cannot carry a gauge symmetry.  However, if no $\mathfrak{e}_6$-type node is attached, then the $-3$ curve may be enhanced.  If a $D$-type $-4$ curve is attached at either the left or the right, then the $-3$ curve may be enhanced to $\mathfrak{g}_2$ or $\mathfrak{so}(N)$, with $N=7,8,9,10,11,12$.  If it is $\mathfrak{g}_2$, then one of the adjacent $-1$ curves can carry either $\mathfrak{sp}(1)$ or no gauge symmetry, but the other must be empty. If it is $\mathfrak{so}(7)$, then the left $-1$ curve can hold $\mathfrak{sp}(M_L)$ and the right can hold $\mathfrak{sp}(M_R)$ with $M_L=0,1,2$, $M_R=0,1,2$, and $M_L + M_R \leq 2$.  If it is $\mathfrak{so}(8)$, then the then the left $-1$ curve can hold $\mathfrak{sp}(1)$ or be empty and the right can hold $\mathfrak{sp}(1)$ or be empty.  The $-3$ curve may also hold $\mathfrak{so}(N)$, with $N=9,10,11,12$.  The left $-1$ curve can hold $\mathfrak{sp}(M_L)$ and the right can hold $\mathfrak{sp}(M_R)$ with $M_L=0,1,2,...$, $M_R=0,1,2,...$ with $M_L + M_R \leq N-7$.  However, we cannot in these cases have $M_L=0$ if the left $-1$ curve is attached to a D-type $-4$ curve, and similarly we cannot have $M_R=0$ if the right $-1$ curve is attached to a D-type $-4$ curve.  In the special case of $N=12$, we cannot even have $M_L=1$ if the left $-1$ curve is attached to a D-type node, and similarly $M_R \neq 1$ if the right $-1$ curve is attached to a D-type node.

\item The interior link 1 can only attach to $D$-type nodes.  As such, it has already been included in the list of classical configurations.  If it is used as a side link for any $E$-type node, it must be empty.

\end{itemize}

\begin{figure}[ptb!]
\begin{center}
\includegraphics[trim=20mm 40mm 20mm 35mm,clip,width=90mm]{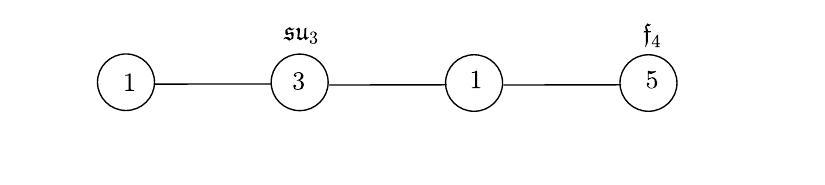}
\includegraphics[trim=20mm 20mm 20mm 20mm,clip,width=90mm]{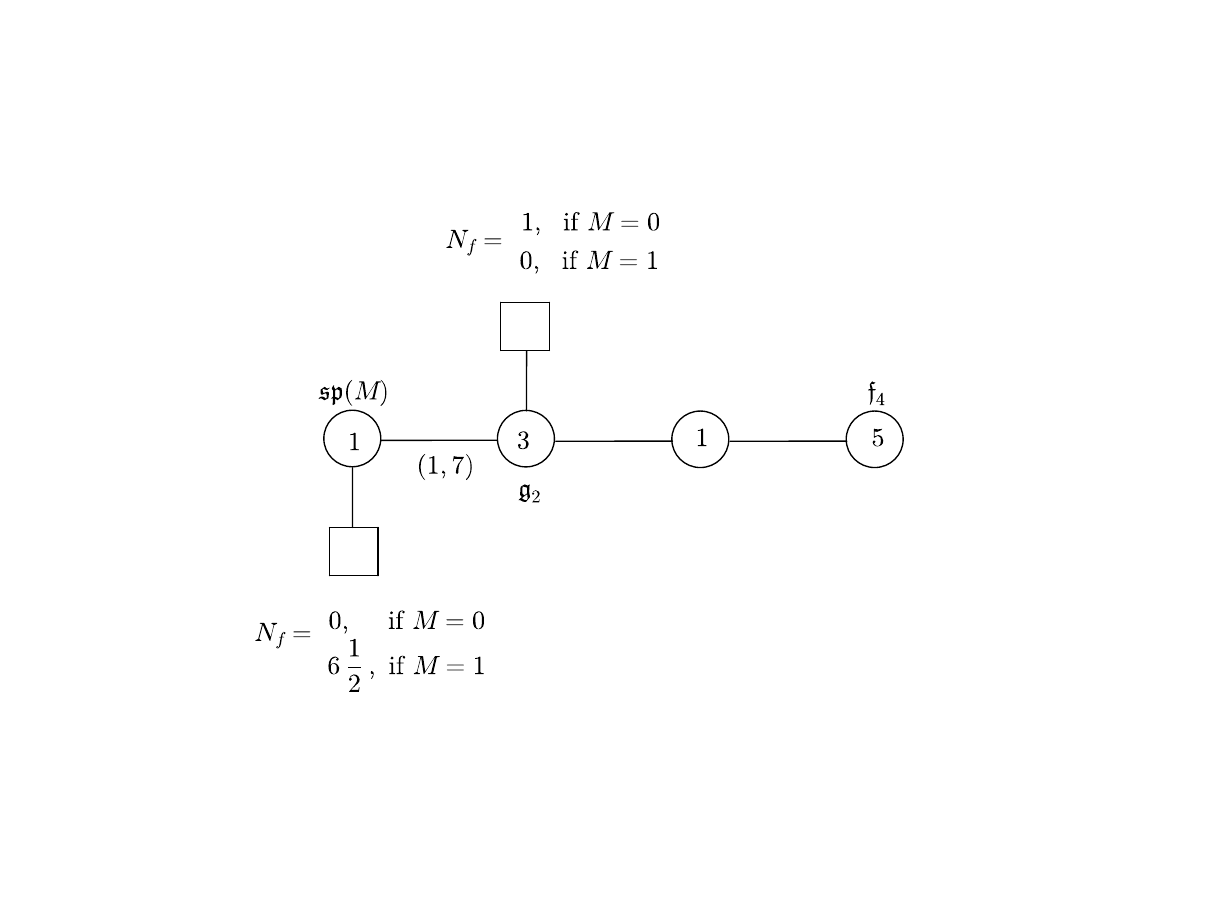}
\includegraphics[trim=20mm 40mm 20mm 28mm,clip,width=90mm]{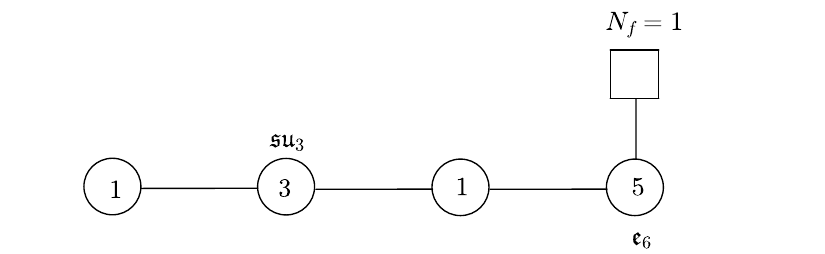}
\end{center}
\caption{Allowed fiber decorations of the left half of the $\overset{3,3}\bigcirc \simeq 1315131$ interior link.  The right half is simply the mirror image.}
\end{figure}

\begin{figure}[ptb!]
\begin{center}
\includegraphics[trim=15mm 42mm 15mm 30mm,clip,width=90mm]{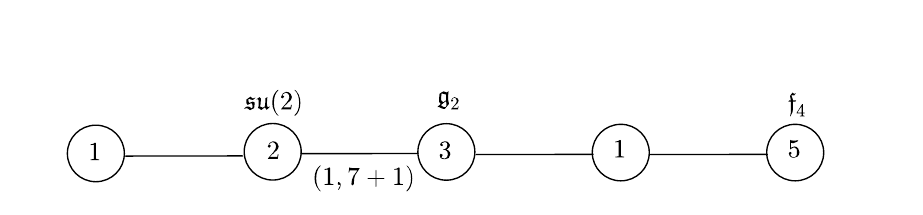}
\end{center}
\caption{Allowed fiber decorations of the left half of the $\overset{4,4}\oplus \simeq 123151321$ interior link.  The right half is simply the mirror image.}
\end{figure}

\begin{figure}[ptb!]
\begin{center}
\includegraphics[trim=5mm 40mm 5mm 40mm,clip,width=90mm]{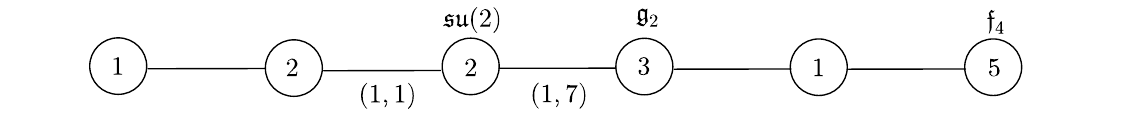}
\end{center}
\caption{Allowed fiber decoration of the left half of the $\overset{5,5}\oplus \simeq 12231513221$ interior link.  The right half is simply the mirror image.}
\end{figure}

\begin{figure}[ptb!]
\begin{center}
\includegraphics[trim=15mm 47mm 15mm 35mm,clip,width=90mm]{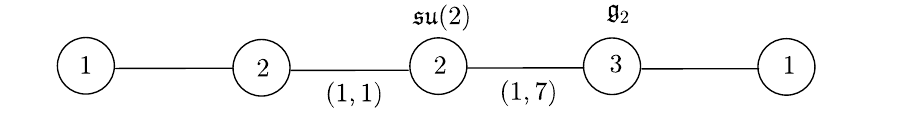}
\end{center}
\caption{Allowed fiber decoration of the $\overset{4,2}\oplus \simeq 12231$ interior link.}
\end{figure}

\begin{figure}[ptb!]
\begin{center}
\includegraphics[trim=15mm 30mm 25mm 40mm,clip,width=90mm]{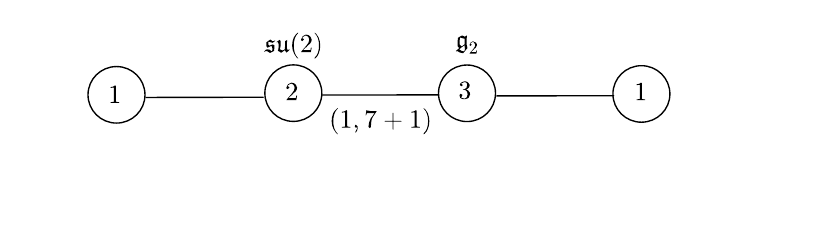}
\includegraphics[trim=15mm 22mm 30mm 25mm,clip,width=90mm]{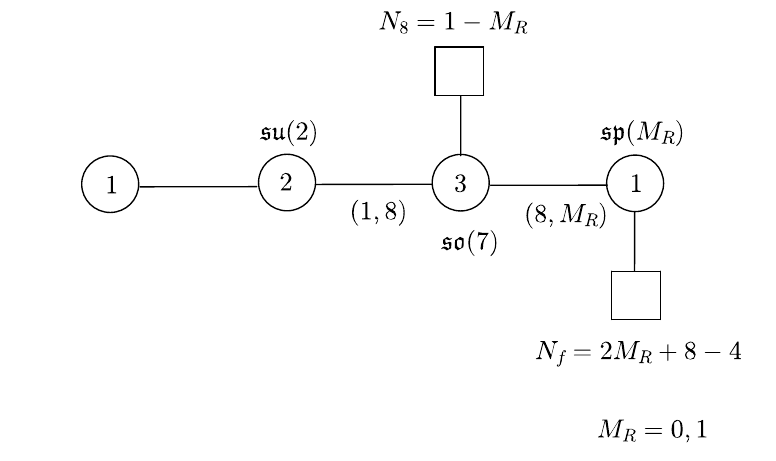}
\end{center}
\caption{Allowed fiber decorations of the $\overset{3,2}\oplus \simeq 1231$ interior link.}
\end{figure}

\begin{figure}[ptb!]
\begin{center}
\includegraphics[trim=30mm 30mm 30mm 25mm,clip,width=70mm]{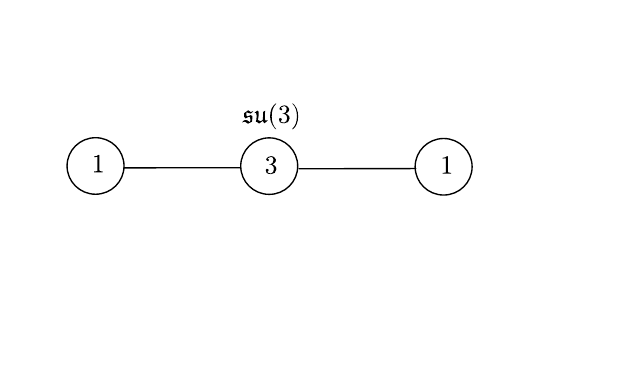}
\includegraphics[trim=25mm 25mm 35mm 20mm,clip,width=70mm]{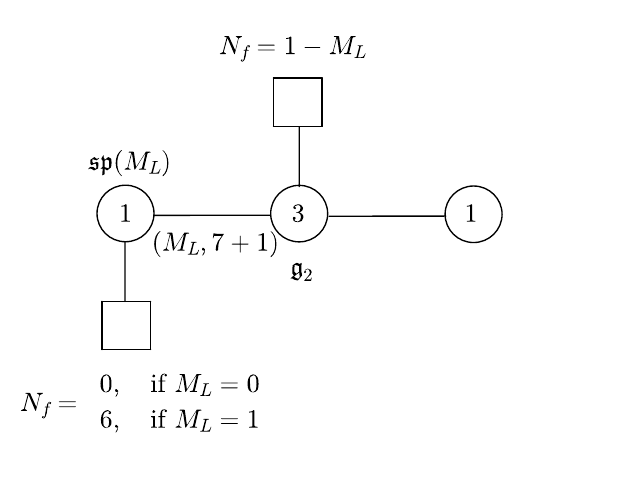}
\includegraphics[trim=28mm 20mm 28mm 17mm,clip,width=70mm]{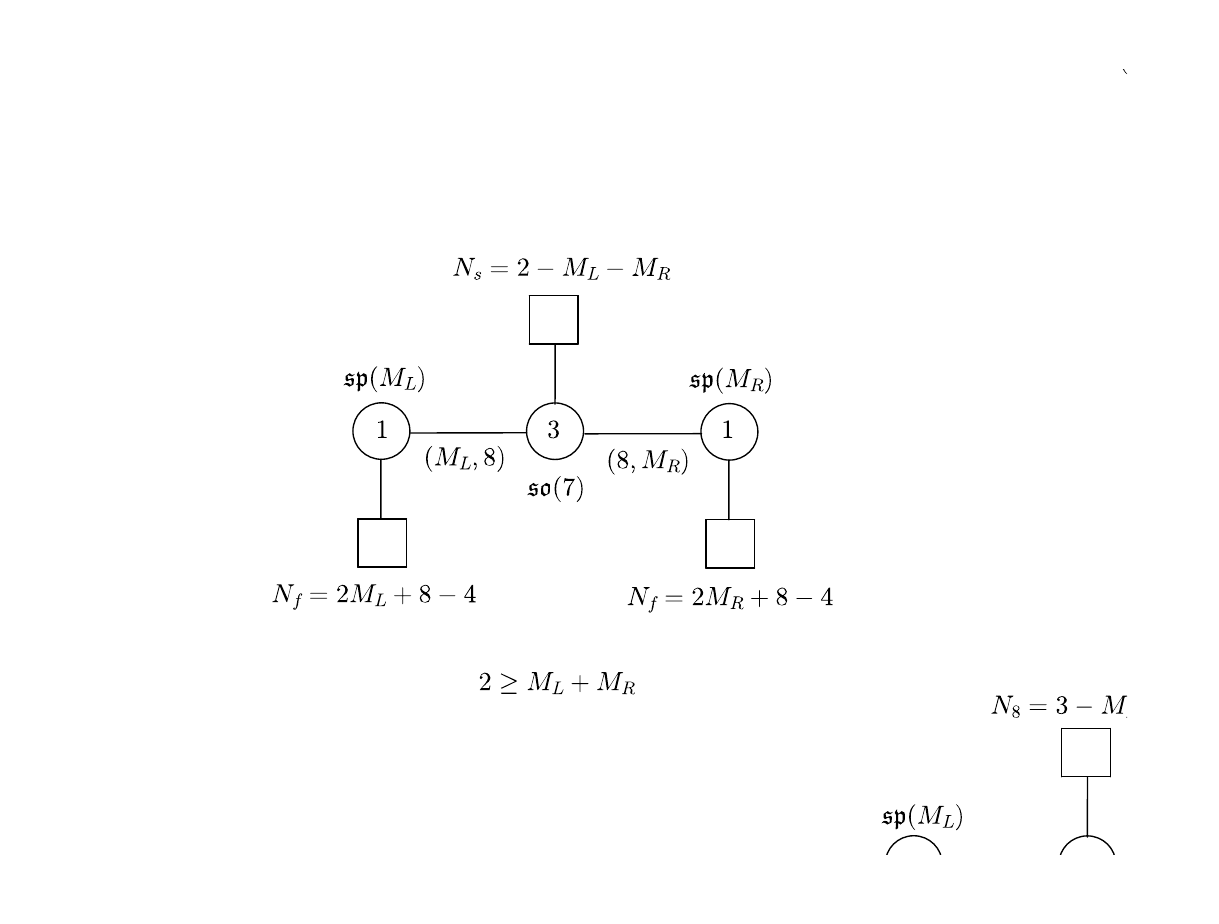}
\includegraphics[trim=28mm 25mm 28mm 18mm,clip,width=70mm]{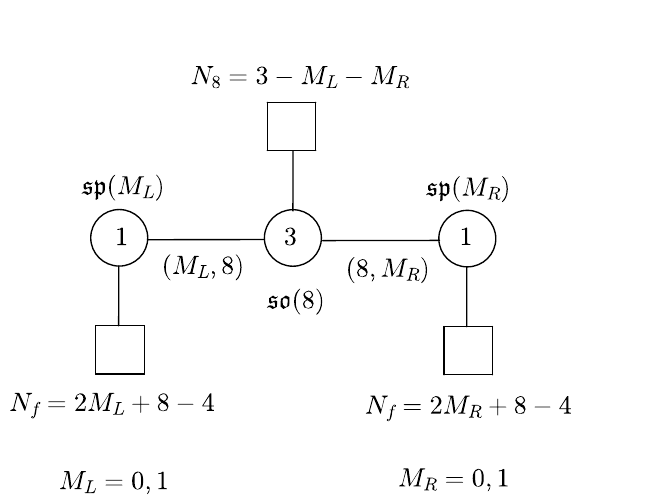}
\includegraphics[trim=25mm 19mm 25mm 10mm,clip,width=70mm]{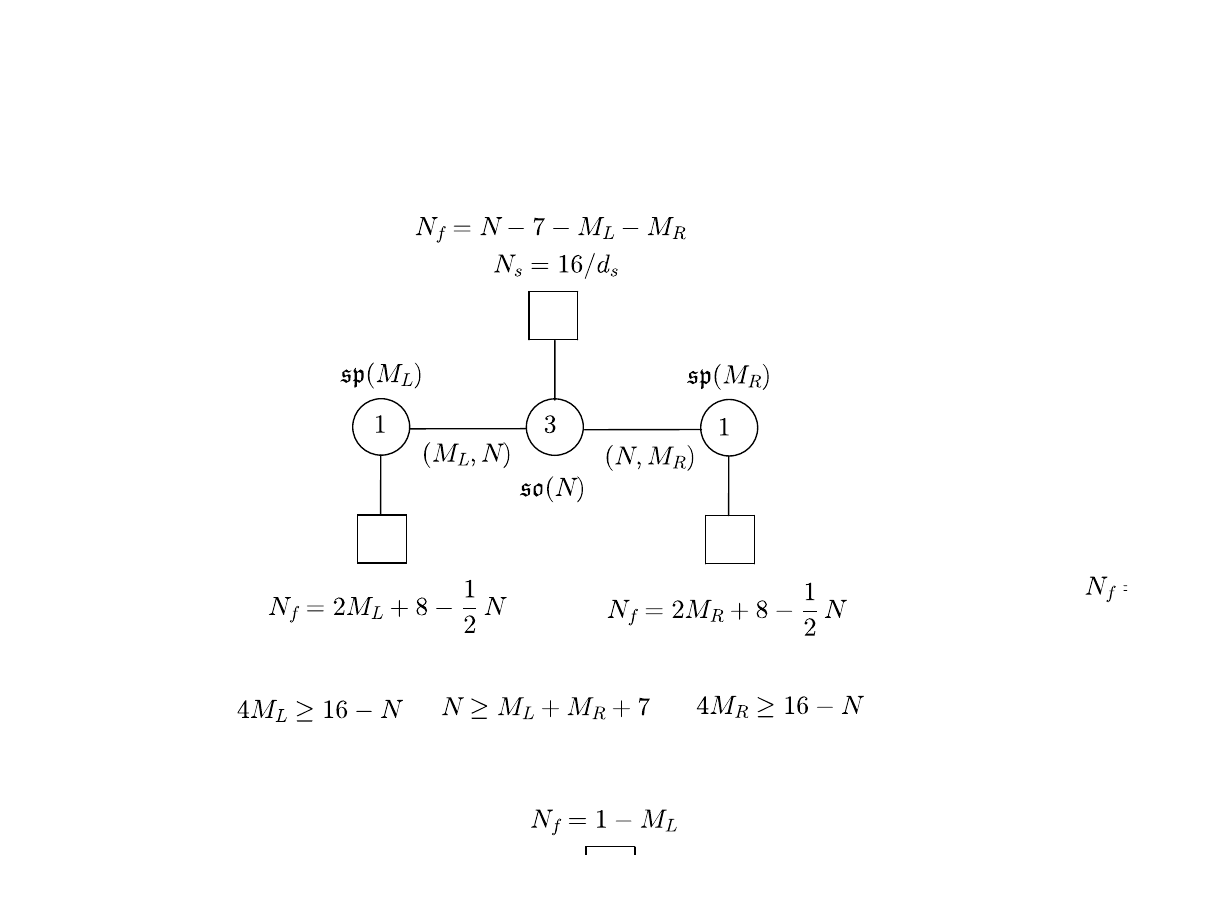}
\end{center}
\caption{Allowed fiber decorations of the $\overset{2,2}\oplus \simeq 131$ interior link.}
\end{figure}

\begin{figure}[ptb!]
\begin{center}
\includegraphics[trim=15mm 40mm 15mm 30mm,clip,width=90mm]{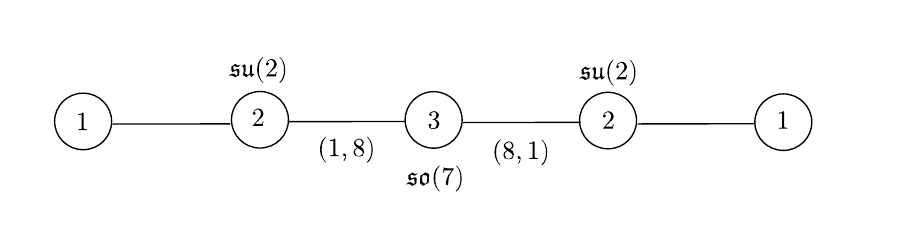}
\end{center}
\caption{Allowed fiber decoration of the $\overset{3,3}\oplus \simeq 12321$ interior link.}
\end{figure}

\section{Boundary Conditions \label{sec:BOUNDARY}}

Our analysis has revealed a rather simple structure for all of the
geometric phases of F-theory which can generate a 6D\ SCFT. Essentially, the base of the elliptic
Calabi-Yau threefold has the structure
of a linear chain of curves, with a small amount of decoration on the ends.
This decoration also includes the various types of tree-like structures, and
non-minimal fiber types. Putting this all together, we have provided a rather
complete picture of all ways to generate 6D\ SCFTs from geometry.

Now, starting from a geometry, it is natural to ask whether there are any
further ways to supplement such a theory. This can occur in 6D\ SCFTs which
possess a flavor symmetry. In these cases, it is possible to consider boundary
conditions \textquotedblleft at infinity\textquotedblright\ for our
configuration \cite{DelZotto:2014hpa} (see also \cite{Gaiotto:2014lca}).

In reference \cite{DelZotto:2014hpa}, two sorts of boundary conditions were identified. One is
associated with T-brane data for intersecting seven-branes, and can be tracked to the choice of a
nilpotent residue for the Higgs field of a flavor symmetry. The other boundary condition arises
(in heterotic language) for small $E_8$ instantons probing an orbifold singularity.

In the specific case of SCFTs, we will present evidence that both kinds of data are already
captured by \textit{geometric} phases of an F-theory compactification. Said differently,
specifying these boundary data simply provide an alternative way to catalogue some of
the theories we have already encountered.

Consider first the case of T-branes \cite{Donagi:2003hh, TBRANES, glueI, glueII, FCFT, Chiou:2011js, HVW, D3gen, Anderson:2013rka, Collinucci:2014qfa, Collinucci:2014taa}. This comes about from non-abelian intersections of seven-branes, and is invisible at the level of the
Weierstrass model, but does manifest itself in the limiting behavior of the intermediate Jacobian of a
smooth Calabi-Yau threefold as it approaches a singular limit \cite{Anderson:2013rka}. In the 6D SCFT a choice of T-brane background
corresponds to activating a vev for some operators of the
SCFT \cite{DelZotto:2014hpa}, which in turn specifies the location of a pole in the associated
Hitchin system for the flavor symmetry \cite{BHVI, DelZotto:2014hpa, Heckman:2014qba}.
The resulting pattern of possible elliptic fibrations can then be viewed either as a specific sequence
of gauge groups, or alternatively, as a choice of nilpotent element (and its orbit) in
$\mathfrak{g}_{\mathbb{C}}$, the complexification of the flavor
symmetry algebra.

Consider next the case of boundary data for small instantons. In the case of a $-1$ curve with a flavor symmetry, the theory
locally behaves like a small instanton of heterotic strings compactified on a
non-compact K3 surface, including K3 surfaces with orbifold singularities \cite{Aspinwall:1997ye}. As explained in \cite{DelZotto:2014hpa}, from the heterotic perspective, there could be additional boundary data for our small
instanton. For an instanton probing an orbifold singlarity $\mathbb{C}^{2}/\Gamma$ with
$\Gamma$ a discrete subgroup of $SU(2)$, this amounts to a choice of
homomorphism from $\pi_{1}(S^{3}/\Gamma)$ into the group $E_{8}$. Since we also have an isomorphism
$\pi_{1}(S^{3}/\Gamma) \simeq \Gamma$, these boundary data are captured by $\mathrm{Hom}(\Gamma , E_{8})$.

As opposed to the situation with T-branes, these boundary data are not
directly linked to the vev of an operator. Rather, they are more
analogous to data such as a discrete theta angle. Experience with other
systems, e.g.\ \cite{Douglas:1996xp}, suggests that this should also be reflected
in purely geometric terms. So, it is natural to conjecture that these cases
can also be covered by a specific pattern of gauge groups.

We will indeed present a very precise extension of heterotic / F-theory duality in which the algebraic / non-geometric data of the heterotic side, namely the elements of $\mathrm{Hom}(\Gamma , E_8)$ will be matched to \textit{purely geometric} data on the F-theory side, namely fiber enhancements of a specific collection of linear curves. To perform this match, we will show that we can exactly match the unbroken flavor symmetries on both sides of the correspondence. On the heterotic side, we specify a choice of $\rho \in \mathrm{Hom}(\Gamma , E_8)$, and then determine the commutant $\rho(\Gamma) \subset E_8$. This commutant specifies the residual flavor symmetry left unbroken by our instanton configuration. On the F-theory side, we will identify linear chains of curves and fiber enhancements on compact curves which lead to 6D SCFTs with very specific flavor symmetry groups. These flavor symmetries will turn out to precisely match to the ones obtained on the heterotic side of the correspondence.

Our plan in this subsection will be to first explain the general contours of our proposal, and in particular the F-theory description
of heterotic small instantons probing an ADE singularity. Then, we illustrate how our proposal works for some of the ADE singularities.
The evidence we present will not be at the level of a proof, but rather at the level of a match which is so specific that it will leave
little doubt that the proposal is correct. With this in place, we will have shown that all such ``boundary data'' in an F-theory description of
a 6D SCFT are captured by purely geometric data. With this in place, we see that the results of the previous sections (i.e., the classification
of bases and fiber enhancements) serves to classify \textit{all} 6D SCFTs.

\subsection{The General Correspondence}

To frame the discussion to follow, we first review some further details of how heterotic small instantons probing an ADE singularity are realized in F-theory. Our discussion here follows \cite{Aspinwall:1997ye} as well as reference \cite{DelZotto:2014hpa}. We then extend this discussion by making a general proposal for how to realize the boundary data of the heterotic theory in terms of an F-theory model.

In this section we consider in detail the case of boundary data for small instantons probing an ADE singularity. In M-theory terms, we have an $E_8$ nine-brane near the singularity $\mathbb{C}^2 / \Gamma$ for $\Gamma$ an ADE subgroup of $SU(2)$. Into this theory we introduce $N$ M5-branes. In heterotic terms, this provides a realization of $N$ small instantons probing an ADE singularity. We can remove the instantons from the system, i.e., move onto the tensor branch by pulling the M5-branes off of the $E_8$ wall.

The F-theory realization amounts to decorating the basic case of an instanton link, namely a chain of curves $1,2,...,2$, and the volume of the curves dictates the (relative) positions of the M5-branes from the $E_8$ wall. To get to the case of an ADE singularity, we introduce the corresponding ADE gauge group, and decorate each fiber by the corresponding algebra:
\begin{equation}
[E_8] \overset{\mathfrak{g}}{1} \overset{\mathfrak{g}}{2}...\overset{\mathfrak{g}}{2} [G].
\end{equation}
There is then a minimal resolution of the collision of singularities which we can perform to reach the fully resolved tensor branch. For further details, see, e.g., references \cite{Aspinwall:1997ye} and \cite{DelZotto:2014hpa}.

Now, we would like to extend this geometric correspondence to the case where we incorporate the boundary data of a small instanton. In heterotic terms, our choice of instanton also requires us to specify a flat connection ``at infinity'', i.e., on $S^3 / \Gamma$. This is classified by a choice of group homomorphism $\pi_1 (S^3 / \Gamma) \rightarrow E_8 $. Note that activating this flat homomorphism breaks the original $E_8$ flavor symmetry. Said differently, the unbroken flavor symmetry is the commutant of $\rho(\Gamma) \subset E_8$, where $\rho$ is our choice of group homomorphism $\rho: \Gamma \rightarrow E_8$. Our plan will be to match the flavor symmetries for these heterotic theories to specific F-theory duals.

The F-theory realization of these heterotic theories proceeds from the following rules:
\begin{itemize}
\item Step 1: Begin with a base of the form $1,2,...,2$. This is the basic example of a small instanton theory.
\item Step 2: For instantons probing a $\Gamma_G$-type orbifold singularity, decorate each fiber to reach the configuration: $\overset{\mathfrak{g}}{1} \overset{\mathfrak{g}}{2}...\overset{\mathfrak{g}}{2} [G]$. Also, perform all forced blowups as required to maintain the existence of
    an elliptic Calabi-Yau. Note, however, that we do \textit{not} assume the presence of a non-compact $E_8$ touching the $-1$ curve.
\item Step 3: Next, introduce additional blowups of the corresponding base. The primary condition is that performing a blowdown of these extra curves takes us back to a small instanton configuration $\overset{\mathfrak{g}}{1} \overset{\mathfrak{g}}{2}...\overset{\mathfrak{g}}{2} [G]$.
\item Step 4: Finally, decorate the fibers of the new base. This will define a 6D SCFT, and anomaly cancellation will dictate a very specific choice of flavor symmetry group.

\end{itemize}
Finally, as a point of notation, we shall often indicate the type $II$ fiber explicitly, reserving $\mathfrak{su}_1$ for the $I_1$ fiber. The reason for this is that (as we explain more fully in Appendix~\ref{appendixF}), the matter content associated with these two fibers can be different, and consequently, they can give rise to 6D theories with different flavor symmetries.

As we shall make heavy use of it later, we now pause to briefly review the finite subgroups of $SU(2)$.  A convenient reference
is chapter six of \cite{MR584445}, which gives explicit descriptions
of the subgroups (up to conjugacy) in terms of the matrices:

\begin{align}
\zeta_{n}  & \equiv\left[
\begin{array}
[c]{cc}%
e^{2\pi i/n} & \\
& e^{-2\pi i/n}%
\end{array}
\right]  \text{, \ \ }\delta\equiv\left[
\begin{array}
[c]{cc}
& 1\\
-1 &
\end{array}
\right]  \text{, \ \ }\tau\equiv\frac{1}{\sqrt{2}}\left[
\begin{array}
[c]{cc}%
e^{-2\pi i/8} & e^{-2\pi i/8}\\
e^{10\pi i/8} & e^{2\pi i/8}%
\end{array}
\right]  \\
\iota & \equiv\frac{1}{e^{4\pi i/5}-e^{6\pi i/5}}\left[
\begin{array}
[c]{cc}%
e^{2\pi i/5}+e^{-2\pi i/5} & 1\\
1 & -e^{2\pi i/5}-e^{-2\pi i/5}%
\end{array}
\right].
\end{align}
These descriptions are shown in Table~\ref{subgroups-table}, along with
the orders of the groups, some convenient isomorphisms, and all possible
nontrivial quotient groups of each group.  In the Table, in
addition to the cyclic groups $\mathbb{Z}_n$, the symmetric groups
$\mathfrak{S}_n$ and the alternating groups $\mathfrak{A}_n$,
we find the dihedral groups
$\mathrm{Dih}_{2k}\subset SO(3)$ of order $2k$
which are the symmetry groups of a regular $k$-gon in $3$-dimensional space,
as well as the ``binary dihedral'' or ``dicyclic'' groups
$\mathbb{D}_k$ of order $4k$ which are the lifts of $\mathrm{Dih}_{2k}$
to the covering group $SU(2)$ of $SO(3)$.

Each of these latter groups $\mathrm{Dih}_{2k}$ and $\mathbb{D}_k$
has two generators $x$ and $y$ with the basic relations
$y^{2k}=e$, $x^{-1}yx=y^{-1}$; for the dihedral groups, we have $x^2=e$ while
for the binary dihedral groups we have $x^2=y^k$.  The center of $\mathbb{D}_k$
is order two, and the quotient by that center gives $\mathrm{Dih}_{2k}$.
Notice that $\mathbb{D}_1 \cong \mathbb{Z}_4$ while $\mathrm{Dih}_2\cong
\mathbb{Z}_2 \times \mathbb{Z}_2$.  Every $\Gamma_{D_n}$ includes one of
these latter two as a quotient, depending on whether $n$ is odd or even.

\begin{table}
\begin{center}
\begin{tabular}{l|c|c|c}
Group & Order & Generators & Quotient groups \\ \hline
$\Gamma_{A_{n-1}}= \mathbb{Z}_n$ & $n$ &
$\zeta_n$
& $\mathbb{Z}_k$ if $k \mathrel{|} n$ \\
$\Gamma_{D_{n}} = \mathbb{D}_{n-2}$ & $4(n{-}2)$ & $\zeta_{2n-4}$,
$\delta$ & $\mathbb{Z}_2$, $\mathrm{Dih}_{2k}$ if $k \mathrel{|} (n{-}2)$,\\
&&&$\mathbb{D}_\ell$ if $\ell \mathrel{|} (n{-}2)$ but $2\ell \mathrel{\not|} (n{-}2)$\\
$\Gamma_{E_6} = \mathbb{T} $
& $24$ & $\zeta_4$, $\delta$,
$\tau$ & $\mathbb{Z}_3$, $\mathfrak{A}_4$ \\
\quad $\cong SL(2,\mathbb{F}_3)$  &&&\\
$\Gamma_{E_7} = \mathbb{O}$ & $48$ & $\zeta_8$, $\delta$, $\tau$ &
$\mathbb{Z}_2$, $\mathfrak{S}_3$, $\mathfrak{S}_4$ \\
$\Gamma_{E_8} = \mathbb{I} $
& $120$ &
$-(\zeta_5)^3$, $\iota$ & $\mathfrak{A}_5$\\
\quad $\cong SL(2,\mathbb{F}_5)$ &&&\\
\end{tabular}
\end{center}
\caption{Finite subgroups of $SU(2)$}
\label{subgroups-table}
\end{table}

The reason that quotient groups are important is that for every homomorphism
$\rho:\Gamma\to E_8$, the image $\rho(\Gamma)$ is a quotient group of $\Gamma$,
which is then embedded into $E_8$.  In the mathematics literature, the
problem of embedding finite groups into $E_8$ has been addressed directly
(rather than the problem of arbitrary homomorphisms into $E_8$).

When the image is a cyclic group, the problem reduces to finding all
elements of a fixed finite order $k$ in $E_8$, and the commutant of
each such element.  This problem was solved
by Kac \cite{MR739850} (for the complexified group $E_8(\mathbb{C})$),
and we will give (and use) his result below.  The classification
(up to conjugacy) of subgroups of $E_8$ isomorphic to
$\mathbb{Z}_2\times\mathbb{Z}_2$ is also known (lemma 3.7 of
\cite{cohen-griess}), and we shall use that result as well.

For more complicated groups,
in reference \cite{FREY} D.D. Frey has determined the embeddings into $E_8(\mathbb{C})$ for the dihedral groups $\mathrm{Dih}_6$, $\mathrm{Dih}_{10}$, and the binary octahedral group $\Gamma_{E_8} \simeq SL(2,5)$, giving the
commutants in each case.
Since we are dealing with the compact real Lie group $E_8$, the passage from complex to real
is of no consequence,\footnote{We thank D.D. Frey for correspondence on this point.} and we shall henceforth refer to the continuous group simply as $E_8$.
Frey's list, together with Kac's analysis, therefore provides us with
a rich set of example homomorphisms on which to test our correspondence with
F-theory.

We will present very strong evidence that the boundary data of the heterotic description are captured by purely geometric data of an F-theory compactification. More precisely, we show that every choice of commutant flavor symmetry on the heterotic side has a direct match to a flavor symmetry on the F-theory side. For each choice of homomorphism, there is one (and only one) configuration of linear chains where we decorate the fibers. Put together, our checks will amount to overwhelming evidence that the correspondence is true.

Let us now proceed to the various choices of discrete subgroups and their homomorphisms into $E_8$.

\subsection{The A-Series Subgroups of $SU(2)$ \label{ssec:ASERIES}}

In this subsection we consider the A-series of discrete subgroups $\Gamma_{A_{N-1}} \subset SU(2)$, namely the case of homomorphisms $\mathbb{Z}_N \rightarrow E_8$. As the notation suggests, in the case of a trivial flat connection, the theory of small instantons would have an $E_8 \times SU(N)$ global symmetry. Once we introduce a non-trivial flat connection, however, the global $E_8$ flavor symmetry will be broken further. Our aim will be to match these unbroken flavor symmetries to possible decorations of fibers in the F-theory setting.

In this vein, we first recall from \cite{MR739850} Kac's  way of characterizing the group homomorphisms $\mathbb{Z}_N \rightarrow E_8$. The image of each homomorphism defines a subgroup of $E_8$. We shall be interested in the commutant subgroup of this image. Now, one way to identify the continuous part of this commutant subgroup is to delete a particular set of nodes from the affine $E_8$ Dynkin diagram.  In particular, we assign each node of the affine $E_8$ Dynkin diagram a value:
\begin{figure}[H]
\begin{center}
\includegraphics[trim= 2mm 24mm 2mm 24mm, clip, width=60mm]{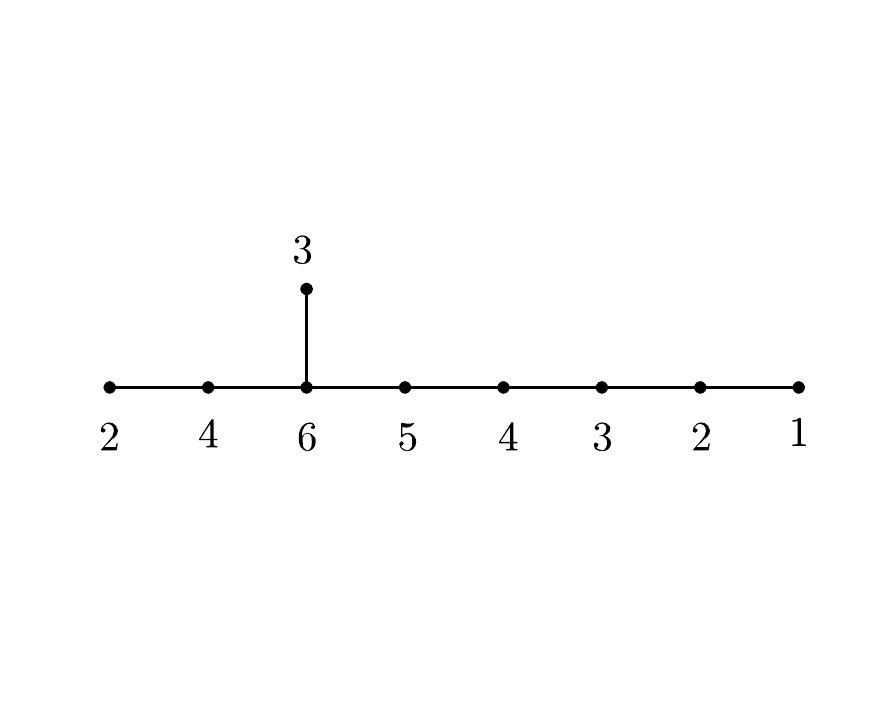}
\end{center}
\end{figure}
Each homomorphism $\mathbb{Z}_N \rightarrow E_8$ then corresponds to a partition of $N$ into the numbers in the diagram.
For each such partition, we delete the nodes whose numbers appear in the partition to find the symmetry that
remains after activating the boundary conditions for the instanton.

As an example, we take our right flavor symmetry group to be $G=SU(4)$, corresponding to the case $\Gamma = \mathbb{Z}_4$.  We must consider partitions of 4 into the numbers of the diagram.  For instance, we can partition as $4=2+1+1$ and so remove the nodes on the far left and far right to get the $D_7$ Dynkin diagram, corresponding to a flavor symmetry of $SO(14)$ after Higgsing.  For that same partition $4=2+1+1$, we can also remove the two nodes on the far right to get the $E_7$ Dynkin diagram, corresponding to an $E_7$ flavor symmetry after Higgsing.  Altogether, there are precisely 10 homomorphisms $\mathbb{Z}_4 \rightarrow E_8$, and hence 10 ways to remove nodes from the affine Dynkin diagram consistent with Higgsing.

Each of these is associated with a particular decoration of the fibers of a particular resolution of the F-theory base $12...2$, which we now illustrate.  Note that in some cases, there is a leftover fundamental representation, which we have labelled ``$N_f=1$."  Ordinarily, there is a $U(1)$ flavor symmetry associated with this extra matter, but it is anomalous.  Hence we do not concern ourselves further with $U(1)$ factors.
Here is the list of theories for this case:

$E_8$:
\begin{dynkin}
    \foreach \x in {1,...,6}
    {
        \dynkindot{\x}{0}
    }
    \dynkindot{7}{0}
    \dynkindot{3}{1}
    \dynkinline{1}{0}{7}{0}
    \dynkinline{3}{0}{3}{1}
  \end{dynkin}
$\Rightarrow [E_8]\,\, 1  \,\, \overset{\mathfrak{su_1}}2 \,\, \overset{\mathfrak{su_2}}2 \,\,\overset{\mathfrak{su_3}}2 \,\, \overset{\mathfrak{su_4}}2 \,\, ... \overset{\mathfrak{su_4}}2 \,\, [SU(4)]$

$E_7$:
\begin{dynkin}
    \foreach \x in {1,...,6}
    {
        \dynkindot{\x}{0}
    }
    \dynkindot{3}{1}
    \dynkinline{1}{0}{6}{0}
    \dynkinline{3}{0}{3}{1}
  \end{dynkin}
$\Rightarrow [E_7]\,\, 1 \,\,\underset{[N_f=1]}{\overset{\mathfrak{su_2}}2} \,\,\overset{\mathfrak{su_3}}2 \,\, \overset{\mathfrak{su_4}}2 \,\, ... \overset{\mathfrak{su_4}}2 \,\, [SU(4)]$

$SO(14)$:
\begin{dynkin}
    \foreach \x in {1,...,6}
    {
        \dynkindot{\x}{0}
    }
    \dynkindot{2}{1}
    \dynkinline{1}{0}{6}{0}
    \dynkinline{2}{0}{2}{1}
  \end{dynkin}
$\Rightarrow [SO(14)]\,\, \overset{\mathfrak{sp_1}}1 \,\,\overset{\mathfrak{su_3}}2 \,\,\overset{\mathfrak{su_4}}2 \,\, ... \overset{\mathfrak{su_4}}2 \,\, [SU(4)]$

$E_7 \times SU(2)$:
\begin{dynkin}
    \foreach \x in {1,...,6}
    {
        \dynkindot{\x}{0}
    }
    \dynkindot{8}{0}
    \dynkindot{3}{1}
    \dynkinline{1}{0}{6}{0}
    \dynkinline{3}{0}{3}{1}
  \end{dynkin}
$\Rightarrow [E_7]\,\, 1 \,\,\overset{\mathfrak{su_2}}{2} \,\,\underset{[SU(2)]}{\overset{\mathfrak{su_4}}2} \,\, ... \overset{\mathfrak{su_4}}2 \,\, [SU(4)]$

$SO(16)$:
\begin{dynkin}
    \foreach \x in {1,...,7}
    {
        \dynkindot{\x}{0}
    }
    \dynkindot{2}{1}
    \dynkinline{1}{0}{7}{0}
    \dynkinline{2}{0}{2}{1}
  \end{dynkin}
$\Rightarrow [SO(16)]\,\, \overset{\mathfrak{sp_2}}1 \,\,\overset{\mathfrak{su_4}}2  \,\, ... \overset{\mathfrak{su_4}}2 \,\, [SU(4)]$

$SO(12) \times SU(2)$:
\begin{dynkin}
    \foreach \x in {1,...,5}
    {
        \dynkindot{\x}{0}
    }
    \dynkindot{7}{0}
    \dynkindot{2}{1}
    \dynkinline{1}{0}{5}{0}
    \dynkinline{2}{0}{2}{1}
  \end{dynkin}
$\Rightarrow [SO(12)]\,\, \overset{\mathfrak{sp_1}}1 \,\,\overset{\mathfrak{su_4}}{\underset{[SU(2)]}2}  \,\, ... \overset{\mathfrak{su_4}}2 \,\, [SU(4)]$

$E_6 \times SU(2)$:
\begin{dynkin}
    \foreach \x in {1,...,5}
    {
        \dynkindot{\x}{0}
    }
    \dynkindot{7}{0}
    \dynkindot{3}{1}
    \dynkinline{1}{0}{5}{0}
    \dynkinline{3}{0}{3}{1}
  \end{dynkin}
$\Rightarrow [E_6]\,\, 1 \,\,\overset{\mathfrak{su_3}}{\underset{[SU(2)]}2} \,\,\overset{\mathfrak{su_4}}2 \,\, ... \overset{\mathfrak{su_4}}2 \,\, [SU(4)]$

$SU(8)$:
\begin{dynkin}
    \foreach \x in {1,...,7}
    {
        \dynkindot{\x}{0}
    }
    \dynkinline{1}{0}{7}{0}
  \end{dynkin}
$\Rightarrow [SU(8)]\,\, \overset{\mathfrak{su_3}}1 \,\,\overset{\mathfrak{su_4}}2  \,\, ... \overset{\mathfrak{su_4}}2 \,\, [SU(4)]$

$SO(10) \times SU(4)$:
\begin{dynkin}
    \foreach \x in {1,...,4}
    {
        \dynkindot{\x}{0}
    }
    \dynkindot{6}{0}
  \dynkindot{7}{0}
  \dynkindot{8}{0}
    \dynkindot{3}{1}
    \dynkinline{1}{0}{4}{0}
   \dynkinline{6}{0}{8}{0}
    \dynkinline{3}{0}{3}{1}
  \end{dynkin}
$\Rightarrow [SO(10)]\,\, 1 \,\,\overset{\mathfrak{su_4}}{\underset{[SU(4)]}2}  \,\, ... \overset{\mathfrak{su_4}}2 \,\, [SU(4)]$

$SU(8) \times Sp(1)$:
\begin{dynkin}
    \foreach \x in {1,...,6}
    {
        \dynkindot{\x}{0}
    }
    \dynkindot{8}{0}
    \dynkindot{1}{1}
    \dynkinline{1}{0}{6}{0}
    \dynkinline{1}{0}{1}{1}
  \end{dynkin}
$\Rightarrow [SU(8)]\,\,  \overset{\mathfrak{su_4}}{\underset{[Sp(1)]}1} \,\,\overset{\mathfrak{su_4}}2  \,\, ... \overset{\mathfrak{su_4}}2 \,\, [SU(4)]$

The last $Sp(1)$ flavor symmetry here arises because of the anti-symmetric tensor multiplet of $\mathfrak{su}_4$ lives on the $-1$ curve, which is alternatively thought of as the fundamental of $\mathfrak{so_6}$ and hence posesses an $Sp(1)$ flavor symmetry.  In some of the diagrams, there would be $U(1)$ factors at the classical level, but these are all anomalous in six dimensions.

We have explicitly verified that this correspondence proceeds as expected for $\Gamma = \mathbb{Z}_N$ for $N \leq 6$. Beginning at $N = 5$, there are some degeneracies in the list of consistent symmetry breaking patterns, e.g.\ two different homomorphisms give rise to $E_7$ left flavor symmetry.  These degeneracies are echoed in the F-theory language, so that there are indeed two distinct F-theory geometries that give rise to a flavor symmetry group $E_{7,L} \times SU(5)_R$.  This provides strong evidence in support of our proposed correspondence.

\subsection{The D-Series Subgroups of $SU(2)$}

In this subsection we consider the D-series of discrete subgroups $\Gamma_{D_{N}} \subset SU(2)$, namely the case of homomorphisms $\mathbb{D}_{N-2} \rightarrow E_8$, where $\mathbb{D}_{N-2}$ is the binary dihedral finite group of order $4N - 8$. We focus on the specific instances discussed in reference \cite{FREY}, namely those of $\Gamma_{D_5}$ and $\Gamma_{D_7}$, and homomorphisms
which factor through the projection to the dihedral groups.  Here and in subsequent sections, we ignore global $U(1)$ factors, which are expected to be absent in the low energy effective field theory.

\subsubsection{Example: $\Gamma_{D_5}$}

$\Gamma_{D_5}$ is the double cover of Dih$_6$.  There are 7 embeddings of Dih$_6$ into $E_8$, with commutants given (up to anomalous $U(1)$ factors) by $SO(9)$, $F_4 \times SU(2)$, $Sp(4) \times SU(2)$, $SO(11) \times SU(2)$, $SO(7) \times SO(7)$, $SU(8)$, and $E_6$.  We now show how each of these shows up as the left global symmetry of an appropriate F-theory geometry:

$SO(9)$:
$$
\,\,  \overset{II}2  \,\, \underset{[SO(9)]}{\overset{\mathfrak{sp_1}}1} \,\, \overset{\mathfrak{so_{10}}}4 \,\,\overset{\mathfrak{sp_1}}1 \,\, \overset{\mathfrak{so_{10}}}4 \,\, ... \overset{\mathfrak{sp_1}}1 \,\, [SO(10)]$$

$F_4 \times SU(2)$:
$$
[F_4] \,\, 1 \,\, \underset{[SU(2)]}{\overset{\mathfrak{g_2}}3}  \,\,1\,\, \overset{\mathfrak{so_{9}}}4 \,\,\underset{[N_f=\frac{1}{2}]}{\overset{\mathfrak{sp_1}}1} \,\, \overset{\mathfrak{so_{10}}}4 \,\, ... \overset{\mathfrak{sp_1}}1 \,\, [SO(10)]$$

$Sp(4) \times SU(2)$:
$$
 {[Sp(4) \times SU(2)]} \,\, {\overset{\mathfrak{so_7}}2}  \,\,1 \,\, \overset{\mathfrak{so_{9}}}4 \,\,\underset{[N_f=\frac{1}{2}]}{\overset{\mathfrak{sp_1}}1} \,\, \overset{\mathfrak{so_{10}}}4 \,\, ... \overset{\mathfrak{sp_1}}1 \,\, [SO(10)]$$

$SO(11) \times SU(2)$:
$$
 {[SO(11) ]} \,\, \overset{\mathfrak{sp_1}}1\,\, \underset{[SU(2)]}{\overset{\mathfrak{so_9}}3}   \,\,\underset{[N_f=\frac{1}{2}]}{\overset{\mathfrak{sp_1}}1} \,\, \overset{\mathfrak{so_{10}}}4 \,\, ... \overset{\mathfrak{sp_1}}1 \,\, [SO(10)]$$

$SO(7) \times SO(7)$:
$$
 {[SO(7) ]} \,\, 1 \,\, \underset{[SO(7)]}{\underset{1}{\overset{\mathfrak{so_{9}}}4}} \,\,\underset{[N_f=\frac{1}{2}]}{\overset{\mathfrak{sp_1}}1} \,\, \overset{\mathfrak{so_{10}}}4 \,\, ... \overset{\mathfrak{sp_1}}1 \,\, [SO(10)]$$

$SU(8)$:
$$
 {[SU(8) ]} \,\, \overset{\mathfrak{su_4}}2  \,\, 1 \,\, \overset{\mathfrak{so_{9}}}4 \,\,\underset{[N_f=\frac{1}{2}]}{\overset{\mathfrak{sp_1}}1} \,\,\overset{\mathfrak{so_{10}}}4 \,\, ... \overset{\mathfrak{sp_1}}1 \,\, [SO(10)]$$

$E_6$:
$$
 {[E_6 ]}\,\, 1 \,\, \overset{\mathfrak{su_3}}3  \,\, 1 \,\,\overset{\mathfrak{so_{9}}}4 \,\,\underset{[N_f=\frac{1}{2}]}{\overset{\mathfrak{sp_1}}1} \,\, \overset{\mathfrak{so_{10}}}4 \,\, ... \overset{\mathfrak{sp_1}}1 \,\, [SO(10)]$$
Note that in the $SO(9)$ case, the fiber above the $-2$ curve is of Kodaira type $I_1$.  This node eats up one of the flavors of the adjacent $-1$ curve, so that instead of the $SO(10)$ flavor symmetry one would get without the $-2$ curve, one gets an $SO(9)$ flavor symmetry.

\subsubsection{Example: $\Gamma_{D_7} $}

$\Gamma_{D_{7}}$ is the double cover of Dih$_{10}$.  There are 13 embeddings of Dih$_{10}$ into $E_8$, with commutants given (up to anomalous $U(1)$ factors) by $SO(8)$, $SO(7)$, $SO(5) \times SO(5)$, $SO(9) \times SU(2)$, $F_4$, $SO(5) \times SO(5)\times SU(2)$, $Sp(4)$, $SO(11) \times SU(2)$, $SO(7) \times SO(7)$, $SO(10)$, $SU(4) \times SU(4)$, $SU(8)$, and $E_6$.  We now show how each of these shows up as the left global symmetry of an appropriate F-theory geometry:

$SO(8)$:
$$
\,\,  \overset{II}2  \,\, \underset{[SO(8)]}{\overset{\mathfrak{sp_1}}1} \,\, \overset{\mathfrak{so_{11}}}4 \,\,\overset{\mathfrak{sp_2}}1 \,\, \overset{\mathfrak{so_{13}}}4 \,\, \underset{[N_f=\frac{1}{2}]}{\overset{\mathfrak{sp_3}}1} \,\, \overset{\mathfrak{so_{14}}}4 \,\,... \overset{\mathfrak{sp_3}}1 \,\, [SO(14)]
$$

$SO(7)$:
$$
 \,\,  \overset{II}2 \,\, \underset{[SO(7)]}{\overset{\mathfrak{sp_1}}1} \,\, \overset{\mathfrak{so_{12}}}4  \,\, \underset{[N_f=1]}{\overset{\mathfrak{sp_3}}1} \,\, \overset{\mathfrak{so_{14}}}4 \,\,... \overset{\mathfrak{sp_3}}1 \,\, [SO(14)]
$$

$SO(9) \times SU(2)$:
$$
[SO(9)]\,\, 1 \,\, \underset{[SU(2)]}{\overset{\mathfrak{so_7}}3} \,\, \underset{[N_f=\frac{1}{2}]}{\overset{\mathfrak{sp_1}}1} \,\, \overset{\mathfrak{so_{11}}}4 \,\,\overset{\mathfrak{sp_2}}1 \,\, \overset{\mathfrak{so_{13}}}4 \,\,\underset{[N_f=\frac{1}{2}]}{\overset{\mathfrak{sp_3}}1} \,\, \overset{\mathfrak{so_{14}}}4 ...\,\, \overset{\mathfrak{sp_3}}1 \,\, [SO(14)]
$$

$F_4$:
$$
[F_4]\,\, 1 \,\, {\overset{\mathfrak{g_2}}3} \,\, \underset{[N_f=\frac{1}{2}]}{\overset{\mathfrak{sp_1}}1} \,\, \overset{\mathfrak{so_{11}}}4 \,\,\overset{\mathfrak{sp_2}}1 \,\, \overset{\mathfrak{so_{13}}}4 \,\, \underset{[N_f=\frac{1}{2}]}{\overset{\mathfrak{sp_3}}1} \,\, \overset{\mathfrak{so_{14}}}4 \,\,... \overset{\mathfrak{sp_3}}1 \,\, [SO(14)]
$$

$SO(5) \times SO(5) \times SU(2)$:
$$
[SO(5)]\,\, 1 \,\, \underset{[SU(2)]}{\overset{[SO(5)]}{\overset{1}{\overset{\mathfrak{so_{11}}}4}}} \,\,\overset{\mathfrak{sp_2}}1 \,\, \overset{\mathfrak{so_{13}}}4 \,\, \underset{[N_f=\frac{1}{2}]}{\overset{\mathfrak{sp_3}}1} \,\, \overset{\mathfrak{so_{14}}}4 \,\, ... \overset{\mathfrak{sp_3}}1 \,\, [SO(14)]
$$

$Sp(4)$:
$$
[Sp(4)] \,\, \overset{g_2}2 \,\, 1 \,\, {\overset{\mathfrak{so_9}}4} \,\, {\overset{\mathfrak{sp_1}}1} \,\, \overset{\mathfrak{so_{11}}}4 \,\,\overset{\mathfrak{sp_2}}1 \,\, \overset{\mathfrak{so_{13}}}4 \,\, \underset{[N_f=\frac{1}{2}]}{\overset{\mathfrak{sp_3}}1} \,\, \overset{\mathfrak{so_{14}}}4 \,\, ...\overset{\mathfrak{sp_3}}1 \,\, [SO(14)]
$$

$SO(11) \times SU(2)$:
$$
[SO(11)]\,\, \overset{\mathfrak{sp_1}}1 \,\, \underset{[SU(2)]}{\overset{\mathfrak{so_9}}3} \,\, {\overset{\mathfrak{sp_1}}1} \,\, \overset{\mathfrak{so_{11}}}4 \,\,\overset{\mathfrak{sp_2}}1 \,\, \overset{\mathfrak{so_{13}}}4 \,\, \underset{[N_f=\frac{1}{2}]}{\overset{\mathfrak{sp_3}}1} \,\, \overset{\mathfrak{so_{14}}}4 \,\,... \overset{\mathfrak{sp_3}}1 \,\, [SO(14)]
$$

$SO(7) \times SO(7)$:
$$
[SO(7)]\,\, 1 \,\, \underset{[SO(7)]}{\underset{1}{\overset{\mathfrak{so_9}}4}}  \,\, {\overset{\mathfrak{sp_1}}1} \,\, \overset{\mathfrak{so_{11}}}4 \,\,\overset{\mathfrak{sp_2}}1 \,\, \overset{\mathfrak{so_{13}}}4 \,\, \underset{[N_f=\frac{1}{2}]}{\overset{\mathfrak{sp_3}}1} \,\, \overset{\mathfrak{so_{14}}}4 \,\,... \overset{\mathfrak{sp_3}}1 \,\, [SO(14)]
$$

$SO(10)$:
$$
[SO(10)]\,\, \overset{\mathfrak{sp_1}}1 \,\, \underset{[N_s=1]}{\overset{\mathfrak{so_{10}}}3}  \,\,\underset{[N_f=\frac{1}{2}]}{\overset{\mathfrak{sp_2}}1}  \,\, \overset{\mathfrak{so_{13}}}4 \,\, \underset{[N_f=\frac{1}{2}]}{\overset{\mathfrak{sp_3}}1} \,\, \overset{\mathfrak{so_{14}}}4 \,\,... \overset{\mathfrak{sp_3}}1 \,\, [SO(14)]
$$

$SU(4) \times SU(4)$:
$$
[SU(4)]\,\, 1 \,\, \underset{[SU(4)]}{\underset{1}{\overset{\mathfrak{so_{10}}}4}} \,\, \underset{[N_f=\frac{1}{2}]}{\overset{\mathfrak{sp_2}}1} \,\, \overset{\mathfrak{so_{13}}}4 \,\, \underset{[N_f=\frac{1}{2}]}{\overset{\mathfrak{sp_3}}1} \,\, \overset{\mathfrak{so_{14}}}4 \,\,... \overset{\mathfrak{sp_3}}1 \,\, [SO(14)]
$$

$SU(8)$:
$$
[SU(8)] \,\, \overset{\mathfrak{su_4}}2 \,\, 1 \,\, {\overset{\mathfrak{so_9}}4} \,\, {\overset{\mathfrak{sp_1}}1} \,\, \overset{\mathfrak{so_{11}}}4 \,\,\overset{\mathfrak{sp_2}}1 \,\, \overset{\mathfrak{so_{13}}}4 \,\, \underset{[N_f=\frac{1}{2}]}{\overset{\mathfrak{sp_3}}1} \,\, \overset{\mathfrak{so_{14}}}4 \,\, ...\overset{\mathfrak{sp_3}}1 \,\, [SO(14)]
$$

$E_6$:
$$
[E_6]\,\,1 \,\, {\overset{\mathfrak{su_{3}}}3}  \,\, 1 \,\,\overset{\mathfrak{so_{9}}}4 \,\,\overset{\mathfrak{sp_1}}1 \,\, \overset{\mathfrak{so_{11}}}4 \,\,\overset{\mathfrak{sp_2}}1 \,\, \overset{\mathfrak{so_{13}}}4\,\, \underset{[N_f=\frac{1}{2}]}{\overset{\mathfrak{sp_3}}1} \,\, \overset{\mathfrak{so_{14}}}4 \,\,... \overset{\mathfrak{sp_3}}1 \,\, [SO(14)]
$$
Once again, the $SO(8)$ and $SO(7)$ cases arises because the type $I_1$ fiber on the $-2$ curve reduces the flavor symmetry living on the adjacent $-1$ curve from $SO(9)$ to $SO(8)$ and $SO(8)$ to $SO(7)$, respectively.

\subsubsection{Example: $\Gamma_{D_n} \rightarrow \mathbb{Z}_k \rightarrow E_8$}

All binary dihedral groups $\Gamma_{D_n} $ admit a quotient group isomorphic to $\mathbb{Z}_2$.  For $n$ odd, $\Gamma_{D_n} $ also admits a quotient group isomorphic to $\mathbb{Z}_4$.  For $n$ even, it admits a quotient group isomorphic to $\mathbb{Z}_2 \times \mathbb{Z}_2$.  Each homomorphism from one of these quotients into $E_8$ gives rise to a homomorphism $\Gamma_{D_n} \rightarrow E_8$.  Using the classification of homomorphisms from these groups into $E_8$ discussed in subsection \ref{ssec:ASERIES}, we may easily determine the list of commutants which must be realized as flavor symmetries in the superconformal field theory.  We now show these theories arise in F-theory.  We begin with the $\mathbb{Z}_2$ case, which gives commutants $E_7 \times SU(2)$ and $SO(16)$:

$E_7 \times SU(2)$:
$$
[E_7]\,\, 1 \,\, \overset{\mathfrak{su_2}}2 \,\, \underset{[SU(2)]}{\overset{\mathfrak{so_7}}3} \,\, 1 \,\, \overset{\mathfrak{so_{9}}}4 \,\, {\overset{\mathfrak{sp_1}}1} \,\, \overset{\mathfrak{so_{11}}}4 \,\,\overset{\mathfrak{sp_2}}1 \,\, \overset{\mathfrak{so_{13}}}4 \,\, \overset{\mathfrak{sp_3}}1 \,\, \overset{\mathfrak{so_{15}}}4 \,\,... \overset{\mathfrak{sp_{n-4}}}1 \,\, [SO(2n)]
$$

$SO(16)$:
$$
[SO(16)]\,\, \overset{\mathfrak{sp_2}}1  \,\,{\overset{\mathfrak{so_7}}3} \,\, 1 \,\, \overset{\mathfrak{so_{9}}}4 \,\, {\overset{\mathfrak{sp_1}}1} \,\, \overset{\mathfrak{so_{11}}}4 \,\,\overset{\mathfrak{sp_2}}1 \,\, \overset{\mathfrak{so_{13}}}4 \,\, \overset{\mathfrak{sp_3}}1 \,\, \overset{\mathfrak{so_{15}}}4 \,\,... \overset{\mathfrak{sp_{n-4}}}1 \,\, [SO(2n)]
$$
For the $m=4$ case, corresponding to $SO(8)$, it is understood that all of the $-4$ curves should have $\mathfrak{so_8}$ gauge algebra and all of the $-1$ curves adjacent to these $-4$ curves should be empty.

For $n$ odd, we also have homomorphisms $\Gamma_{D_n} \rightarrow \mathbb{Z}_4 \rightarrow E_8$.  This introduces commutants $E_7$, $E_6 \times SU(2)$, $SU(8)$, $SO(10) \times SU(4)$, $SU(8) \times SU(2)$, and $SU(12) \times SU(2)$.  These are realized in F-theory as follows:

$E_7$:
$$
[E_7]\,\, 1 \,\, \overset{\mathfrak{su_2}}2 \,\, {\overset{\mathfrak{g_2}}3} \,\, 1 \,\, \overset{\mathfrak{so_{9}}}4 \,\, {\overset{\mathfrak{sp_1}}1} \,\, \overset{\mathfrak{so_{11}}}4 \,\,\overset{\mathfrak{sp_2}}1 \,\, \overset{\mathfrak{so_{13}}}4 \,\, \overset{\mathfrak{sp_3}}1 \,\, \overset{\mathfrak{so_{15}}}4 \,\,... \overset{\mathfrak{sp_{n-4}}}1 \,\, [SO(2n)]
$$

$E_6 \times SU(2)$:
$$
[E_6]\,\, 1 \,\, \overset{\mathfrak{su_3}}3 \,\, 1 \,\, \overset{\mathfrak{so_{10}}}4 \,\, {\overset{\mathfrak{sp_2}}1} \,\, \overset{\mathfrak{so_{14}}}4 \,\,... \overset{\mathfrak{so_{2n-4}}}4 \,\, \overset{\mathfrak{sp_{n-5}}}1 \,\,\underset{[SU(2)]}{\overset{\mathfrak{so_{2n}}}4} \,\, \overset{\mathfrak{sp_{n-4}}}1\,\,...\overset{\mathfrak{sp_{n-4}}}1 \,\, [SO(2n)]
$$

$SU(8)$:
$$
[SU(8)]\,\, \overset{\mathfrak{su_{n}}}2 \,\, \overset{\mathfrak{sp_{n-4}}}1 \,\, {\overset{\mathfrak{so_{2n}}}4} \,\, \overset{\mathfrak{sp_{n-4}}}1\,\,...\overset{\mathfrak{sp_{n-4}}}1 \,\, [SO(2n)]
$$

$SO(10) \times SU(4)$:
$$
[SO(10)] \,\, \overset{\mathfrak{sp_{\lceil \frac{n-4}{2} \rceil}}}1 \,\, \underset{[SU(4)]}{\underset{1}{\underset{\mathfrak{sp_{\lfloor \frac{n-4}{2} \rfloor}}}{\overset{\mathfrak{so_{2n}}}4}}} \,\, \overset{\mathfrak{sp_{n-4}}}1\,\,...\overset{\mathfrak{sp_{n-4}}}1 \,\, [SO(2n)]
$$

$SU(8) \times SU(2)$:
$$
[SU(8)]\,\, \overset{\mathfrak{su_{4}}}2 \,\, 1 \,\, \overset{\mathfrak{so_{10}}}4 \,\, {\overset{\mathfrak{sp_2}}1} \,\, \overset{\mathfrak{so_{14}}}4 \,\,... \overset{\mathfrak{so_{2n-4}}}4 \,\, \overset{\mathfrak{sp_{n-5}}}1 \,\,\underset{[SU(2)]}{\overset{\mathfrak{so_{2n}}}4} \,\, \overset{\mathfrak{sp_{n-4}}}1\,\,...\overset{\mathfrak{sp_{n-4}}}1 \,\, [SO(2n)]
$$

$SO(12) \times SU(2)$:
$$
[SO(12)]\,\,  \overset{\mathfrak{sp_{1}}}1 \,\,  \underset{[SU(2)]}{\overset{\mathfrak{so_7}}3} \,\, 1 \,\, \overset{\mathfrak{so_{9}}}4 \,\, {\overset{\mathfrak{sp_1}}1} \,\, \overset{\mathfrak{so_{11}}}4 \,\,\overset{\mathfrak{sp_2}}1 \,\, \overset{\mathfrak{so_{13}}}4 \,\, \overset{\mathfrak{sp_3}}1 \,\, \overset{\mathfrak{so_{15}}}4 \,\,... \overset{\mathfrak{sp_{n-4}}}1 \,\, [SO(2n)]
$$

For $n$ even, there are homomorphisms $\Gamma_{D_n} \rightarrow \mathbb{Z}_2 \times \mathbb{Z}_2 \rightarrow E_8$.  This introduces commutants $SU(8)$, $SO(8) \times SO(8)$, $E_6$, and $SO(12) \times SU(2) \times SU(2)$ (as shown
in lemma 3.7 of \cite{cohen-griess}).  These may be realized in F-theory as follows:

$SU(8)$:
$$
[SU(8)]\,\, \overset{\mathfrak{su_{n}}}2 \,\, \overset{\mathfrak{sp_{n-4}}}1 \,\, {\overset{\mathfrak{so_{2n}}}4} \,\, \overset{\mathfrak{sp_{n-4}}}1\,\,...\overset{\mathfrak{sp_{n-4}}}1 \,\, [SO(2n)]
$$

$SO(8) \times SO(8)$:
$$
[SO(8)] \,\, \overset{\mathfrak{sp_{\frac{n-4}{2} }}}1 \,\, \underset{[SO(8)]}{\underset{1}{\underset{\mathfrak{sp_{ \frac{n-4}{2} }}}{\overset{\mathfrak{so_{2n}}}4}}} \,\, \overset{\mathfrak{sp_{n-4}}}1\,\,...\overset{\mathfrak{sp_{n-4}}}1 \,\, [SO(2n)]
$$

$E_6$:
$$
[E_6]\,\, 1 \,\, {\overset{\mathfrak{su_3}}3} \,\, 1 \,\, \overset{\mathfrak{so_{9}}}4 \,\, {\overset{\mathfrak{sp_1}}1} \,\, \overset{\mathfrak{so_{11}}}4 \,\,\overset{\mathfrak{sp_2}}1 \,\, \overset{\mathfrak{so_{13}}}4 \,\, \overset{\mathfrak{sp_3}}1 \,\, \overset{\mathfrak{so_{14}}}4 \,\,... \overset{\mathfrak{sp_{n-4}}}1 \,\, [SO(2n)]
$$

$SO(12) \times SU(2) \times SU(2)$:
$$
[SO(12)]\,\,  \overset{\mathfrak{sp_{1}}}1 \,\,  \underset{[SU(2)]}{\overset{\mathfrak{so_8}}3} \,\, \overset{\mathfrak{sp_{1}}}1 \,\, \overset{\mathfrak{so_{12}}}4 \,\, {\overset{\mathfrak{sp_3}}1} \,\, \overset{\mathfrak{so_{16}}}4 \,\,... \overset{\mathfrak{so_{2n-4}}}4 \,\, \overset{\mathfrak{sp_{n-5}}}1 \,\,\underset{[SU(2)]}{\overset{\mathfrak{so_{2n}}}4} \,\, \overset{\mathfrak{sp_{n-4}}}1\,\,...\overset{\mathfrak{sp_{n-4}}}1 \,\, [SO(2n)]
$$
In this latest case for $n=4$, all of the $-4$ curves carry $\mathfrak{so_8}$ gauge algebra, all of the adjacent $-1$ curves carry no gauge algebra, and there is an $SU(2) \times SU(2)$ flavor symmetry living on the $-3$ curve.

\subsection{The E-Series Subgroups of $SU(2)$}

In this subsection we consider the E-series of discrete subgroups $\Gamma_{E_{N}} \subset SU(2)$ for $N = 6,7,8$, which respectively refer to the binary tetrahedral (order $24$), binary octahedral (order $48$), and binary icosahedral (order $120$) finite groups. We shall confine our analysis to the case of $\Gamma_{E_8}$, the most non-trivial case.  Once again, we ignore global $U(1)$ factors, which are expected to be absent in the low energy effective field theory.

\subsubsection{Example: $\Gamma_{E_8}$}

We now turn to the case of homomorphisms $\Gamma_{E_8} \rightarrow E_8$ which are embeddings. The resulting commutant subgroups have been obtained in \cite{FREY}. Our aim will be to show that each case has a match to an F-theory compactification.  In many cases, there is more than one F-theory compactification which could be associated to a single homomorphism (we do not display these extra geometries for the sake of brevity).  This indicates the presence of non-trivial dualities between the various theories. Here is the list of theories for this case:

$SU(3) \times SU(2)$:
$$
[SU(2)]\,\, \overset{\mathfrak{su_1}}2\,\, \underset{[SU(3)]}1\,\, \overset{\mathfrak{e_6}}6\,\,1 \,\, \overset{\mathfrak{su_3}}3 \,\, 1 \,\, \overset{\mathfrak{f_4}}5 \,\, 1 \,\, \overset{\mathfrak{g_2}}3 \,\, {\overset{\mathfrak{su_2}}2} \,\, 2 \,\, 1 \,\, \overset{\mathfrak{e_8}}{(12)} \,\, ...[E_8]
$$

$SU(3) \times SU(2)$:
$$
[SU(3)]\,\, \overset{\mathfrak{su_2}}2 \,\, \overset{\mathfrak{su_1}}2 \,\, \underset{[SU(2)]}1\,\, \overset{\mathfrak{e_7}}8\,\,1 \,\, \overset{\mathfrak{su_2}}2 \,\, \overset{\mathfrak{g_2}}3 \,\, 1 \,\, \overset{\mathfrak{f_4}}5 \,\, 1 \,\, \overset{\mathfrak{g_2}}3 \,\, {\overset{\mathfrak{su_2}}2} \,\, 2 \,\, 1 \,\, \overset{\mathfrak{e_8}}{(12)} \,\, ...[E_8]
$$

$SU(2) \times SU(2)$:
$$
[SU(2)]\,\,  \overset{\mathfrak{so_9}}4 \,\, 1 \,\, \underset{[SU(2)]}{\overset{\mathfrak{g_2}}3} \,\, 1 \,\, \overset{\mathfrak{f_4}}5 \,\, 1 \,\, \overset{\mathfrak{g_2}}3 \,\, {\overset{\mathfrak{su_2}}2} \,\, 2 \,\, 1 \,\, \overset{\mathfrak{e_8}}{(12)} \,\, ...[E_8]
$$

$SU(2) \times SU(2)$:
$$
[SU(2)] \,\,  \overset{\mathfrak{su_1}}2 \,\, 1 \,\, \overset{2}{\overset{1}{\underset{[SU(2)]}{\underset{2}{\underset{\mathfrak{su_1}}{\underset{1}{\overset{\mathfrak{e_8}}{(12)}}}}}}} \,\, ...[E_8]
$$

$SU(3)$:
$$
[SU(3)]\,\,  1\,\, \overset{\mathfrak{e_6}}6\,\,1 \,\, {\overset{\mathfrak{su_2}}2} \,\, \overset{\mathfrak{g_2}}3 \,\, 1 \,\, \overset{\mathfrak{f_4}}5 \,\, 1 \,\, \overset{\mathfrak{g_2}}3 \,\, {\overset{\mathfrak{su_2}}2} \,\, 2 \,\, 1 \,\, \overset{\mathfrak{e_8}}{(12)} \,\, ...[E_8]
$$

$SU(2) \times SU(2) \times SU(2)$:
$$
[SU(2) \times SU(2) \times SU(2)]\,\,  \overset{\mathfrak{so_8}}3 \,\,  1 \,\, \overset{\mathfrak{g_2}}3 \,\,{\overset{\mathfrak{su_2}}2} \,\, 2 \,\, 1 \,\, \overset{\mathfrak{e_8}}{(11)} \,\, ...[E_8]
$$

$SU(2) \times SU(2) \times SU(2)$:
$$
[SU(2) \times SU(2) \times SU(2)] \,\,  \overset{\mathfrak{so_8}}3 \,\, 1 \,\, \overset{\mathfrak{so_8}}4 \,\,  1 \,\, {\overset{\mathfrak{g_2}}3} \,\, \overset{\mathfrak{su_2}}2 \,\, 2 \,\, 1 \,\, \overset{\mathfrak{e_8}}{(12)} \,\, ...[E_8]
$$

$SU(2)$:
$$
[SU(2)]\,\,  \overset{\mathfrak{so_9}}4 \,\, 1 \,\, {\overset{\mathfrak{su_3}}3} \,\, 1 \,\, \overset{\mathfrak{f_4}}5 \,\, 1 \,\, \overset{\mathfrak{g_2}}3 \,\,{\overset{\mathfrak{su_2}}2} \,\, 2 \,\, 1 \,\, \overset{\mathfrak{e_8}}{(12)} \,\, ...[E_8]
$$

$SU(2) $:
$$
[SU(2)]\,\,  \overset{\mathfrak{so_7}}3 \,\, \overset{\mathfrak{su_2}}2 \,\,  1 \,\, \overset{\mathfrak{f_4}}5 \,\, 1 \,\, \overset{\mathfrak{g_2}}3 \,\,{\overset{\mathfrak{su_2}}2} \,\, 2 \,\, 1 \,\, \overset{\mathfrak{e_8}}{(12)} \,\, ...[E_8]
$$

$SU(2) \times SU(2)$:
$$
\overset{II}2 \,\, \underset{[SU(2)]}{\overset{\mathfrak{su_2}}2} \,\, \underset{[SU(2)]}{\overset{\mathfrak{su_2}}2} \,\, \overset{II}2 \,\, 1 \,\,\overset{\mathfrak{e_8}}{(11)} \,\, ...[E_8]
$$

$SU(2)$:
$$
\overset{\mathfrak{so_8}}4 \,\, 1 \,\, \underset{[SU(2)]}{\overset{\mathfrak{g_2}}3} \,\, 1 \,\, \overset{\mathfrak{f_4}}5 \,\, 1 \,\, \overset{\mathfrak{g_2}}3 \,\,{\overset{\mathfrak{su_2}}2} \,\, 2 \,\, 1 \,\, \overset{\mathfrak{e_8}}{(12)} \,\, ...[E_8]
$$

$SU(2)$:
$$
\underset{[N_f=1]}{\overset{\mathfrak{f_4}}4} \,\, 1 \,\, \underset{[SU(2)]}{\overset{\mathfrak{g_2}}3} \,\, 1 \,\, \overset{\mathfrak{f_4}}5 \,\, 1 \,\, \overset{\mathfrak{g_2}}3 \,\,{\overset{\mathfrak{su_2}}2} \,\, 2 \,\, 1 \,\, \overset{\mathfrak{e_8}}{(12)} \,\, ...[E_8]
$$

$SU(2)$:
$$
[SU(2)] \,\, 1 \,\, \underset{[N_f=\frac{1}{2}]}{\overset{\mathfrak{e_7}}7} \,\, 1 \,\, 2 \,\, {\overset{\mathfrak{su_2}}2} \,\, {\overset{\mathfrak{g_2}}3} \,\, 1 \,\, \overset{\mathfrak{f_4}}5 \,\, 1 \,\, \overset{\mathfrak{g_2}}3 \,\,{\overset{\mathfrak{su_2}}2} \,\, 2 \,\, 1 \,\, \overset{\mathfrak{e_8}}{(12)} \,\, ...[E_8]
$$

$SU(2) \times SU(2)$:
$$
[SU(2)] \,\, \overset{\mathfrak{g_2}}3 \,\,  1 \,\, \underset{[SU(2)]}{\overset{\mathfrak{so_9}}4} \,\,1 \,\, {\overset{\mathfrak{g_2}}3} \,\, \overset{\mathfrak{su_2}}2 \,\, 2 \,\, 1 \,\, \overset{\mathfrak{e_8}}{(12)} \,\, ...[E_8]
$$

$SU(2) $:
$$
\overset{\mathfrak{su_3}}3 \,\,  \underset{[SU(2)]}1 \,\, {\overset{\mathfrak{g_2}}3} \,\, {\overset{\mathfrak{su_2}}2} \,\, 2 \,\, 1 \,\, \overset{\mathfrak{e_8}}{(11)} \,\, ...[E_8]
$$

$SU(2) $:
$$
 [SU(2)]\,\,\overset{\mathfrak{g_2}}3 \,\,  1 \,\, \overset{\mathfrak{g_2}}3 \,\,{\overset{\mathfrak{su_2}}2} \,\, 2 \,\, 1 \,\, \overset{\mathfrak{e_8}}{(11)} \,\, ...[E_8]
$$

$SU(2) \times SU(2)$:
$$
[SU(2)] \,\, \overset{\mathfrak{so_7}}3 \,\,  \underset{[SU(2)]}{\overset{\mathfrak{sp_1}}1} \,\, \overset{\mathfrak{so_9}}4 \,\,1 \,\, {\overset{\mathfrak{g_2}}3} \,\, \overset{\mathfrak{su_2}}2 \,\, 2 \,\, 1 \,\, \overset{\mathfrak{e_8}}{(12)} \,\, ...[E_8]
$$

$G_2 \times SU(2)$:
$$
[G_2]\,\, 1 \,\, \overset{\mathfrak{f_4}}5 \,\, 1 \,\, \underset{[SU(2)]}{\overset{\mathfrak{g_2}}3} \,\, 1 \,\, \overset{\mathfrak{f_4}}5 \,\, 1 \,\, \overset{\mathfrak{g_2}}3 \,\,{\overset{\mathfrak{su_2}}2} \,\, 2 \,\, 1 \,\, \overset{\mathfrak{e_8}}{(12)} \,\, ...[E_8]
$$

$SU(2) \times SU(2) \times SU(2)$:
$$
[SU(2)] \,\, 1\,\, \underset{[SU(2)]}{\underset{1}{\overset{[SU(2)]}{\overset{1}{\overset{\mathfrak{e_7}}8}}}} \,\,1 \,\, {\overset{\mathfrak{su_2}}2} \,\, \overset{\mathfrak{g_2}}3 \,\, 1 \,\, \overset{\mathfrak{f_4}}5 \,\, 1 \,\, \overset{\mathfrak{g_2}}3 \,\,{\overset{\mathfrak{su_2}}2} \,\, 2 \,\, 1 \,\, \overset{\mathfrak{e_8}}{(12)} \,\, ...[E_8]
$$

$SU(3)$:
$$
[SU(3)]\,\,\overset{\mathfrak{su_3}}2 \,\, {\overset{\mathfrak{su_3}}2} \,\, \overset{\mathfrak{su_3}}2 \,\, \overset{\mathfrak{su_2}}2 \,\, \overset{\mathfrak{su_1}}2 \,\, 1 \,\, ...[E_8]
$$

$SU(4) \times SU(2)$:
$$
 [SU(4)] \,\, 1 \,\, \underset{[SU(2)]}{\overset{\mathfrak{so_{10}}}4} \,\, \overset{\mathfrak{sp_1}}1 \,\, \overset{\mathfrak{so_9}}4  \,\, 1 \,\, {\overset{\mathfrak{g_2}}3}  \,\,  \overset{\mathfrak{su_2}}2 \,\, 2 \,\, 1 \,\, \overset{\mathfrak{e_8}}{(12)} \,\, ...[E_8]
$$

$SU(2)$:
$$
 2 \,\, 2 \,\, \underset{[SU(2)]}1\,\, \overset{\mathfrak{e_7}}8 \,\,1 \,\, {\overset{\mathfrak{su_2}}2} \,\, \overset{\mathfrak{g_2}}3 \,\, 1 \,\, \overset{\mathfrak{f_4}}5 \,\, 1 \,\, \overset{\mathfrak{g_2}}3 \,\,{\overset{\mathfrak{su_2}}2} \,\, 2 \,\, 1 \,\, \overset{\mathfrak{e_8}}{(12)} \,\, ...[E_8]
$$

$SU(2) \times SU(2)$:
$$
[SU(2)] \,\, 1\,\, \underset{[SU(2)]}{\underset{1}{\underset{[N_f=\frac{1}{2}]}{\overset{\mathfrak{e_7}}7}}} \,\, 1 \,\, {\overset{\mathfrak{su_2}}2} \,\, \overset{\mathfrak{g_2}}3 \,\, 1 \,\, \overset{\mathfrak{f_4}}5 \,\, 1 \,\, \overset{\mathfrak{g_2}}3 \,\,{\overset{\mathfrak{su_2}}2} \,\, 2 \,\, 1 \,\, \overset{\mathfrak{e_8}}{(12)} \,\, ...[E_8]
$$

$SU(2) \times SU(2) $:
$$
[SU(2)] \,\, 1\,\, \underset{[SU(2)]}{\underset{1}{\overset{\mathfrak{e_7}}8}} \,\,1 \,\, 2 \,\, {\overset{\mathfrak{su_2}}2} \,\, \overset{\mathfrak{g_2}}3 \,\, 1 \,\, \overset{\mathfrak{f_4}}5 \,\, 1 \,\, \overset{\mathfrak{g_2}}3 \,\,{\overset{\mathfrak{su_2}}2} \,\, 2 \,\, 1 \,\, \overset{\mathfrak{e_8}}{(12)} \,\, ...[E_8]
$$

$SU(2) $:
$$
[SU(2)] \,\, 1\,\, \underset{[N_f=1]}{\overset{\mathfrak{e_7}}6} \,\, 1 \,\, {\overset{\mathfrak{su_2}}2} \,\, \overset{\mathfrak{g_2}}3 \,\, 1 \,\, \overset{\mathfrak{f_4}}5 \,\, 1 \,\, \overset{\mathfrak{g_2}}3 \,\,{\overset{\mathfrak{su_2}}2} \,\, 2 \,\, 1 \,\, \overset{\mathfrak{e_8}}{(12)} \,\, ...[E_8]
$$

$G_2 \times SU(2)$:
$$
[G_2]  1 \,\,{\overset{\mathfrak{f_4}}5} \,\, 1 \,\, \underset{[SU(2)]}{\overset{\mathfrak{g_2}}3} \,\, 1 \,\, \overset{\mathfrak{f_4}}5 \,\, 1 \,\, \overset{\mathfrak{g_2}}3 \,\,{\overset{\mathfrak{su_2}}2} \,\, 2 \,\, 1 \,\, \overset{\mathfrak{e_8}}{(12)} \,\, ...[E_8]
$$

$G_2 \times SU(2)$:
$$
[SU(2)] \,\, \overset{\mathfrak{g_2}}3 \,\, 1 \,\, \underset{[G_2]}{\underset{1}{\overset{\mathfrak{f_4}}5}} \,\, 1 \,\, \overset{\mathfrak{g_2}}3 \,\,{\overset{\mathfrak{su_2}}2} \,\, 2 \,\, 1 \,\, \overset{\mathfrak{e_8}}{(12)} \,\, ...[E_8]
$$

$F_4 \times SU(2)$:
$$
[F_4]\,\, 1 \,\, \underset{[SU(2)]}{\overset{\mathfrak{g_2}}3} \,\, 1 \,\, \overset{\mathfrak{f_4}}5 \,\, 1 \,\, \overset{\mathfrak{g_2}}3 \,\,{\overset{\mathfrak{su_2}}2} \,\, 2 \,\, 1 \,\, \overset{\mathfrak{e_8}}{(12)} \,\, ...[E_8]
$$

$SU(2) \times SU(2) \times SU(2)$:
$$
[SU(2) \times SU(2)] \,\,  \overset{\mathfrak{so_8}}3 \,\,  \underset{[SU(2)]}{\overset{\mathfrak{sp_1}}1} \,\, \overset{\mathfrak{so_9}}4 \,\,  1 \,\, {\overset{\mathfrak{g_2}}3} \,\, \overset{\mathfrak{su_2}}2 \,\, 2 \,\, 1 \,\, \overset{\mathfrak{e_8}}{(12)} \,\, ...[E_8]
$$

$SU(3)$:
$$
\overset{\mathfrak{su_2}}2 \,\, \underset{[SU(3)]}{\overset{\mathfrak{su_4}}2} \,\, \overset{\mathfrak{su_3}}2 \,\, \overset{\mathfrak{su_2}}2 \,\, \overset{\mathfrak{su_1}}2 \,\, 1 \,\, ...[E_8]
$$

$SU(6)$:
$$
[SU(6)]\,\, \overset{\mathfrak{su_3}}2\,\, 1\,\, \overset{\mathfrak{e_6}}6\,\,1 \,\, \overset{\mathfrak{su_3}}3 \,\, 1 \,\, \overset{\mathfrak{f_4}}5 \,\, 1 \,\, \overset{\mathfrak{g_2}}3 \,\,{\overset{\mathfrak{su_2}}2} \,\, 2 \,\, 1 \,\, \overset{\mathfrak{e_8}}{(12)} \,\, ...[E_8]
$$

$SU(6)$:
$$
[SU(6)]\,\, \overset{\mathfrak{su_4}}2 \,\, \overset{\mathfrak{su_2}}2\,\, 1\,\, \overset{\mathfrak{e_7}}8\,\,1 \,\, \overset{\mathfrak{su_2}}2\,\, \overset{\mathfrak{g_2}}3 \,\, 1 \,\, \overset{\mathfrak{f_4}}5 \,\, 1 \,\, \overset{\mathfrak{g_2}}3 \,\,{\overset{\mathfrak{su_2}}2} \,\, 2 \,\, 1 \,\, \overset{\mathfrak{e_8}}{(12)} \,\, ...[E_8]
$$

$Sp(4)$:
$$
[Sp(4)]\,\, \overset{\mathfrak{g_2}}2 \,\, 1 \,\, \overset{\mathfrak{f_4}}5 \,\, 1 \,\, \overset{\mathfrak{g_2}}3 \,\,{\overset{\mathfrak{su_2}}2} \,\, 2 \,\, 1 \,\, \overset{\mathfrak{e_8}}{(11)} \,\, ...[E_8]
$$

$E_7$:
$$
[E_7]\,\, 1 \,\,{\overset{\mathfrak{su_2}}2} \,\, \overset{\mathfrak{g_2}}3 \,\, 1 \,\, \overset{\mathfrak{f_4}}5 \,\, 1 \,\, \overset{\mathfrak{g_2}}3 \,\,{\overset{\mathfrak{su_2}}2} \,\, 2 \,\, 1 \,\, \overset{\mathfrak{e_8}}{(11)} \,\, ...[E_8]
$$

$\emptyset$:
$$
\overset{\mathfrak{so_8}}4 \,\, 1 \,\, \overset{\mathfrak{su_3}}3 \,\, 1 \,\, \overset{\mathfrak{f_4}}5 \,\, 1 \,\, \overset{\mathfrak{g_2}}3 \,\,{\overset{\mathfrak{su_2}}2} \,\, 2 \,\, 1 \,\, \overset{\mathfrak{e_8}}{(12)} \,\, ...[E_8]
$$

$\emptyset$:
$$
 \overset{\mathfrak{g_2}}3 \,\, {\overset{\mathfrak{su_2}}2} \,\,  1 \,\, \overset{\mathfrak{f_4}}5 \,\, 1 \,\, \overset{\mathfrak{g_2}}3 \,\,{\overset{\mathfrak{su_2}}2} \,\, 2 \,\, 1 \,\, \overset{\mathfrak{e_8}}{(12)} \,\, ...[E_8]
$$

$SO(7) \times SU(2)$:
$$
[SO(7)]\,\,   1\,\, \underset{[SU(2)]}{\overset{\mathfrak{so_9}}4}\,\,1 \,\, \overset{\mathfrak{g_2}}3 \,\,{\overset{\mathfrak{su_2}}2} \,\, 2 \,\, 1 \,\, \overset{\mathfrak{e_8}}{(11)} \,\, ...[E_8]
$$

$SU(4)$:
$$
[SU(4)]\,\,  \overset{\mathfrak{su_4}}2  \,\, {\overset{\mathfrak{su_4}}2}  \,\, {\overset{\mathfrak{su_3}}2} \,\, {\overset{\mathfrak{su_2}}2} \,\, \overset{\mathfrak{su_1}}2 \,\, 1 \,\, \overset{\mathfrak{e_8}}{(12)} \,\, ...[E_8]
$$

$SU(4)$:
$$
[SU(4)]\,\,   {\overset{\mathfrak{su_3}}2} \,\, {\overset{\mathfrak{su_2}}2} \,\, \overset{\mathfrak{su_1}}2 \,\, 1 \,\, \overset{\mathfrak{e_8}}{(10)} \,\, ...[E_8]
$$

$SO(12)$:
$$
[SO(12)] \,\,  \overset{\mathfrak{sp_1}}1  \,\, \overset{\mathfrak{g_2}}3 \,\, 1  \,\, \overset{\mathfrak{f_4}}5 \,\, 1 \,\,   \overset{\mathfrak{g_2}}3 \,\, \overset{\mathfrak{su_2}}2 \,\, 2 \,\, 1 \,\, \overset{\mathfrak{e_8}}{(10)} \,\, ...[E_8]
$$

$SO(11)$:
$$
[SO(11)]\,\,   \overset{\mathfrak{sp_1}}1\,\, \overset{\mathfrak{so_9}}4\,\,1 \,\, \overset{\mathfrak{g_2}}3 \,\,{\overset{\mathfrak{su_2}}2} \,\, 2 \,\, 1 \,\, \overset{\mathfrak{e_8}}{(11)} \,\, ...[E_8]
$$

$SO(13)$:
$$
[SO(13)]\,\,   \overset{\mathfrak{sp_2}}1\,\, \overset{\mathfrak{so_{11}}}4 \,\, \overset{\mathfrak{sp_1}}1\,\, \overset{\mathfrak{so_{9}}}4 \,\, {1} \,\, \overset{\mathfrak{g_2}}3 \,\, \overset{\mathfrak{su_2}}2 \,\, 2 \,\, 1 \,\, \overset{\mathfrak{e_8}}{(11)} \,\, ...[E_8]
$$

$SU(2) \times SU(2)$:
$$
[SU(2) \times SU(2)] \,\, \overset{\mathfrak{so_9}}3 \,\, \overset{\mathfrak{sp_1}}1 \,\, \overset{\mathfrak{so_9}}4  \,\, 1 \,\, \overset{\mathfrak{g_2}}3 \,\,{\overset{\mathfrak{su_2}}2} \,\, 2 \,\, 1 \,\, \overset{\mathfrak{e_8}}{(12)} \,\, ...[E_8]
$$

$SU(2) \times SU(2)$:
$$
[SU(2)]\,\,  \overset{\mathfrak{su_3}}2 \,\,  \underset{[SU(2)]}{\overset{\mathfrak{su_4}}2}  \,\,\overset{\mathfrak{su_3}}2  \,\, {\overset{\mathfrak{su_2}}2} \,\, \overset{\mathfrak{su_1}}2 \,\, 1 \,\, \overset{\mathfrak{e_8}}{(12)} \,\, ...[E_8]
$$

$SU(2) \times SU(2)$:
$$
\overset{\mathfrak{su_2}}2 \,\, \underset{[SU(2) \times SU(2)]}{\overset{\mathfrak{so_8}}3} \,\,  1  \,\, {\overset{\mathfrak{g_2}}3}  \,\, \overset{\mathfrak{su_2}}2 \,\, 2 \,\, 1 \,\, \overset{\mathfrak{e_8}}{(12)} \,\, ...[E_8]
$$

$SU(2)$:
$$
\overset{\mathfrak{su_2}}2 \,\, \underset{[SU(2) ]}{\overset{\mathfrak{so_7}}3} \,\,  1  \,\, {\overset{\mathfrak{g_2}}3}  \,\, \overset{\mathfrak{su_2}}2 \,\, 2 \,\, 1 \,\, \overset{\mathfrak{e_8}}{(12)} \,\, ...[E_8]
$$

$SU(2)$:
$$
\overset{\mathfrak{su_2}}2  \,\, \underset{[SU(2)]}{\overset{\mathfrak{su_3}}2}  \,\, {\overset{\mathfrak{su_2}}2} \,\, \overset{\mathfrak{su_1}}2 \,\, 1 \,\, \overset{\mathfrak{e_8}}{(11)} \,\, ...[E_8]
$$

$SU(2)$:
$$
[SU(2)] \,\, \overset{\mathfrak{su_2}}2 \,\,  \underset{[N_f=1]}{\overset{\mathfrak{su_2}}2} \,\, \overset{\mathfrak{su_1}}2 \,\, 1 \,\, \overset{\mathfrak{e_8}}{(10)} \,\, ...[E_8]
$$

$SO(5) \times SO(5)$:
$$
[SO(5)]\,\,  1 \,\, \underset{[SO(5)]}{\overset{\mathfrak{so_{11}}}4} \,\, {\overset{\mathfrak{sp_1}}1} \,\, \overset{\mathfrak{so_{9}}}4 \,\, 1 \,\, \overset{\mathfrak{g_2}}3 \,\, \overset{\mathfrak{su_2}}2 \,\, 2 \,\, 1 \,\, \overset{\mathfrak{e_8}}{(12)} \,\, ...[E_8]
$$

$SO(5) $:
$$
[SO(5)] \,\, \overset{\mathfrak{so_{7}}}3 \,\,1 \,\, \overset{\mathfrak{so_{8}}}4 \,\, 1 \,\, {\overset{\mathfrak{g_2}}3} \,\, \overset{\mathfrak{su_2}}2 \,\, 2 \,\, 1 \,\, \overset{\mathfrak{e_8}}{(12)} \,\, ...[E_8]
$$

$SO(5) $:
$$
[SO(5)] \,\, \underset{[N_s=1]}{\overset{\mathfrak{so_{10}}}3} \,\, \underset{[N_f=\frac{1}{2}]}{\overset{\mathfrak{sp_1}}1} \,\, \overset{\mathfrak{so_{9}}}4 \,\, 1 \,\, \overset{\mathfrak{g_2}}3 \,\,{\overset{\mathfrak{su_2}}2} \,\, 2 \,\, 1 \,\, \overset{\mathfrak{e_8}}{(12)} \,\, ...[E_8]
$$

$SO(5) $:
$$
[SO(5)] \,\, \overset{\mathfrak{so_{7}}}3  \,\, 1 \,\, {\overset{\mathfrak{g_2}}3} \,\, \overset{\mathfrak{su_2}}2 \,\, 2 \,\, 1 \,\, \overset{\mathfrak{e_8}}{(12)} \,\, ...[E_8]
$$

$SO(5) $:
$$
 \overset{\mathfrak{su_2}}2 \,\, \underset{[SO(5)]}{\overset{\mathfrak{g_2}}2} \,\, \overset{\mathfrak{su_2}}2 \,\, 2 \,\, 1 \,\, \overset{\mathfrak{e_8}}{(11)} \,\, ...[E_8]
$$

$SO(5)$:
$$
[SO(5)] \,\, \overset{\mathfrak{so_{10}}}4 \,\, 1 \,\, \overset{\mathfrak{su_3}}3 \,\, 1 \,\, \overset{\mathfrak{f_4}}5 \,\, 1 \,\, \overset{\mathfrak{g_2}}3 \,\,{\overset{\mathfrak{su_2}}2} \,\, 2 \,\, 1 \,\, \overset{\mathfrak{e_8}}{(12)} \,\, ...[E_8]
$$

$SO(5) \times SU(2)$:
$$
[SO(5) \times SU(2)] \,\,  {\overset{\mathfrak{so_9}}3}  \,\, 1 \,\, {\overset{\mathfrak{g_2}}3} \,\, \overset{\mathfrak{su_2}}2 \,\, 2 \,\, 1 \,\, \overset{\mathfrak{e_8}}{(11)} \,\, ...[E_8]
$$

$SO(9) \times SU(2)$:
$$
[SO(9)]\,\,  \overset{\mathfrak{sp_1}}1 \,\, \underset{[SU(2)]}{\overset{\mathfrak{so_{11}}}4} \,\, {\overset{\mathfrak{sp_1}}1} \,\, \overset{\mathfrak{so_{9}}}4 \,\, 1 \,\, \overset{\mathfrak{g_2}}3 \,\, \overset{\mathfrak{su_2}}2 \,\, 2 \,\, 1 \,\, \overset{\mathfrak{e_8}}{(12)} \,\, ...[E_8]
$$

The only non-trivial normal subgroup of $\Gamma_{E_8}$ is isomorphic to $\mathbb{Z}_2$, which means that the only only (non-trivial) homomorphisms $\Gamma_{E_8} \rightarrow E_8$ are characterized by embeddings of the order 60 subgroup of $\Gamma_{E_8} \simeq \mathfrak{A}_5$ into $E_8$.  Such homomorphisms have also been characterized by \cite{FREY}.  We now show how one may realize these homomorphisms as F-theory geometries:

$G_2$:
$$
[G_2]\,\, 1 \,\, \overset{\mathfrak{f_4}}5 \,\, 1 \,\, {\overset{\mathfrak{su_3}}3} \,\, 1 \,\, \overset{\mathfrak{f_4}}5 \,\, 1 \,\, \overset{\mathfrak{g_2}}3 \,\,{\overset{\mathfrak{su_2}}2} \,\, 2 \,\, 1 \,\, \overset{\mathfrak{e_8}}{(12)} \,\, ...[E_8]
$$

$SO(8)$:
$$
[SO(8)]\,\, 1 \,\, \overset{\mathfrak{so_8}}4  \,\, 1 \,\, \overset{\mathfrak{g_2}}3 \,\,{\overset{\mathfrak{su_2}}2} \,\, 2 \,\, 1 \,\, \overset{\mathfrak{e_8}}{(11)} \,\, ...[E_8]
$$

$SO(7)$:
$$
[SO(7)]\,\,  1 \,\, \overset{\mathfrak{so_9}}4 \,\,  \underset{[N_f=1]}{\overset{\mathfrak{sp_1}}1} \,\, \overset{\mathfrak{so_9}}4 \,\, 1 \,\, \overset{\mathfrak{g_2}}3 \,\,{\overset{\mathfrak{su_2}}2} \,\, 2 \,\, 1 \,\, \overset{\mathfrak{e_8}}{(12)} \,\, ...[E_8]
$$

$SU(5)$:
$$
[SU(5)]\,\,  \overset{\mathfrak{su_4}}2  \,\,  \overset{\mathfrak{su_3}}2 \,\, {\overset{\mathfrak{su_2}}2} \,\, \overset{\mathfrak{su_1}}2 \,\, 1 \,\, \overset{\mathfrak{e_8}}{(11)} \,\, ...[E_8]
$$

$F_4$:
$$
[F_4]\,\, 1 \,\, \overset{\mathfrak{g_2}}3 \,\,{\overset{\mathfrak{su_2}}2} \,\, 2 \,\, 1 \,\, \overset{\mathfrak{e_8}}{(10)} \,\, ...[E_8]
$$

$SO(10)$:
$$
[SO(10)]\,\,\overset{\mathfrak{sp_1}}1 \,\, \overset{\mathfrak{so_{10}}}4  \,\,\overset{\mathfrak{sp_1}}1 \,\, \overset{\mathfrak{so_9}}4  \,\, 1 \,\, \overset{\mathfrak{g_2}}3 \,\,{\overset{\mathfrak{su_2}}2} \,\, 2 \,\, 1 \,\, \overset{\mathfrak{e_8}}{(12)} \,\, ...[E_8]
$$

$E_6$:
$$
[E_6]\,\,  1 \,\, {\overset{\mathfrak{su_3}}3} \,\, 1 \,\, \overset{\mathfrak{f_4}}5 \,\, 1 \,\, \overset{\mathfrak{g_2}}3 \,\,{\overset{\mathfrak{su_2}}2} \,\, 2 \,\, 1 \,\, \overset{\mathfrak{e_8}}{(12)} \,\, ...[E_8]
$$

$SU(3)$:
$$
[SU(3)]\,\,{\overset{\mathfrak{su_3}}2}\,\,{\overset{\mathfrak{su_3}}2}\,\,{\overset{\mathfrak{su_2}}2}  \,\,  \overset{\mathfrak{su_1}}2 \,\, 1 \,\, \overset{\mathfrak{e_8}}{(12)} \,\, ...[E_8]
$$

$\emptyset$:
$$
{\overset{\mathfrak{su_2}}2}\,\,{\overset{\mathfrak{su_3}}2}\,\,{\overset{\mathfrak{su_2}}2}  \,\,  \overset{\mathfrak{su_1}}2 \,\, 1 \,\, \overset{\mathfrak{e_8}}{(12)} \,\, ...[E_8]
$$

$\emptyset$:
$$
\overset{\mathfrak{su_3}}3 \,\, 1 \,\, \overset{\mathfrak{so_8}}4  \,\, 1 \,\, \overset{\mathfrak{g_2}}3 \,\,{\overset{\mathfrak{su_2}}2} \,\, 2 \,\, 1 \,\, \overset{\mathfrak{e_8}}{(12)} \,\, ...[E_8]
$$

$SU(2) \times SU(2)$:
$$
\overset{\mathfrak{g_2}}3 \,\,  \underset{[SU(2) \times SU(2)]}{\overset{\mathfrak{sp_1}}1} \,\, \overset{\mathfrak{so_9}}4  \,\, 1 \,\, \overset{\mathfrak{g_2}}3 \,\,{\overset{\mathfrak{su_2}}2} \,\, 2 \,\, 1 \,\, \overset{\mathfrak{e_8}}{(12)} \,\, ...[E_8]
$$

$SU(2)$:
$$
[SU(2)]\,\, \overset{\mathfrak{su_1}}2 \,\, 1 \,\, \overset{\mathfrak{e_8}}{(8)} \,\, ...[E_8]
$$

$SU(2)$:
$$
[SU(2)] \,\,  \overset{\mathfrak{so_{10}}}4 \,\, {\overset{\mathfrak{sp_1}}1} \,\, {\overset{\mathfrak{g_2}}3} \,\, 1 \,\, \overset{\mathfrak{f_4}}5 \,\, 1 \,\, \overset{\mathfrak{g_2}}3 \,\,{\overset{\mathfrak{su_2}}2} \,\, 2 \,\, 1 \,\, \overset{\mathfrak{e_8}}{(12)} \,\, ...[E_8]
$$

$SU(2) \times SU(2) \times SU(2)$:
$$
[SU(2)] \,\, \overset{\mathfrak{su_1}}2 \,\, 1\,\, \underset{[SU(2)]}{\underset{2}{\underset{\mathfrak{su_1}}{\underset{1}{\overset{[SU(2)]}{\overset{\mathfrak{su_1}}{\overset{2}{\overset{1}{\overset{\mathfrak{e_8}}{(12)}}}}}}}}} \,\, ...[E_8]
$$

$G_2 \times G_2$:
$$
[G_2] \,\,  1 \,\, \underset{[G_2]}{\underset{1}{\overset{\mathfrak{f_4}}5}} \,\, 1 \,\, \overset{\mathfrak{g_2}}3 \,\,{\overset{\mathfrak{su_2}}2} \,\, 2 \,\, 1 \,\, \overset{\mathfrak{e_8}}{(12)} \,\, ...[E_8]
$$

$SU(3) \times SU(3)$:
$$
[SU(3)] \,\,  1 \,\, \underset{[SU(3)]}{\underset{1}{\overset{\mathfrak{e_6}}6}} \,\, 1 \,\, {\overset{\mathfrak{su_3}}3} \,\,1 \,\, \overset{\mathfrak{f_4}}5 \,\, 1\,\, \overset{\mathfrak{g_2}}3 \,\,{\overset{\mathfrak{su_2}}2} \,\, 2 \,\, 1 \,\, \overset{\mathfrak{e_8}}{(12)} \,\, ...[E_8]
$$

\subsection{The Reverse Correspondence}

It is remarkable that we have found an F-theory realization for all of the breaking patterns expected on the heterotic side. It is
natural to ask, however, whether the F-theory small instanton theories can generate any SCFTs with
a flavor symmetry which cannot be realized by a choice of
breaking pattern controlled by a discrete group homomorphism. Up to
some minor discrepancies with the list obtained in reference \cite{FREY},
we find that the match is onto but not one-to-one. The presence of
multiple F-theory models with the same flavor symmetry on the heterotic side is accounted for by identifying
all theories which can be connected by RG flows. Said differently, for each RG flow along which we preserve the
same flavor symmetry, we expect to get a single embedding on the heterotic side. Detailed
examples of such flows are given in reference \cite{BackToTheFuture}.

In comparing with the breaking patterns obtained in reference \cite{FREY}, we have found that in some cases where the algebra $B_3$ (i.e. $SO(7)$) has been indicated, the F-theory realization instead indicates that the algebra $C_3$ (i.e. $Sp(3)$) should instead appear. More precisely,
we have found exactly three instances of $Sp(3)$ flavor symmetry, two of $Sp(3) \times SU(2)$, one of $SO(7)$, and one of $SO(7) \times SU(2)$.  The lists of \cite{FREY}, on the other hand, contain no instances of either $Sp(3)$ or $Sp(3) \times SU(2)$, four instances of $SO(7)$, and three of $SO(7) \times SU(2)$.  We suspect that there may be slight typos in this list and that three of the instances of $B_3$ should actually be $C_3$, while two of the instances of $B_3 A_1$ should be $C_3 A_1$.\footnote{We are grateful to D.D. Frey for discussions on this point.} It would be most instructive to verify that there is indeed a typo in the list of \cite{FREY}. For completeness, here is the list of theories where an $Sp(3)$ flavor symmetry algebra appears:

$Sp(3)$:
$$
[Sp(3)] \,\, \overset{\mathfrak{so_{11}}}{3} \,\, \overset{\mathfrak{sp_1}}{1} \,\, \overset{\mathfrak{so_9}}4 \,\, 1 \,\, \overset{\mathfrak{g_2}}3 \,\, \overset{\mathfrak{su_2}}2 \,\, 2 \,\, 1 \,\, \overset{\mathfrak{e_8}}{(12)} \,\, ...[E_8]
$$

$Sp(3)$:
$$
[Sp(3)] \,\, \overset{\mathfrak{g_2}}2 \,\, \overset{\mathfrak{su_2}}2  \,\,  2 \,\, 1 \,\, {\overset{\mathfrak{e_8}}{(9)}}  \,\, ...[E_8]
$$

$Sp(3) $:
$$
[Sp(3)] \,\, \overset{\mathfrak{g_2}}2 \,\, {\overset{\mathfrak{su_2}}2}\,\, 1\,\, \overset{\mathfrak{e_7}}8\,\,1 \,\, \overset{\mathfrak{su_2}}2\,\, \overset{\mathfrak{g_2}}3 \,\, 1 \,\, \overset{\mathfrak{f_4}}5 \,\, 1 \,\, \overset{\mathfrak{g_2}}3 \,\,{\overset{\mathfrak{su_2}}2} \,\, 2 \,\, 1 \,\, \overset{\mathfrak{e_8}}{(12)} \,\, ...[E_8]
$$

$Sp(3) \times SU(2)$:
$$
[Sp(3) \times SU(2)] \overset{\mathfrak{so_7}}2 \,\, \overset{\mathfrak{su_2}}{2}\,\, \underset{[SU(2)]}1\,\, \overset{\mathfrak{e_7}}8\,\,1 \,\, \overset{\mathfrak{su_2}}2\,\, \overset{\mathfrak{g_2}}3 \,\, 1 \,\, \overset{\mathfrak{f_4}}5 \,\, 1 \,\, \overset{\mathfrak{g_2}}3 \,\,{\overset{\mathfrak{su_2}}2} \,\, 2 \,\, 1 \,\, \overset{\mathfrak{e_8}}{(12)} \,\, ...[E_8]
$$

$Sp(3) \times SU(2)$:
$$
[Sp(3)]\,\, \overset{\mathfrak{g_2}}2 \,\,  \overset{\mathfrak{su_2}}2  \,\,  2 \,\, 1 \,\, \underset{[SU(2)]}{\underset{2}{\underset{\mathfrak{su_1}}{\underset{1}{\overset{\mathfrak{e_8}}{(12)}}}}}  \,\, ...[E_8]
$$

Additionally, we have also presented F-theory models which have flavor symmetries $SU(3) \times SU(3)$ and $G_2 \times G_2$.
Absent from the list of reference \cite{FREY} is the ``mixed case'' $G_2 \times SU(3)$, which we suspect must also exist. Indeed,
the F-theory realization of this flavor symmetry pattern is:

$G_2 \times SU(3)$:
$$
[SU(3)]\,\, \overset{\mathfrak{su_2}}2  \,\,  \overset{\mathfrak{su_1}}2 \,\, 1 \,\, \underset{[G_2]}{\underset{2}{\underset{\mathfrak{su_2}}{\underset{2}{\underset{II}{\underset{1}{\overset{\mathfrak{e_8}}{(12)}}}}}}} \,\, ...[E_8]
$$

Again, let us stress that (up to these few cases which we expect will be favorably resolved), the
correspondence is so tight as to leave little doubt about the existence of the proposed duality.

\newpage

\section{Conclusions and Future Directions \label{sec:CONC}}

In this paper we have presented a general classification of 6D SCFTs. The
primary tool in our analysis has been a combination of bottom up constraints
for the 6D effective field theory on the tensor branch, and the complementary
perspective of F-theory compactification. Perhaps the most striking outcome
from this classification is that all 6D\ SCFTs have the structure of
generalized quiver theories in which the links are themselves SCFTs. Our
strategy for accomplishing this result has been to first classify all possible
bases in F-theory which can support an SCFT. Next, we have classified the
general ways to enhance the fiber type of such a base. Finally, we have
presented strong evidence that all boundary data decorations of
these configurations can be understood as the limiting behavior of these
geometric phases.  In the remainder of this section we discuss some
potential avenues of future investigation.

In the previous sections we presented a general classification of F-theory
compactifications which can generate a 6D\ SCFT. It is natural to ask to what
extent this stringy input can be viewed in purely field theoretic
terms.  We have seen that many of the stringy ingredients can be explained
using field theory data, but not all of the stringy data has yet found a field theoretic home.
It would be nice to fill this gap.

Our main emphasis in this work has been on giving a full list of 6D\ SCFTs.
With this in place, we can ask whether there are possible redundancies.  Namely--do different tensor branches ultimately
correspond to the same 6D theory? It would be interesting to determine whether there are other such redundancies in our list.

It is also natural to consider detailed properties of these
theories, for example their operator content. The fact that all of these
theories have a rather similar structure as generalized quivers suggests the
possibility of extracting universal lessons for all 6D\ SCFTs. This would also likely shed significant light on the microscopics of M5-branes.  Recently, the elliptic genera of strings in a number of examples of 6D\ SCFTs have been computed \cite{MStrings,OMStrings,KlemmVafa,KimVafa}, but the present classification offers far more complicated examples, and it would be good to understand the strings of these theories as well.

With a classification of 6D theories in place, a next step would be to consider
the compactification of these theories to lower dimensions, and the possible SCFTs generated in this way. In fact,
it is tempting to conjecture that \textit{all} SCFTs can be obtained by compactification, and then further
relevant and marginal deformations of these theories. Providing evidence for or against this conjecture
would be most instructive.

Finally, throughout this paper we have seen that the structure of a 6D SCFT has some striking analogies with that of chemistry. Pushing
this analogy further, one might consider the time-dependent process of building up a 6D SCFT from smaller ingredients. This would
provide a tractable way to study the time-dependent formation of theories in a landscape of vacua, perhaps along the lines of
references \cite{Douglas:2006za, Douglas:2012bu}.

\newpage

\section*{Acknowledgements}

We thank M.~Bertolini, M.~Del Zotto, T.~Dumitrescu, D.~D.\ Frey, P.~Merkx, D.~S.\ Park, and W.~Taylor for helpful discussions. DRM, TR, and
CV also thank the 2014 Summer Workshop at the Simons Center for Geometry and
Physics for hospitality, where some of this work was completed. The work of
DRM is supported by NSF grant PHY-1307513. The work of TR and CV is supported
by NSF grant PHY-1067976.  TR is also supported by the NSF GRF under DGE-1144152.



\appendix

\section{Instructions for Using the \texttt{Mathematica} Notebooks}

Our arXiv submission features two \texttt{Mathematica} notebooks which may be used to compute allowed bases and fiber enhancements.  To access the  \texttt{Mathematica} notebook, proceed to the URL where the arXiv submission and abstract is displayed.  On the righthand side of the webpage, there will be a box labeled
``Download:". Click on the link ``Other format," and then click on the link ``Download source."  In some cases, it may be necessary to append the ending .tar.gz to the end of the file.  The set of submission files along with the  \texttt{Mathematica} notebook can then be accessed by unzipping this file. For further instruction on unzipping such files, see for example http://arxiv.org/help/unpack.

Once the files have finished downloading, they should be moved to a single directory.  The directory of the file  \texttt{Bases.nb} should be set to the directory in which these files reside by editing the path in the `SetDirectory' line at the top of the notebook.   \texttt{Bases.nb} relies upon the .txt files  \texttt{DE$\_$Bases.txt} and  \texttt{output$\_$template$\_$file.txt} to produce its output in the form of a .tex file, so if the directory is not set to the location of these .txt files, the program will not be able to run successfully.

 \texttt{Bases.nb} requires three inputs: a left side link and a right side link (see Appendix \ref{linkappendix}) as well as a number $n$ of nodes.  The program will take these input data and output a list of all bases with the specified number $n$ of DE nodes and the specified links on the left and right.  Note that the program is also equipped to handle tree-shaped side links.  These should be entered as discussed in the comments at the top of the notebook.

The name of the output .tex file will be given at the end of the program.  The .tex file contains lines of the form,
\begin{equation}
E_8''' \op E_8' \op E_8 \op 122315.
\end{equation}
The notation used here is the same as that which was introduced in section \ref{ssec:twoLinks} and appears in Appendices \ref{infiniteappendix} and \ref{outlierappendix}.  Recall that a ``primed node" indicates that the self-intersection of the curve has been increased by $1$, so that $E_7'$ indicates a curve with self-intersection $-7$, $E_8'$ indicates a curve with self-intersection $-11$, $E_8''$ indicates a curve with self-intersection $-10$, and $E_8'''$ indicates a curve with self-intersection $-9$.  The previous line therefore corresponds to the base,
\begin{equation}
(9)12231513221(11)12231513221(12)122315.
\end{equation}

The notebook  \texttt{Fiber$\_$Enhancements.nb} takes an input link from the table in Appendix \ref{linkappendix} and outputs all possible sequences of gauge symmetries that can live on the fiber above each specified curve.  For instance, on the fiber above the link $223$, the allowed gauge algebras are,
\begin{equation}
\{1\} \oplus \mathfrak{su_2} \oplus \mathfrak{g_2}\,,~~~\{1\} \oplus \mathfrak{su_2} \oplus \mathfrak{so_7} \,,~~~\{1\} \oplus\mathfrak{su_2} \oplus \mathfrak{so_8}.
\end{equation}
Any link which does not contain a curve of self-intersection $-3$ or below can support an infinite number of fiber enhancements.  In these cases, the user must specify the gauge groups appearing on the left and right curves of the base.  The notebook will then compute all ways of filling in the interior of the base.  Note, however, that the notebook does not include the three exceptional fiber enhancements listed in (\ref{threebonuscases}).

\section{The Long Bases \label{infiniteappendix}}

In this Appendix we give all long bases. These are all sequences of
nodes which can be continued to arbitrary size. Since they can always be
incorporated at a later stage of analysis, we do not consider the addition of
side links to the left and right of the quiver or instanton links.
Additionally, for expository purposes (and for the sake of brevity), in this
Appendix we only consider the case of nodes which are not primed. The case of
primed nodes is a specific subset, though the combinatorics of where we can
place a primed node is better presented in the companion \texttt{Mathematica}
programs (which are included in the \texttt{arXiv} submission). However, to
see how to introduce these additional ingredients, we do list the
self-intersections of all curves after blowing down all interior links.

We use the notation introduced in subsection \ref{ssec:twoLinks}. The only
caveat is that now, for the sake of brevity, we have not included superscripts
on $\oplus$ symbols for the minimal link. For example, whenever the symbol
$\oplus$ appears between two $E_{8}$ nodes, it should be interpreted as the
symbol $\overset{5,5}{\oplus}$, which is the minimal conformal matter between
such nodes. The expression $E_{8}^{\oplus3}$ is also equivalent to the
expression $E_{8}\oplus E_{8}\oplus E_{8}$.

Recall that the minimal link between two nodes is as follows:
\begin{equation}
D\oplus D\simeq D\overset{1,1}{\oplus}D
\end{equation}%
\begin{equation}
D\oplus E_{6}\simeq D\overset{2,2}{\oplus}E_{6}%
\end{equation}%
\begin{equation}
D\oplus E_{7}\simeq D\overset{2,3}{\oplus}E_{7}%
\end{equation}%
\begin{equation}
D\oplus E_{8}\simeq D\overset{2,4}{\oplus}E_{8}%
\end{equation}%
\begin{equation}
E_{6}\oplus E_{6}\simeq E_{6}\overset{2,2}{\oplus}E_{6}%
\end{equation}%
\begin{equation}
E_{6}\oplus E_{7}\simeq E_{6}\overset{3,3}{\oplus}E_{7}%
\end{equation}%
\begin{equation}
E_{6}\oplus E_{8}\simeq E_{6}\overset{3,5}{\oplus}E_{8}%
\end{equation}%
\begin{equation}
E_{7}\oplus E_{7}\simeq E_{7}\overset{3,3}{\oplus}E_{7}%
\end{equation}%
\begin{equation}
E_{7}\oplus E_{8}\simeq E_{7}\overset{4,5}{\oplus}E_{8}%
\end{equation}%
\begin{equation}
E_{8}\oplus E_{8}\simeq E_{8}\overset{5,5}{\oplus}E_{8}%
\end{equation}

Using these conventions, we now list the possible bases which can support an
arbitrary number of nodes. We also include the resulting configuration of
curves from blowing down all interior links (which we assume are minimal). In
all cases, we take the integer $n\geq1$, but only display the pattern after
blowdown for generic $n$. Small values of $n$ can readily be reconstructed
from the given data.

First, the configurations involving only one of $E_{8}$, $E_{7}$, $E_{6}$, or
$D$:
\begin{align}
&  E_{8}^{\oplus n}\overset{L}{\rightarrow}\underset{n}{\underbrace{72...27}%
}\\
&  E_{7}^{\oplus n}\overset{L}{\rightarrow}\underset{n}{\underbrace{52...25}%
}\\
&  E_{6}^{\oplus n}\overset{L}{\rightarrow}\underset{n}{\underbrace{42...24}%
}\\
&  D^{\oplus n}\overset{L}{\rightarrow}\underset{n}{\underbrace{32...23}}%
\end{align}

The configurations with $E_{7}$ and $E_{8}$:
\begin{align}
&  E_{7}\oplus E_{8}^{\oplus n}\overset{L}{\rightarrow}%
4\underset{n}{\underbrace{2...27}}\\
&  E_{7}\overset{5,5}{\oplus}E_{8}^{\oplus n}\overset{L}{\rightarrow
}3\underset{n}{\underbrace{2...27}}\\
&  E_{7}\oplus E_{8}^{\oplus n}\oplus E_{7}\overset{L}{\rightarrow}%
4\underset{n}{\underbrace{2...2}}4\\
&  E_{7}\overset{5,5}{\oplus}E_{8}^{\oplus n}\oplus E_{7}\overset{L}{\rightarrow
}3\underset{n}{\underbrace{2...2}}4\\
&  E_{7}\overset{5,5}{\oplus}E_{8}^{\oplus n}\overset{5,5}{\oplus}E_{7}%
\overset{L}{\rightarrow}3\underset{n}{\underbrace{2...2}}3
\end{align}

The configurations with $E_{6}$ and $E_{8}$:
\begin{align}
&  E_{6}\oplus E_{8}^{\oplus n}\overset{L}{\rightarrow}%
3\underset{n}{\underbrace{2...27}}\\
&  E_{6}\overset{4,5}{\oplus}E_{8}^{\oplus n}\,\overset{L}{\rightarrow
}2\underset{n}{\underbrace{2...27}}\\
&  E_{6}\overset{5,5}{\oplus}E_{8}^{\oplus n}\,\overset{L}{\rightarrow
}1\underset{n}{\underbrace{2...27}}\\
&  E_{6}\oplus E_{8}^{\oplus n}\oplus E_{6}\overset{L}{\rightarrow
}3\underset{n}{\underbrace{2...2}}3\\
&  E_{6}\overset{4,5}{\oplus}E_{8}^{\oplus n}\oplus E_{6}%
\,\overset{L}{\rightarrow}2\underset{n}{\underbrace{2...2}}3\\
&  E_{6}\overset{5,5}{\oplus}E_{8}^{\oplus n}\oplus E_{6}%
\overset{L}{\rightarrow}1\underset{n}{\underbrace{2...2}}3\\
&  E_{6}\overset{4,5}{\oplus}E_{8}^{\oplus n}\overset{5,4}{\oplus}%
E_{6}\overset{L}{\rightarrow}2\underset{n}{\underbrace{2...2}}2\\
&  E_{6}\overset{5,5}{\oplus}E_{8}^{\oplus n}\overset{5,4}{\oplus}%
E_{6}\overset{L}{\rightarrow}1\underset{n}{\underbrace{2...2}}2
\end{align}

The configurations with $E_{6}$ and $E_{7}$:
\begin{align}
&  E_{6}\oplus E_{7}^{\oplus n}\overset{L}{\rightarrow}%
3\underset{n}{\underbrace{2...25}}\\
&  E_{6}\oplus E_{7}^{\oplus n}\oplus E_{6}\overset{L}{\rightarrow
}3\underset{n}{\underbrace{2...2}}3
\end{align}

The configurations with $E_{6}$, $E_{7}$, and $E_{8}$:
\begin{align}
&  E_{6}\oplus E_{8}^{\oplus n}\oplus E_{7}\overset{L}{\rightarrow
}3\underset{n}{\underbrace{2...2}}4\\
&  E_{6}\overset{4,5}{\oplus}E_{8}^{\oplus n}\oplus E_{7}%
\overset{L}{\rightarrow}2\underset{n}{\underbrace{2...2}}4\\
&  E_{6}\overset{5,5}{\oplus}E_{8}^{\oplus n}\oplus E_{7}%
\overset{L}{\rightarrow}1\underset{n}{\underbrace{2...2}}4\\
&  E_{6}\oplus E_{8}^{\oplus n}\overset{5,5}{\oplus}E_{7}%
\,\overset{L}{\rightarrow}3\underset{n}{\underbrace{2...2}}3\\
&  E_{6}\overset{4,5}{\oplus}E_{8}^{\oplus n}\overset{5,5}{\oplus}%
E_{7}\overset{L}{\rightarrow}2\underset{n}{\underbrace{2...2}}3\\
&  E_{6}\overset{5,5}{\oplus}E_{8}^{\oplus n}\overset{5,5}{\oplus}%
E_{7}\overset{L}{\rightarrow}1\underset{n}{\underbrace{2...2}}3
\end{align}

The configurations with $D$ and $E_{8}$:
\begin{align}
&  D\oplus E_{8}^{\oplus n}\,\overset{L}{\rightarrow}%
2\underset{n}{\underbrace{32...27}}\\
&  D\overset{3,5}{\oplus}E_{8}^{\oplus n}\overset{L}{\rightarrow
}1\underset{n}{\underbrace{22...27}}\\
&  D^{\oplus2}\oplus E_{8}^{\oplus n}\,\overset{L}{\rightarrow}%
31\underset{n}{\underbrace{32...27}}\\
&  D\oplus E_{8}^{\oplus n}\oplus D\overset{L}{\rightarrow}%
2\underset{n}{\underbrace{32...23}}2\\
&  D\overset{3,5}{\oplus}E_{8}^{\oplus n}\oplus D\overset{L}{\rightarrow
}1\underset{n}{\underbrace{22...23}}2\\
&  D^{\oplus2}\oplus E_{8}^{\oplus n}\oplus D\overset{L}{\rightarrow
}31\underset{n}{\underbrace{32...23}}2\\
&  D^{\oplus2}\oplus E_{8}^{\oplus n}\overset{5,3}{\oplus}%
D\overset{L}{\rightarrow}31\underset{n}{\underbrace{32...22}}1\\
&  D^{\oplus2}\oplus E_{8}^{\oplus n}\oplus D^{\oplus2}\overset{L}{\rightarrow
}31\underset{n}{\underbrace{32...23}}13
\end{align}

The configurations with $D$ and $E_{7}$:
\begin{align}
&  D\oplus E_{7}^{\oplus n}\,\overset{L}{\rightarrow}%
2\underset{n}{\underbrace{2...2}}\\
&  D\overset{3,3}{\oplus}E_{7}^{\oplus n}\overset{L}{\rightarrow
}1\underset{n}{\underbrace{2...2}}\\
&  D\oplus E_{7}^{\oplus n}\oplus D\overset{L}{\rightarrow}%
2\underset{n}{\underbrace{2...2}}2\\
&  D\overset{3,3}{\oplus}E_{7}^{\oplus n}\oplus D\overset{L}{\rightarrow
}1\underset{n}{\underbrace{2...2}}2
\end{align}

The configurations with $D$ and $E_{6}$:
\begin{align}
&  D\oplus E_{6}^{\oplus n}\,\overset{L}{\rightarrow}%
2\underset{n}{\underbrace{2...2}}\\
&  D\oplus E_{6}^{\oplus n}\oplus D\overset{L}{\rightarrow}%
2\underset{n}{\underbrace{2...2}}2
\end{align}

The configurations with $D$, $E_{7}$, and $E_{8}$:
\begin{align}
&  E_{7}\oplus E_{8}^{\oplus n}\oplus D\overset{L}{\rightarrow}%
4\underset{n}{\underbrace{2...23}}2\\
&  E_{7}\overset{5,5}{\oplus}E_{8}^{\oplus n}\oplus D\overset{L}{\rightarrow
}3\underset{n}{\underbrace{2...23}}2\\
&  E_{7}\oplus E_{8}^{\oplus n}\overset{5,3}{\oplus}D\overset{L}{\rightarrow
}4\underset{n}{\underbrace{2...2}}1\\
&  E_{7}\overset{5,5}{\oplus}E_{8}^{\oplus n}\overset{5,3}{\oplus
}D\overset{L}{\rightarrow}3\underset{n}{\underbrace{2...2}}1\\
&  E_{7}\oplus E_{8}^{\oplus n}\oplus D^{\oplus2}\overset{L}{\rightarrow
}4\underset{n}{\underbrace{2...23}}13\\
&  E_{7}\overset{5,5}{\oplus}E_{8}^{\oplus n}\oplus D^{\oplus2}%
\overset{L}{\rightarrow}3\underset{n}{\underbrace{2...23}}13
\end{align}

The configurations with $D$, $E_{6}$, and $E_{8}$:
\begin{align}
&  E_{6}\oplus E_{8}^{\oplus n}\oplus D\,\overset{L}{\rightarrow
}3\underset{n}{\underbrace{2...23}}2\\
&  E_{6}\overset{4,5}{\oplus}E_{8}^{\oplus n}\oplus D\overset{L}{\rightarrow
}2\underset{n}{\underbrace{2...23}}2\\
&  E_{6}\overset{5,5}{\oplus}E_{8}^{\oplus n}\oplus D\overset{L}{\rightarrow
}1\underset{n}{\underbrace{2...23}}2\\
&  E_{6}\oplus E_{8}^{\oplus n}\overset{5,3}{\oplus}D\overset{L}{\rightarrow
}3\underset{n}{\underbrace{2...22}}1\\
&  E_{6}\overset{4,5}{\oplus}E_{8}^{n}\overset{5,3}{\oplus}%
D\overset{L}{\rightarrow}2\underset{n}{\underbrace{2...22}}1\\
&  E_{6}\oplus E_{8}^{\oplus n}\oplus D^{\oplus2}\overset{L}{\rightarrow
}3\underset{n}{\underbrace{2...23}}13\\
&  E_{6}\overset{4,5}{\oplus}E_{8}^{\oplus n}\oplus D^{\oplus2}%
\overset{L}{\rightarrow}2\underset{n}{\underbrace{2...23}}13\\
&  E_{6}\overset{5,5}{\oplus}E_{8}^{\oplus n}\oplus D^{\oplus2}%
\overset{L}{\rightarrow}1\underset{n}{\underbrace{2...23}}13
\end{align}

The configurations with $D$, $E_{6}$, and $E_{7}$:
\begin{align}
&  E_{6}\oplus E_{7}^{\oplus n}\oplus D\overset{L}{\rightarrow}%
3\underset{n}{\underbrace{2...22}}2\\
&  E_{6}\oplus E_{7}^{\oplus n}\overset{3,3}{\oplus}D\,\overset{L}{\rightarrow
}3\underset{n}{\underbrace{2...22}}1
\end{align}

\section{The Short Bases \label{outlierappendix}}

In this Appendix we list all short chains bases. These are
all sequences of nodes in the base which cannot be continued to arbitrarily
long size. As in our Appendix on long bases, we do not consider the addition
of side links to the left and right of the quiver or small instantons arising
on an $E_{7}$ or $E_{8}$ curve. Again, here we list the configuration of
nodes, as well as the resulting configuration of curves after blowing down the
links between nodes. In this case, there can still be repeating patterns,
albeit ones which cannot be continued to an arbitrary number of nodes.
Nevertheless, when we list the blowdowns of links, we shall state the
\textquotedblleft generic\textquotedblright\ result, as the other case can
also be readily extracted from these general considerations. To indicate that
we are dealing with the generic situation, we shall often write an underbrace,
but may sometimes omit the number of curves included in the underbrace. For
the case of brevity, we omit the posisbility of primed nodes, which can be
found in the companion \texttt{Mathematica} files.

The configurations with only one of $E_{8}$, $E_{7}$, $E_{6}$, and $D$:
\begin{align}
E_{7}^{\oplus n}\overset{4,4}{\oplus}E_{7}\overset{L}{\rightarrow
}\underset{n}{\underbrace{52...1}}4,~~~~n  &  \leq4\\
E_{7}^{\oplus n}\overset{4,5}{\oplus}E_{7}\,\overset{L}{\rightarrow
}\underset{n}{\underbrace{52...1}}3,~~~~n  &  \leq3
\end{align}%
\begin{align}
&  E_{7}\overset{5,5}{\oplus}E_{7}\overset{L}{\rightarrow}33\\
&  E_{6}\overset{3,3}{\oplus}E_{6}\overset{L}{\rightarrow}33\\
&  E_{6}\overset{3,4}{\oplus}E_{6}\overset{L}{\rightarrow}32\\
&  E_{6}\overset{4,4}{\oplus}E_{6}\overset{L}{\rightarrow}22\\
&  E_{6}\overset{3,5}{\oplus}E_{6}\overset{L}{\rightarrow}31\\
&  E_{6}\overset{4,5}{\oplus}E_{6}\overset{L}{\rightarrow}21
\end{align}%
\begin{align}
&  E_{6}\overset{3,3}{\bigcirc}E_{6}\overset{L}{\rightarrow}33\\
&  E_{6}\oplus E_{6}\overset{3,3}{\oplus}E_{6}\overset{L}{\rightarrow}413\\
&  E_{6}\oplus E_{6}\overset{3,4}{\oplus}E_{6}\overset{L}{\rightarrow}412\\
&  E_{6}\oplus E_{6}\overset{3,3}{\bigcirc}E_{6}\overset{L}{\rightarrow}413
\end{align}%
\begin{align}
&  E_{6}\oplus E_{6}\oplus E_{6}\overset{3,3}{\oplus}E_{6}%
\overset{L}{\rightarrow}4213\\
&  E_{6}\oplus E_{6}\oplus E_{6}\overset{3,3}{\bigcirc}E_{6}%
\overset{L}{\rightarrow}4213
\end{align}%
\begin{equation}
D\overset{2,2}{\oplus}D^{\oplus n}\,\overset{L}{\rightarrow}2\underset{n}{\underbrace{13}}%
,~~~~n\leq2
\end{equation}%
\begin{equation}
D\overset{2,3}{\oplus}D\overset{L}{\rightarrow}21
\end{equation}

The configurations with $E_{7}$ and $E_{8}$:
\begin{align}
E_{7}^{\oplus2}\oplus E_{8}^{\oplus n}\,\overset{L}{\rightarrow}%
51\underset{n}{\underbrace{2...2}},~~~~n  &  \leq4\\
E_{7}^{\oplus n}\oplus E_{8}\overset{L}{\rightarrow}%
\underset{n}{\underbrace{52...21}}7,~~~~n  &  \leq7\\
E_{7}^{\oplus2}\oplus E_{8}^{\oplus n}\oplus E_{7}\,\overset{L}{\rightarrow
}51\underset{n}{\underbrace{2...2}}4,~~~n  &  \leq3\\
E_{7}^{\oplus2}\oplus E_{8}^{\oplus n}\overset{5,5}{\oplus}E_{7}%
\overset{L}{\rightarrow}51\underset{n}{\underbrace{2...2}}3\,,~~~n  &  \leq3
\end{align}

The configurations with $E_{6}$ and $E_{8}$:
\begin{align}
E_{6}^{\oplus2}\oplus E_{8}^{\oplus n}\overset{L}{\rightarrow}%
41\underset{n}{\underbrace{2...27}},~~~~n  &  \leq3\\
E_{6}^{\oplus n}\oplus E_{8}\overset{L}{\rightarrow}\underset{n}{\underbrace{42...21}}%
7\,,~~~~n  &  \leq7\\
E_{6}^{\oplus2}\oplus E_{8}^{\oplus n}\oplus E_{6}\overset{L}{\rightarrow
}41\underset{n}{\underbrace{22}3},~~~~n  &  \leq2
\end{align}%
\begin{equation}
E_{6}^{\oplus2}\oplus E_{8}\overset{5,4}{\oplus}E_{6}\overset{L}{\rightarrow
}4122
\end{equation}

The configurations with $E_{6}$ and $E_{7}$:
\begin{align}
E_{6}\overset{3,4}{\oplus}E_{7}^{\oplus n}\overset{L}{\rightarrow
}3\underset{n}{\underbrace{125}},~~~~n  &  \leq3\\
E_{6}\oplus E_{7}^{\oplus n}\overset{4,4}{\oplus}E_{7}\overset{L}{\rightarrow
}3\underset{n}{\underbrace{221}4},~~~~n  &  \leq3\\
E_{6}\oplus E_{7}^{\oplus n}\overset{4,5}{\oplus}E_{7}\overset{L}{\rightarrow
}3\underset{n}{\underbrace{21}3},~~~~n  &  \leq2
\end{align}%
\begin{align}
&  E_{6}\overset{3,5}{\oplus}E_{7}\overset{L}{\rightarrow}33\\
&  E_{6}\overset{4,5}{\oplus}E_{7}\overset{L}{\rightarrow}23\\
&  E_{6}\overset{5,4}{\oplus}E_{7}\overset{L}{\rightarrow}14\\
&  E_{6}\overset{5,5}{\oplus}E_{7}\overset{L}{\rightarrow}13
\end{align}%
\begin{align}
E_{6}^{\oplus2}\oplus E_{7}^{\oplus n}\overset{L}{\rightarrow}%
41\underset{n}{\underbrace{225}}\,,~~~~n  &  \leq3\\
E_{6}^{\oplus n}\oplus E_{7}\overset{L}{\rightarrow}%
\underset{n}{\underbrace{42..21}}5\,,~~~~n  &  \leq5\\
E_{6}^{\oplus n}\overset{3,4}{\oplus}E_{7}\overset{L}{\rightarrow
}\underset{n}{\underbrace{3221}}4,~~~~n  &  \leq4\\
E_{6}^{\oplus n}\overset{3,5}{\oplus}E_{7}\overset{L}{\rightarrow
}\underset{n}{\underbrace{321}}3,~~~~n  &  \leq3\\
E_{6}\overset{3,4}{\oplus}E_{7}^{\oplus n}\oplus E_{6}%
\,\overset{L}{\rightarrow}3\underset{n}{\underbrace{12}}3,~~~~n  &  \leq2
\end{align}%
\begin{equation}
E_{6}\overset{4,4}{\oplus}E_{7}\oplus E_{6}\overset{L}{\rightarrow}213
\end{equation}%
\begin{equation}
E_{6}^{\oplus2}\oplus E_{7}^{\oplus n}\oplus E_{6}\overset{L}{\rightarrow
}41\underset{n}{\underbrace{22}}4\,,~~~~n\leq2
\end{equation}

The configurations with $E_{6}$, $E_{7}$, and $E_{8}$:
\begin{align}
E_{6}^{\oplus2}\oplus E_{8}^{\oplus n}\oplus E_{7}\,\overset{L}{\rightarrow
}41\underset{n}{\underbrace{22}}4,~~~~n  &  \leq2\\
E_{6}^{\oplus2}\oplus E_{8}^{\oplus n}\overset{5,5}{\oplus}E_{7}%
\overset{L}{\rightarrow}41\underset{n}{\underbrace{22}}3,~~~~n  &  \leq2\\
E_{6}\oplus E_{8}^{\oplus n}\oplus E_{7}^{\oplus2}\overset{L}{\rightarrow
}3\underset{n}{\underbrace{222}}15\,,~~~~n  &  \leq3\\
E_{6}\overset{4,5}{\oplus}E_{8}^{\oplus n}\oplus E_{7}^{\oplus2}%
\overset{L}{\rightarrow}2\underset{n}{\underbrace{22}}15\,,~~~~n  &  \leq2\\
E_{6}\oplus E_{7}\oplus E_{8}^{\oplus n}\overset{L}{\rightarrow}%
31\underset{n}{\underbrace{27}},~~~~n  &  \leq2\\
E_{6}\oplus E_{7}^{\oplus n}\oplus E_{8}\overset{L}{\rightarrow}%
3\underset{n}{\underbrace{2...21}}7\,,~~~~n  &  \leq6
\end{align}%
\begin{align}
&  E_{6}\oplus E_{7}\oplus E_{8}\oplus E_{7}\overset{L}{\rightarrow}3124\\
&  E_{6}\oplus E_{7}\oplus E_{8}\overset{5,5}{\oplus}E_{7}%
\overset{L}{\rightarrow}3123\\
&  E_{6}\oplus E_{7}\oplus E_{8}\oplus E_{6}\overset{L}{\rightarrow}3123
\end{align}

The configurations with $D$ and $E_{8}$:
\begin{align}
D^{\oplus3}\oplus E_{8}^{\oplus n}\overset{L}{\rightarrow}321\underset{n}{\underbrace{37}}%
\,,~~~~n  &  \leq2\\
D^{\oplus n}\oplus E_{8}\overset{L}{\rightarrow}%
\underset{n}{\underbrace{32...21}}8,~~~~n  &  \leq8\\
D^{\oplus3}\oplus E_{8}^{\oplus n}\oplus D\overset{L}{\rightarrow
}321\underset{n}{\underbrace{33}}1\,,~~~~n  &  \leq2
\end{align}

The configurations with $D$ and $E_{7}$:
\begin{align}
&  D\overset{3,4}{\oplus}E_{7}\overset{L}{\rightarrow}14\\
&  D\overset{3,5}{\oplus}E_{7}\overset{L}{\rightarrow}13
\end{align}%
\begin{align}
D\overset{2,4}{\oplus}E_{7}^{\oplus n}\overset{L}{\rightarrow}2\underset{n}{\underbrace{14}}%
,~~~~n  &  \leq2\\
D\oplus E_{7}^{\oplus n}\overset{4,4}{\oplus}E_{7}\overset{L}{\rightarrow
}2\underset{n}{\underbrace{21}}4,~~~~n  &  \leq2
\end{align}%
\begin{equation}
\end{equation}%
\begin{equation}
D\oplus E_{7}\overset{4,5}{\oplus}E_{7}\rightarrow213
\end{equation}%
\begin{align}
D^{\oplus2}\oplus E_{7}^{\oplus n}\overset{L}{\rightarrow}31\underset{n}{\underbrace{25}}%
\,,~~~~n  &  \leq2\\
D^{\oplus n}\oplus E_{7}\overset{L}{\rightarrow}%
\underset{n}{\underbrace{32..21}\,}5,~~~~n  &  \leq5
\end{align}

The configurations with $D$ and $E_{6}$:
\begin{equation}
D\overset{2,3}{\oplus}E_{6}^{\oplus n}\,\overset{L}{\rightarrow}%
2\underset{n}{\underbrace{13}},~~~~n\leq2
\end{equation}%
\begin{align}
&  D\overset{3,3}{\oplus}E_{6}\overset{L}{\rightarrow}13\\
&  D\overset{3,3}{\bigcirc}E_{6}\overset{L}{\rightarrow}13\\
&  D\overset{2,4}{\oplus}E_{6}\overset{L}{\rightarrow}22\\
&  D\overset{3,4}{\oplus}E_{6}\overset{L}{\rightarrow}12\\
&  D\oplus E_{6}\overset{3,3}{\oplus}E_{6}\overset{L}{\rightarrow}213\\
&  D\oplus E_{6}\overset{3,3}{\bigcirc}E_{6}\overset{L}{\rightarrow}213
\end{align}%
\begin{equation}
D^{\oplus2}\oplus E_{6}^{\oplus n}\overset{L}{\rightarrow}31\underset{n}{\underbrace{24}}%
\,,~~~~n\leq2
\end{equation}%
\begin{align}
&  D\oplus D\overset{2,3}{\oplus}E_{6}\overset{L}{\rightarrow}313\\
&  D\oplus D\overset{2,4}{\oplus}E_{6}\overset{L}{\rightarrow}312
\end{align}%
\begin{equation}
D^{\oplus n}\oplus E_{6}\overset{L}{\rightarrow}\underset{n}{\underbrace{3221}}4,~~~~n\leq4
\end{equation}

The configurations with $D$, $E_{7}$, and $E_{8}$:
\begin{equation}
D\oplus E_{7}^{\oplus n}\oplus E_{8}\overset{L}{\rightarrow}%
2\underset{n}{\underbrace{2...21}}7\,,~~~~n\leq5
\end{equation}%
\begin{equation}
D\oplus E_{7}\oplus E_{8}\oplus D\overset{L}{\rightarrow}2132
\end{equation}%
\begin{equation}
E_{7}^{\oplus2}\oplus E_{8}^{\oplus n}\oplus D\overset{L}{\rightarrow
}51\underset{n}{\underbrace{2...23}}2,~~~~n\leq4
\end{equation}%
\begin{equation}
E_{7}^{\oplus3}\oplus E_{8}\oplus D\overset{L}{\rightarrow}52132
\end{equation}%
\begin{equation}
E_{7}^{\oplus2}\oplus E_{8}^{\oplus n}\oplus D^{\oplus2}%
\overset{L}{\rightarrow}51\underset{n}{\underbrace{23}}13,~~~~n\leq2
\end{equation}%
\begin{align}
&  E_{7}\oplus E_{8}\oplus D^{\oplus3}\overset{L}{\rightarrow}43123\\
&  E_{7}\overset{5,5}{\oplus}E_{8}\oplus D^{\oplus3}\overset{L}{\rightarrow
}33123
\end{align}

The configurations with $D$, $E_{6}$, and $E_{8}$:
\begin{equation}
D\oplus E_{6}^{\oplus n}\oplus E_{8}\overset{L}{\rightarrow}%
2\underset{n}{\underbrace{2...21}}7,~~~~n\leq5
\end{equation}%
\begin{equation}
D\oplus E_{6}\oplus E_{8}\oplus D\overset{L}{\rightarrow}2132
\end{equation}%
\begin{equation}
E_{6}^{\oplus n}\oplus E_{8}\oplus D\overset{L}{\rightarrow}\underset{n}{\underbrace{421}}%
32,~~~~n\leq3
\end{equation}%
\begin{align}
&  E_{6}\oplus E_{8}\oplus D^{\oplus3}\overset{L}{\rightarrow}33123\\
&  E_{6}^{\oplus2}\oplus E_{8}\oplus D^{\oplus2}\overset{L}{\rightarrow}41313
\end{align}

The configurations with $D$, $E_{6}$, and $E_{7}$:
\begin{align}
D\oplus E_{6}^{\oplus n}\oplus E_{7}\overset{L}{\rightarrow}2\underset{n}{\underbrace{221}}%
5\,,~~~~n  &  \leq3\\
D\oplus E_{6}^{\oplus n}\overset{3,4}{\oplus}E_{7}\overset{L}{\rightarrow
}2\underset{n}{\underbrace{21}}4,~~~~n  &  \leq2
\end{align}%
\begin{align}
&  E_{6}\overset{3,4}{\oplus}E_{7}\oplus D\overset{L}{\rightarrow}312\\
&  E_{6}^{\oplus2}\oplus E_{7}\oplus D\overset{L}{\rightarrow}4122\\
&  E_{6}\oplus E_{7}\oplus D^{\oplus2}\overset{L}{\rightarrow}3213
\end{align}

And finally, the configurations with $D$, $E_{6}$, $E_{7}$, and $E_{8}$:
\begin{equation}
D\oplus E_{8}^{\oplus n}\oplus E_{7}\oplus E_{6}\overset{L}{\rightarrow
}2\underset{n}{\underbrace{32}}13,~~~~n\leq2
\end{equation}

\newpage

\section{Classification of Links \label{linkappendix}}

The following tables give a list of the conformal matter links which are linearly shaped. The ``interior" links are those that can be placed in the interior of a quiver diagram, used as a side link, or stand alone as a base by themselves.  The ``alkali links" can only be used as side links or in isolation, appearing either on one side of the linear quiver or standing alone rather than stretching between two DE nodes.  By convention, we list the alkali links that appear on the left side of one of the quiver diagrams (the links that can be placed on the right side are simply the reverse of these links).  Finally, the links which cannot attach to any DE nodes are referred to as ``noble molecules."  These are similar to a noble gas in that they are ``inert" and cannot touch anything else.

The first column of the tables lists the links.  The second column gives the resulting links upon blowing down all $-1$ curves, and the third column gives (minus) the number of blowdowns that this inflicts upon the adjacent matter.  The fourth column indicates which nodes are permitted to lie at the sides of this link.  The tables are listed in order of the number of $-5$ curves that appear in the link.
\begin{longtable}{ |c|c|c|c|}
\hline
\multicolumn{4}{ |c| }{Interior Links with no $-5$ Curves} \\
\endfirsthead
\hline
Link & After Blowdown  & Blowdowns Induced& Adjacent to\\ \hline
\endhead \hline
Link & After Blowdown  & Blowdowns Induced& Adjacent to\\ \hline
1 &   $\{ \}$& (-1,-1) & $(D,D)$ \\ \hline
131 &  $\{ \}$ & (-2,-2) & $(D,D), (D,E_6), (E_6,E_6)$ \\ \hline
1231 &   $\{ \}$& (-3,-2) & $(D,D),(E_6,D),(E_7,D)$ \\ \hline
12321 &   $\{ \}$& (-3,-3) & $(D,E_6),(D,E_7),(E_6,E_6),$\\
 & & & $(E_6,E_7),(E_7,E_7)$ \\ \hline
12231 &   $\{ \}$& (-4,-2) &  $(E_6,D),(E_7,D),(E_8,D)$\\ \hline

\multicolumn{4}{ |c| }{Alkali 2-Links with no $-5$ Curves} \\
\hline
Link & After Blowdown  & Blowdowns Induced& Adjacent to\\ \hline
$2\ov{1}31$ & $\{ \}$  & -3 & $D$ \\ \hline
$12\ov{2}31$ &  $\{ \}$ & -3 & $D$ \\ \hline
$2\ov{1}321$ &  $\{ \}$ & -4 & $E_6,E_7$ \\ \hline

\multicolumn{4}{ |c| }{Alkali 1-Links with no $-5$ Curves} \\
\hline
Link & After Blowdown  & Blowdowns Induced& Adjacent to\\ \hline
$3\ov{2}21$ & $\{ \}$ & -4 &  $E_6,E_7$ \\ \hline
$2\ov{2}31$ & $2\ov{2}2$ & -1 &  $D$ \\ \hline
3221 & 2 & -3 & $D,E_6,E_7,E_8$ \\ \hline
2313221 & $\{ \}$  & -6 & $E_7, E_8$\\ \hline
22313221 &  $\{ \}$ & -7 & $E_7, E_8$\\ \hline
313221 &  $\{ \}$ & -5 & $E_6,E_7, E_8$\\ \hline
321 & 2 & -2& $D,E_6,E_7$ \\ \hline
2321 & 22 & -2& $D,E_6,E_7$ \\ \hline
231321 &  $\{ \}$ & -5& $D,E_6,E_7$ \\ \hline
2231321 &  $\{ \}$ & -6& $E_7$ \\ \hline
31321 &   $\{ \}$& -4& $E_6,E_7$ \\ \hline
31 & 2 & -1 & $D,E_6$ \\ \hline
23131 &   $\{ \}$& -4 & $E_6$ \\ \hline
223131 &  $\{ \}$ & -5 & $E_6$ \\ \hline
3131 &  $\{ \}$ & -3 & $D,E_6$ \\ \hline
231 & 22 & -1 & $D$ \\ \hline
2231 & 222 & -1 & $D$ \\ \hline
2...21 & $\{ \}$ &  $-n_2+1, n_2=1,...,10$ &  $E_8$ \\ \hline
2...21 &  $\{ \}$ & $-n_2+1, n_2=1,...,6$ &  $E_7$ \\ \hline
2...21 & $\{ \}$ & $-n_2+1, n_2=1,...,4$ &  $E_6$ \\ \hline
2...21 &  $\{ \}$& $-n_2+1, n_2=1,2$ &  $D$ \\ \hline

\multicolumn{4}{ |c| }{Noble 2-Molecules with no $-5$ Curves} \\
\hline
Link & After Blowdown  & Blowdowns Induced& Adjacent to\\ \hline
$1\ov{1}322$ &  $\{ \}$&  &  \\ \hline

\multicolumn{4}{ |c| }{Noble 1-Molecules with no $-5$ Curves} \\
\hline
2...21,$n_2 > 10$  & & &  \\ \hline

\multicolumn{4}{ |c| }{Noble 0-Molecules with no $-5$ Curves} \\
\hline
Link & After Blowdown  & Blowdowns Induced& Adjacent to\\ \hline
$2\ov{2}313$ &  $2\ov{2}22$ & &  \\ \hline
$2\ov{2}3132$ & $2\ov{2}222$ & &  \\ \hline
$2\ov{2}31322$ & $2\ov{2}2222$& &  \\ \hline
$2\ov{2}22...$ &  $2\ov{2}22...$ & &  \\ \hline
$22\ov{2}222$ & $22\ov{2}222$ & & \\ \hline
$22\ov{2}2222$ & $22\ov{2}2222$ & & \\ \hline
$22\ov{2}22222$ & $22\ov{2}22222$ & &  \\ \hline

23132 & 2222 & & \\ \hline

223132 & 22222 & & \\ \hline

3132 & 222 & & \\ \hline

2132 & $\{ \}$ & & \\ \hline

3123 & $\{ \}$ & & \\ \hline

2231322 & 222222 & & \\ \hline

31322 & 2222 & & \\ \hline

21322 & $\{ \}$ & & \\ \hline

313 & 22 & & \\ \hline

23213 & $\{ \}$ & & \\ \hline

213 & $\{ \}$ & & \\ \hline

2...2 & 2...2 &  &  \\ \hline

\end{longtable}

\begin{longtable}{ |c|c|c|c|}
\hline
\multicolumn{4}{ |c| }{Interior Links with one $-5$ Curve} \\
\endfirsthead
\hline
Link & After Blowdown  & Blowdowns Induced& Adjacent to\\ \hline
\endhead \hline
Link & After Blowdown  & Blowdowns Induced& Adjacent to\\ \hline
1315131 &  $\{ \}$ & (-3,-3) & $(D,E_6),(E_6,E_6)$  \\ \hline
12315131 & $\{ \}$  & (-4,-3) & $(E_6,D),(E_6,E_6),(E_7,D),(E_7,E_6)$  \\ \hline
123151321 &   $\{ \}$& (-4,-4) &  $(E_6,E_6),(E_7,E_6),(E_7,E_7)$  \\ \hline
122315131 & $\{ \}$  & (-5,-3) & $(E_6,E_6),(E_7,E_6),(E_8,E_6)$ \\ \hline
1223151321 &  $\{ \}$ & (-5,-4) & $(E_6,E_6),(E_7,E_6),(E_6,E_7),(E_7,E_7)$ \\
& & & $(E_8,E_6),(E_7,E_7),(E_8,E_7)$  \\ \hline
12231513221 & $\{ \}$  & (-5,-5) & $(E_6,E_6),(E_6,E_7),(E_7,E_7),(E_6,E_8)$ \\
& & & $(E_7,E_7),(E_7,E_8),(E_8,E_8)$  \\ \hline

\multicolumn{4}{ |c| }{Alkali 3-Links with one $-5$ Curve} \\
\hline
Link & After Blowdown  & Blowdowns Induced& Adjacent to\\ \hline
$1\ov{1}5131$ & $\{ \}$  & -3& $D,E_6$\\ \hline
$1\ov{1}51321$ &  $\{ \}$ & -4& $E_6,E_7$\\ \hline
$1\ov{1}513221$ &   $\{ \}$& -5& $E_6,E_7,E_8$\\ \hline

\multicolumn{4}{ |c| }{Alkali 2-Links with one $-5$ Curve} \\
\hline
Link & After Blowdown  & Blowdowns Induced& Adjacent to\\ \hline
$31\ov{1}5131$ &  $\{ \}$& -4& $E_6$\\ \hline
$31\ov{1}51321$ & $\{ \}$ & -5& $E_6,E_7$\\ \hline
$31\ov{1}513221$ & $\{ \}$ & -6& $E_7,E_8$\\ \hline
$231\ov{1}51321 $& $\{ \}$ & -6& $E_7$\\ \hline
$231\ov{1}513221 $& $\{ \}$ & -7& $E_7,E_8$\\ \hline
$2231\ov{1}513221 $& $\{ \}$ & -8& $E_{8}$\\ \hline
$231\ov{1}5131 $&$\{ \}$  & -5& $E_6$\\ \hline
$2231\ov{1}51321$ & $\{ \}$ & -7& $E_7$\\ \hline
1513221 & 2 & -4& $E_6,E_7,E_8 $\\ \hline
151321 &  2 & -3 & $SO,E_6,E_7$ \\ \hline
1512321 &  $\{ \}$ & -4 & $E_6,E_7$ \\ \hline

\multicolumn{4}{ |c| }{Alkali 1-Links with one $-5$ Curve} \\
\hline
Link & After Blowdown  & Blowdowns Induced& Adjacent to\\ \hline
513221 & 3 & -4& $E_6,E_7,E_8 $\\ \hline
321513221 &$\{ \}$ & -6& $E_7,E_8$ \\ \hline
231513221 & 222 & -4 & $E_6,E_7,E_8$\\ \hline
2231513221 & 2222 & -4& $E_6,E_7,E_8 $\\ \hline
31513221 &  22 & -4 & $E_6,E_7,E_8$ \\ \hline
2321513221 & $\{ \}$  & -7 & $E_7,E_8 $\\ \hline
51321 & 3  & -3 & $D,E_6,E_7$ \\ \hline
512321 &  2 & -3 & $D,E_6,E_7$ \\ \hline
32151321 &  $\{ \}$ & -5 & $E_6,E_7 $\\ \hline
23151321 &  222 & -3 & $D,E_6,E_7$ \\ \hline
223151321 &   2222 & -3 & $D,E_6,E_7$ \\ \hline
3151321 & 22  & -3 & $D,E_6,E_7$ \\ \hline
232151321 & $\{ \}$ & -6 & $E_7 $\\ \hline
231512321 & $\{ \}$ & -6 & $E_7$ \\ \hline
2231512321 & $\{ \}$ & -7 & $E_7$ \\ \hline
31512321 & $\{ \}$ & -5 & $E_6,E_7$ \\ \hline
5131 & 3 & -2& $D,E_6 $\\ \hline
3215131 &  $\{ \}$& -4 & $E_6 $\\ \hline
2315131 & 222 & -2 & $D,E_6 $\\ \hline
22315131 & 2222 & -2 & $D,E_6 $\\ \hline
315131 & 22 & -2 & $D,E_6 $\\ \hline
23215131 & $\{ \}$ & -5 & $E_6 $\\ \hline
51231 & 2 & -2 & $D $\\ \hline
151231 & $\{ \}$ & -3 & $D $\\ \hline
512231 & $\{ \}$ & -3 & $D $\\ \hline

215131 & $\{ \}$& -3 &$ D,E_6 $\\ \hline
2151321 & $\{ \}$& -4 & $E_6, E_7$ \\ \hline
21513221 & $\{ \}$& -5 & $E_6,E_7,E_8$ \\ \hline

\multicolumn{4}{ |c| }{Noble 4-Molecules with one $-5$ Curve} \\
\hline
$1\ov{1}{\underset{1}5}1$ & $\{ \}$ & & \\ \hline

\multicolumn{4}{ |c| }{Noble 3-Molecules with one $-5$ Curve} \\
\hline
$1\ov{1}{\underset{1}5}13$ & $\{ \}$ & & \\ \hline
$1\ov{1}{\underset{1}5}132$ & $\{ \}$ & & \\ \hline
$1\ov{1}{5}1$ & 2 & & \\ \hline

\multicolumn{4}{ |c| }{Noble 2-Molecules with one $-5$ Curve} \\
\hline
$151\ov{1}32$ & $\{ \}$ & & \\ \hline
$1\ov{1}512$ & $\{ \}$ & & \\ \hline
$31\ov{1}51$ &22  & & \\ \hline
$231\ov{1}51$ &222  & & \\ \hline
$2231\ov{1}51$ &2222  & & \\ \hline
$2231\ov{1}5131$ & $\{ \}$ & & \\ \hline

\multicolumn{4}{ |c| }{Noble 1-Molecules with one $-5$ Curve} \\
\hline
Link & After Blowdown  & Blowdowns Induced& Adjacent to\\ \hline
$51\ov{1}32$ & 2 & & \\ \hline
$3151\ov{1}32$ & $\{ \}$ & & \\ \hline
$23151\ov{1}32$ & $\{ \}$ & & \\ \hline
$223151\ov{1}32$ & $\{ \}$ & & \\ \hline
$51\ov{1}322$ & $\{ \}$ & & \\ \hline
$512\ov{1}32$ & $\{ \}$ & & \\ \hline
$31\ov{1}512$ & $\{ \}$ & & \\ \hline
$231\ov{1}512$ & $\{ \}$ & & \\ \hline
$2231\ov{1}512$ & $\{ \}$ & & \\ \hline
$31\ov{1}513$ &  222 & & \\ \hline
$231\ov{1}513$ &2222  & & \\ \hline
$2231\ov{1}513$ &22222  & & \\ \hline
$231\ov{1}5132$ &22222  & & \\ \hline
$2231\ov{1}5132$ &222222  & & \\ \hline
$2231\ov{1}51322$ &2222222  & & \\ \hline

13215132 & $\{ \}$ & & \\ \hline

223151231 & $\{ \}$ & & \\ \hline

3151231 & $\{ \}$ & & \\ \hline

\multicolumn{4}{ |c| }{Noble 0-Molecules with one $-5$ Curve} \\
\hline
Link & After Blowdown  & Blowdowns Induced& Adjacent to\\ \hline

$31\ov{\ov{3}1}513$ &$2\ov{2}22$  & & \\ \hline
$231\ov{\ov{3}1}513$ &$22\ov{2}22$  & & \\ \hline
$2231\ov{\ov{3}1}513$ &$222\ov{2}22$  & & \\ \hline
$231\ov{\ov{3}1}5132$ &$22\ov{2}222$  & & \\ \hline
$2231\ov{\ov{3}1}5132$ &$222\ov{2}222$  & & \\ \hline

3215 & 23 & & \\ \hline

2315 & 224 & & \\ \hline

32215 & 22 & & \\ \hline

22315 & 2224 & & \\ \hline

315 & 24 & & \\ \hline

23215 & 223 & & \\ \hline

215 & 3 & & \\ \hline

2215 & 2 & & \\ \hline

22215 & $\{ \}$ & & \\ \hline

313215 & $\{ \}$ & & \\ \hline

231315 & $\{ \}$ & & \\ \hline

31315 & 2 & & \\ \hline

3215132 & 2222 & & \\ \hline

2315132 & 22322 & & \\ \hline

22315132 & 222322 & & \\ \hline

315132 & 2322 & & \\ \hline

23215132 & 22222 & & \\ \hline

215132 & 222 & & \\ \hline

2215132 & $\{ \}$ & & \\ \hline

22315123 & 22222 & & \\ \hline

315123 & 222 & & \\ \hline

215123 & $\{ \}$ & & \\ \hline

223151322 & 2223222 & & \\ \hline

3151322 & 23222 & & \\ \hline

232151322 & 222222 & & \\ \hline

2151322 & 2222 & & \\ \hline

22151322 & $\{ \}$ & & \\ \hline

31513 & 232 & & \\ \hline

2321513 & 2222 & & \\ \hline

21513 & 22 & & \\ \hline

221513 & $\{ \}$ & & \\ \hline

2151232 & $\{ \}$ & & \\ \hline

21512 & $\{ \}$ & & \\ \hline

31315132 & $\{ \}$ & & \\ \hline

313151322 & $\{ \}$ & & \\ \hline

3131513 & $\{ \}$ & & \\ \hline

\end{longtable}

\begin{longtable}{ |c|c|c|c|}
\hline
\multicolumn{4}{ |c| }{Interior Links with two $-5$ Curves} \\
\hline
\multicolumn{4}{ |c| }{Alkali 2-Links with two $-5$ Curves} \\
\endfirsthead
\hline
Link & After Blowdown  & Blowdowns Induced& Adjacent to\\ \hline
\endhead \hline
Link & After Blowdown  & Blowdowns Induced& Adjacent to\\ \hline
15131513221 &$\{ \}$   & -6 & $E_7,E_8$ \\ \hline
1513151321 &  $\{ \}$& -5 & $E_6,E_7$ \\ \hline
151315131 & $\{ \}$ & -4 & $E_6$ \\ \hline

\multicolumn{4}{ |c| }{Alkali 1-Links with two $-5$ Curves} \\
\hline
Link & After Blowdown  & Blowdowns Induced& Adjacent to\\ \hline
51231513221 & $\{ \}$  & -6 & $E_7,E_8$ \\ \hline
5131513221 & 2  & -5 & $E_6,E_7,E_8$ \\ \hline
2315131513221 & $\{ \}$  & -8 & $E_8$ \\ \hline
22315131513221 &  $\{ \}$ & -9 & $E_8$ \\ \hline
315131513221 &$\{ \}$   & -7 & $E_7,E_8$ \\ \hline
5123151321 & $\{ \}$ & -5 & $E_6,E_7 $\\ \hline
513151321 &  2& -4 & $E_6,E_7$ \\ \hline
231513151321 &$\{ \}$   & -7 & $E_7$ \\ \hline
31513151321 &$\{ \}$   & -6 & $E_7$ \\ \hline
512315131 & $\{ \}$ & -4 & $E_6$ \\ \hline
51315131 & 2 & -3 & $D,E_6$ \\ \hline
3151315131 &  $\{ \}$& -5 & $E_6$ \\ \hline

\multicolumn{4}{ |c| }{Noble 1-Molecules with two $-5$ Curves} \\
\hline
Link & After Blowdown  & Blowdowns Induced& Adjacent to\\ \hline

2231513151321 & $\{ \}$ & & \\ \hline

13151315132 & $\{ \}$ & & \\ \hline

131513151322 & $\{ \}$ & & \\ \hline

\multicolumn{4}{ |c| }{Noble 0-Molecules with two $-5$ Curves} \\
\hline
Link & After Blowdown  & Blowdowns Induced& Adjacent to\\ \hline

513215 & 32 & & \\ \hline

5132215 & 2 & & \\ \hline

51315 & 33 & & \\ \hline

5123215 & 22 & & \\ \hline

231513215 & 2222 & & \\ \hline

2231513215 & 22222 & & \\ \hline

31513215 & 222 & & \\ \hline

21513215 & $\{ \}$ & & \\ \hline

31512315 & $\{ \}$ & & \\ \hline

2315132215 & $\{ \}$ & & \\ \hline

22315132215 & $\{ \}$ & & \\ \hline

315132215 & $\{ \}$ & & \\ \hline

32151315 & $\{ \}$ & & \\ \hline

23151315 & 2223 & & \\ \hline

223151315 & 22223 & & \\ \hline

3151315 & 223 & & \\ \hline

2151315 & 2 & & \\ \hline

23151315132 & 222222 & & \\ \hline

223151315132 & 2222222 & & \\ \hline

3151315132 & 22222 & & \\ \hline

2151315132 & $\{ \}$ & & \\ \hline

2231513151322 & 22222222 & & \\ \hline

31513151322 & 222222 & & \\ \hline

21513151322 & $\{ \}$ & & \\ \hline

315131513 & 2222 & & \\ \hline

215131513 & $\{ \}$ & & \\ \hline

\end{longtable}

\begin{longtable}{ |c|c|c|c|}
\hline
\multicolumn{4}{ |c| }{Interior Links with three $-5$ Curves} \\
\hline

\hline
\multicolumn{4}{ |c| }{Alkali Links with three $-5$ Curves} \\

\hline
\multicolumn{4}{ |c| }{Noble 0-Molecules with three $-5$ Curves} \\
\endfirsthead
\hline
Link & After Blowdown  & Blowdowns Induced& Adjacent to\\ \hline
\endhead \hline
Link & After Blowdown  & Blowdowns Induced& Adjacent to\\ \hline

5131513215 & $\{ \}$ & & \\ \hline

513151315 & 22 & & \\ \hline

\end{longtable}

\newpage

\section{Some F-theory Considerations} \label{appendixF}

\subsection{Localized vs. Non-Local Matter} \label{subs:nonlocal}

In F-theory constructions, matter in the adjoint representation of the
gauge algebra is not localized, but is rather spread over the curve
in the base \cite{Witten:1996qb}.  There is one other circumstance when matter is not
localized:  in cases that there is monodromy on the Kodaira fiber,
there can be a second representation with non-localized matter.

This issue was considered in \cite{BershadskyPLUS,LieF,Grassi:2011hq}, and
various formulas specifying matter were clarified
and finalized in \cite{Grassi:2011hq}.  However, there is one subtlety which
\cite{Grassi:2011hq} overlooked:  there can be representations
which are partly composed of localized matter and partly composed
of non-local matter. We briefly explain how to supplement \cite{Grassi:2011hq}
in order to take this subtlety into account.

For gauge algebras $\mathfrak{g}$ with non-simply-laced Dynkin diagrams,
the F-theory realization is on a divisor $\Sigma$ on the base which
has a branched cover $\widetilde{\Sigma}\to\Sigma$ given by the monodromy
representation on the Kodaira fiber.  Correspondingly, there is a larger
algebra $\widetilde{\mathfrak{g}}$ with a symmetry $\tau$ such that
$\mathfrak{g}$ is the part of $\widetilde{\mathfrak{g}}$ invariant under
$\tau$.  The adjoint of $\widetilde{\mathfrak{g}}$ decomposes into
the adjoint of $\mathfrak{g}$ plus another representation $\rho$.

In this subsection, we will take $\widetilde{\Sigma}\to\Sigma$ to have
degree $2$.  (This excludes the case of $\mathfrak{g}_2$, which we
will treat in another subsection.)  Then the genus $\tilde{g}$ of
$\widetilde{\Sigma}$ satisfies the formula
\[ \tilde{g}-g = (g-1) + \frac12 b\]
where $g$ is the genus of $\Sigma$.
Since the nonlocalized matter consists of $(g-1)$ copies of the adjoint
plus $\tilde{g}-g$ copies of $\rho$, we can describe the matter as
consisting of $(g-1)$ copies of the adjoint of $\widetilde{\mathfrak{g}}$
together with $\frac12 b$ copies of $\rho$.

However, the branch points may also carry additional copies of $\rho$,
which should be viewed as localized at those points.  In addition,
there may be copies of $\rho$ localized at other points of $\Sigma$.
(Neither of those points was made clear in  \cite{Grassi:2011hq}.)
For Kodaira types $III$ and $I_{2n+1}$ with monodromy there is
additional matter localized at the branch points.
In the former case, Table 9 of \cite{Grassi:2011hq} indicates
that there should be three
half-fundamentals\footnote{These are half-fundamentals
because a branch point is a zero in the cycle $\beta_\Sigma$,
whereas the representations listed in Table 9 are associated to the
cycle $\frac12\beta_\Sigma$.} at each branch point,
but two of those are associated with
non-localized matter (as indicated in the $\rho_\alpha$ column of that
Table).  Thus, each branch point is associated to a localized
half-fundamental in addition to the non-localized matter.

In the case of $I_{2n+1}$, Table 9 of \cite{Grassi:2011hq} omits one piece of
information which was present in Table 2 of \cite{BershadskyPLUS}:
the residual discriminant vanishes to order (at least) $3$ rather than to order $1$
at each of the branch points.  In the language of \cite{Grassi:2011hq}, the
points of $\operatorname{div}(\beta_\Sigma)$ are all contained in the cycle
$\operatorname{div}(\gamma_\Sigma)$,
and the corresponding line of Table 9 could have been written

\begin{center}
\begin{tabular}{||c|c|c|c|c||} \hline
Type & Algebra & $\rho_\alpha$ & $\rho_{\sqrt{\beta}}$ & $\rho_{\gamma/\beta}$ \\ \hline
$I_{2n+1}$, $n\ge1$ & $\mathfrak{sp}(n)$ &
$\operatorname{adj} + \Lambda^2_{\text{irr}} + 2\cdot \operatorname{fund}$
& $\Lambda^2_{\text{irr}} + 3\cdot \operatorname{fund}$ &
$\operatorname{fund}$\\
\hline
\end{tabular}
\end{center}

\noindent
(note the change in the heading of the final column).
When expressed in these terms, it is clear that we have a localized
half-fundamental
associated to each branch point (in addition to non-localized matter)
in this case as well.  In this case, we also have {\em additional}\/
fundamentals in the spectrum, localized at points of
$\operatorname{div}(\gamma)-\operatorname{div}(\beta) = (6L-(2n+1)\Sigma)|_\Sigma$,
in the language of Table 8 of \cite{Grassi:2011hq}.

It is important to analyze the geometry carefully in examples rather
than blindly following the formulas.  For example, for Kodaira type
$IV$ with monodromy on a curve $\Sigma$ with $\Sigma^2=-2$, there
are generally $4$ branch points and one expects non-localized
matter since $\tilde{g}-g=1$.  However, there a special cases in
which three of the four branch coincide.  The branch divisor is
still odd, and there is still a double cover, but in this case
$\tilde{g}=g=0$.  The ``missing'' matter is now localized at the
point of high multiplicity.  (This happens in particular in the 223
cluster, in which the curve of Kodaira type $IV$ has one half-fundamentals
localized at one intersection point, and seven half-fundamentals localized
at the other point.)

\subsection{Unpaired tensors}

We wish to use F-theory to study the unpaired tensors corresponding
to curves $\Sigma$ of self-intersection $-2$.  Along any curve $\Sigma$
in the base,
there is a ``generic Kodaira type'' specifying how the fiber in
the elliptic fibration appears over the generic point of $\Sigma$.
If that Kodaira type is anything other than $I_0$, $I_1$, or $II$, there
is a gauge algebra associated to $\Sigma$; thus, we can restrict our
attention to those three cases.

Our first claim is that if $\Sigma$ meets another curve $C$ which itself
has an associated gauge algebra, then that gauge algebra must be $\mathfrak{su}(2)$.  This follows by considering how the residual discriminant of $\Sigma$
meets $\Sigma$.  Recall that if the discriminant vanishes to order $m$
along $\Sigma$, then $(-12K_B-m\Sigma)\cdot \Sigma = 2m$ since $\Sigma^2=-2$
and $K_B\cdot\Sigma=0$.  Since $0\le m\le 2$, the discriminant can have
multiplicity at most $4$ along the curve $C$.
Moreover, when $m\le 1$ the discrimimant can have multiplicity at most $2$
and in that case, the only option for a $C$ with a gauge symmetry is $I_2$
which has gauge algebra $\mathfrak{su}(2)$.
In that case, the intersection point contributes a fundamental representation
of $\mathfrak{su}(2)$ to the overall matter representation.

To settle the issue for $\Sigma$ of type $II$ we need a computation.
In the course of making the computation, we will also analyze the
intersection point whenever $C$ {\em does}\/ have an associated
 gauge symmetry.
\begin{enumerate}
\item $\Sigma$ could meet a curve $C$ of Kodaira type $I_m$, $m\le4$.  If $2\le m\le4$,
then (using \cite{Katz:2011qp}), the equation can be put into the form
\[ y^2 = x^3 + tux^2 + tz^kvx^2 + tz^{2k}w ,\]
when $m=2k$, and
\[ y^2 = x^3 + \frac14t\mu\sigma^2x^2 + (\frac12tz\mu\sigma\tau +tz^2\tilde{v})x + \frac14tz^2\mu\tau^2 + tz^3\tilde{w}\]
when $m=3$,
where $\Sigma=\{t=0\}$ and the gauge divisor $C$ is $\{z=0\}$.
In the first case,
the Weierstrass coefficients are
\[ f = -\frac13 t^2u^2 + tz^kv, \quad g = \frac2{27}t^3u^3-\frac13t^2z^kuv+tz^{2k}w,\]
and the discriminant is
\[ \Delta=4f^3 + 27g^2 = t^2z^{2k}\left(4t^2u^3w-t^2u^2v^2-18tz^kuvw+4tz^{k}v^3
+27z^{2k}w^2\right). \]
The first thing to observe is that along $\Sigma=\{t=0\}$, the intersection
number of the residual discriminant with $\Sigma$ is $4k$ due to the monomial
$z^{2k}z^{2k}$ multiplying $27w^2$.  Since that intersection number is
bounded by $4$, it is not possible to have $k=2$.  So we assume that $k=1$.

It also follows that $f|_{z=0}$ has a zero of order $2$ at $t=0$,
$g|_{z=0}$ has a zero of order $3$, and
$(\Delta/z^2)|_{z=0}$ has a zero of order $4$ at $t=0$.
Using Table 7 of \cite{Grassi:2011hq}, we see that
`$\beta_{z=0}$' has a zero of order $1$ at $t=0$ ,
while `$\gamma_{z=0}$'
has a zero of order $2$
(since `$\delta_{z=0}$' has a zero of order $4$).
It then follows from \cite{Grassi:2011hq} (particularly Table 9)
 that  the matter consists of two fundamentals
of $\mathfrak{su}(2)$.
Moreover, since the residual discriminant has intersection number $4$
with $\Sigma$, this is the only intersection point.

In the case $m=3$, the Weierstrass coefficients are
\begin{align*}
 f &= -\frac1{48}t^2\mu^2\sigma^4 + \frac12tz\mu\sigma\tau +tz^2\tilde{v}\\
 g &= \frac1{864} t^3\mu^3\sigma^6
-\frac1{24}t^2z\mu^2\sigma^3\tau -\frac1{12}t^2z^2\mu\sigma^2\tilde{v}
+\frac14tz^2\mu\tau^2 + tz^3\tilde{w}
\end{align*}
and the discrimimant is
\[ \Delta=\frac1{16}t^3z^3\mu^3\sigma^3(t\sigma^3w-t\sigma^2\tau v - \tau^3) + O(z^4)\]
(using formula (4.10) from \cite{Katz:2011qp}).
Again,
 $f|_{z=0}$ has a zero of order $2$ at $t=0$,
and $g|_{z=0}$ has a zero of order $3$, but this time
$(\Delta/z^3)|_{z=0}$ has a zero of order $3$ at $t=0$.
Using Table 7 of \cite{Grassi:2011hq}, we see that
`$\beta_{z=0}$' has a zero of order $1$ at $t=0$
 (which implies that the
gauge algebra is $\mathfrak{sp}(1)$ rather than $\mathfrak{su}(3)$),
while `$\gamma_{z=0}$'
has a zero of order $1$
(since `$\delta_{z=0}$' has a zero of order $3$).
The analysis in Section~\ref{subs:nonlocal} now shows that the localized matter
consists of a half-fundamental of $\mathfrak{sp}(1)$.

\item $\Sigma$ could meet a curve $C$ of Kodaira type $IV$.
The equation takes the form
\[ y^2 = x^3 + tz^2 \varphi x + tz^2 \gamma\]
with discriminant
\[ \Delta = t^2z^4(4tz^2 \varphi^3 + 27 \gamma^2).\]
From this, we can see that the residual discrimimant
$(\Delta/t^2)|_{t=0}$ has a zero of order $4$ at $t=0$, which implies
that this
is the unique intersection point with $\Sigma$.

The gauge algebra associated to $C$ is determined by $(g/z^2)|_{z=0}
= t\gamma|_{z=0}$ which has a zero of order $1$ at $t=0$.  (The order
cannot be higher because this is the unique intersection point.)
This implies that the gauge algebra is $\mathfrak{su}(2)$, since
the order is odd.  As shown in Section~\ref{subs:nonlocal},
there is a half-fundamental associated to this intersection point.

\item $\Sigma$ could meet a curve of Kodaira type $III$.
The equation takes the form
\[ y^2 = x^3 + tz \varphi x + tz^2 \gamma \]
with discriminant
\[ \Delta = t^2z^3(4t \varphi^3 + 27 z \gamma^2).\]
In this case, there is a third component of the discrimimant passing
through $z=t=0$, and the residual discriminant $(\Delta/t^2)|_{t=0}$
again has a zero of order $4$ at $z=0$, making this the unique
intersection point with $\Sigma$.

The matter is determined by $(f/z)|_{z=0}=t\varphi|_{z=0}$ which
has a zero of order $1$ at $t=0$.  This implies that there are
two $\mathfrak{su}(2)$ fundamentals in the matter representation
associated to this intersection point.

\end{enumerate}

We formulate our conclusions by counting the total number of hypermultiplets
transforming under the gauge symmetry, since these are the ones which
become free in $\mathcal{T}_\Sigma$.
The conclusion is that when $\Sigma$ has Kodaira type $II$, there are
either $1$ or $4$ hypermultiplets (corresponding to a half-fundamental
or two fundamental representations), whereas when $\Sigma$ has Kodaira
type $I_1$, there are $2$ hypermultiplets (corresponding to a single
fundamental representation).  The case of Kodaira type $I_0$ has no
gauge symmetry and no associated matter.

\subsection{Gauge algebras for Kodaira fiber type $I_0^*$}

The most delicate question to answer for an F-theory model involving
a divisor $\Sigma$ with
Kodaira type $I_0^*$ is: what is the gauge algebra associated to that
divisor?  The criterion is clear in terms of the Weierstrass equation
\[ y^2 = x^3 + t^2 \varphi x + t^3 \gamma :\]
one must ask whether the auxiliary cubic (in an auxiliary variable $\psi$)
\[ \psi^3 + \varphi|_{t=0} \psi + \gamma|_{t=0} \]
has one, two, or three irreducible factors (which correspond to
gauge algebra $\mathfrak{g}_2$, $\mathfrak{so}(7)$, and
$\mathfrak{so}(8)$, respectively).  However, determining whether
a given cubic factors or not is quite difficult, and algorithms are
not known.

The question can sometimes be answered by means of some necessary
conditions.  If the cubic factors completely, so that the gauge algebra
is $\mathfrak{so}(8)$, then the reduced discriminant must factor as
\begin{equation} (\Delta/t^6)|_{t=0}=A^2B^2C^2, \label{eq:soeight} \end{equation}
and in particular it must be a square.  In this case, the points in
$A=0$, $B=0$,
and $C=0$ represent matter in the vector and the two spinor representations.
However, the factorization of the reduced discriminant
\eqref{eq:soeight} is not sufficient for the auxiliary cubic to factor.

Similarly, if the cubic factors in a linear factor and a quadratic factor,
then the reduced discriminant must factor as
\begin{equation} (\Delta/t^6)|_{t=0}=A^2B, \label{eq:soseven} \end{equation}
and $B$ cannot be a square.  In this case, $B=0$ gives branch points for
a double cover, and the vector representation is non-localized and
determined by that cover.  The points in $A=0$ represent localized matter in
the spinor representation.

These necessary conditions, in combination with other standard F-theory
restrictions, can sometimes directly be used to rule out enhancements of the gauge
algebra beyond $\mathfrak{g}_2$ (or beyond $\mathfrak{so}(7)$).  But in
other cases, we must perform a more extensive analysis.  In particular:
\begin{enumerate}
\item Suppose that the curve $\Sigma =\{t=0\}$ has
Kodaira type $I_0^*$ and that it meets another curve $C=\{z=0\}$ of Kodaira
type $IV$.  Then we can write $\varphi|_{t=0}=z^2\overline{\varphi}$ and
$\gamma|_{t=0}=z^2\overline{\gamma}$.  Suppose the auxiliary cubic factors
as $(\psi-\alpha)(\psi^2+\alpha\psi+\beta)$ (which will be true for
either gauge algebra $\mathfrak{so}(7)$ or gauge algebra $\mathfrak{so}(8)$).
Then we can write
\begin{align*}
z^2\overline{\varphi} &= \beta-\alpha^2 \\
z^2\overline{\gamma} &= -\alpha\beta .
\end{align*}
It follows that $z^2$ divides $\alpha(\beta-\alpha^2)-\alpha\beta=-\alpha^3$
so that $z$ divides $\alpha$.  Then, $z^2$ divides
$(\beta-\alpha^2)+\alpha^2=\beta$ which implies that $z^3$
divides $-\alpha\beta$.  Choosing $\widetilde{\alpha}$ and
$\widetilde{\beta}$ in a neighborhood of $\Sigma$
which restrict to $\alpha/z$ and $\beta/z^2$ on $\Sigma$,
we find that the Weierstrass equation can be written in the form
\[ y^2 = x^3 + ((\widetilde{\beta}-\widetilde{\alpha}^2)z^2t^2+\varphi'z^2t^3)x
+ (-\widetilde{\alpha}\widetilde{\beta}z^3t^3+\gamma'z^2t^4).
\]
Thus we see that the Weierstrass coeffients have multiplicity $4$ and $6$
at $z=t=0$, which means that there are already tensionless strings in this
model, i.e., it is not in the tensor branch of the theory, contrary to
assumption.

The conclusion is that in this case, the gauge symmetry can only be
$\mathfrak{g}_2$.  This applies in particular to the 223 non-Higgsable
cluster.
\item Suppose instead that the curve $\Sigma =\{t=0\}$ of
Kodaira type $I_0^*$ meets  $C=\{z=0\}$ of Kodaira
type $III$.  This time, we can write $\varphi|_{t=0}=z\overline{\varphi}$ and
$\gamma|_{t=0}=z^2\overline{\gamma}$.  Suppose the auxiliary cubic factors
completely into linear factors
as $(\psi-\sigma)(\psi-\tau)(\psi+\sigma+\tau)$ (which will be true for
gauge algebra $\mathfrak{so}(8)$).
Then we can write
\begin{align*}
z\overline{\varphi} &= -\sigma^2-\sigma\tau-\tau^2\\
z^2\overline{\gamma} &= \sigma\tau(\sigma+\tau) .
\end{align*}
It follows that $z$ divides
$\sigma(-\sigma^2-\sigma\tau-\tau^2)+\sigma\tau(\sigma+\tau)=-\sigma^3$
so that $z$ divides $\sigma$,
and also that $z$ divides
$\tau(-\sigma^2-\sigma\tau-\tau^2)+\sigma\tau(\sigma+\tau)=-\tau^3$
so that $z$ divides $\tau$.
Choosing $\widetilde{\sigma}$ and
$\widetilde{\tau}$ in a neighborhood of $\Sigma$
which restrict to $\sigma/z$ and $\tau/z$ on $\Sigma$,
we find that the Weierstrass equation can be written in the form
\[ y^2 = x^3 + ((-\widetilde{\sigma}^2-\widetilde{\sigma}\widetilde{\tau}-\widetilde{\tau}^2)z^3t^2+\varphi'z^2t^3)x
+ (\widetilde{\sigma}\widetilde{\tau}(\widetilde{\sigma}+\widetilde{\tau}z^3t^3+\gamma'z^2t^4)).
\]
Thus we again see that the Weierstrass coeffients have multiplicity $4$ and $6$
at $z=t=0$, which means that there are already tensionless strings in this
model, i.e., it is not in the tensor branch of the theory, contrary to
assumption.

The conclusion is that in this case, the gauge symmetry cannot  be
$\mathfrak{so}(8)$.  This applies in particular to the 23 non-Higgsable
cluster.
\end{enumerate}

\subsection{$\mathfrak{sp}_n$ fibers and enhancement of $\mathfrak{g}_2$ factors}
It was pointed out in \cite{Morrison:2012np} that the mixed representation of the $\mathfrak{g}_2 \oplus \mathfrak{su}_2$ gauge algebra may at times enhance from $\frac{1}{2}(\textbf{7},\textbf{2})$ to $\frac{1}{2}(\textbf{7+1},\textbf{2})$.  In fact, this enhancement depends on the fiber type of the curve carrying the $\mathfrak{su}_2$ factor.  If the $\mathfrak{su}_2$ is associated with a (non-split) fiber of type $I_3^{ns}$ or $IV$, then the representation of $\mathfrak{g}_2$ will be $7$-dimensional.  If the fiber is of type $I_2$ or $III$, then it will be $8$-dimensional.

An important example of this involves the 322 NHC, which has fiber types $I_0^{*ns}$, $IV^{ns}$, $II$, respectively.  There is a half-fundamental of $\mathfrak{su}_2$ localized between the two $-2$ curves, so there are only $7$ half-fundamentals left living on the middle $-2$ curve.  For this configuration to respect anomaly considerations, therefore, the mixed representation of $\mathfrak{g}_2 \oplus \mathfrak{su}_2$ must be $\frac{1}{2}(\textbf{7},\textbf{2})$.  Since the middle $-2$ curve has fiber type $IV^{ns}$, this is indeed the case.  On the other hand, the 32 NHC has fiber types $I_0^*$, $III$ respectively.  Here, the mixed representation of $\mathfrak{g}_2 \oplus \mathfrak{su}_2$ is $\frac{1}{2}(\textbf{7+1},\textbf{2})$.

More generally, a $-1$ curve carrying gauge algebra $\mathfrak{sp}_n$ will have global symmetry $SO(4n+16)$, corresponding to the $4n+16$ half-fundamentals needed for anomaly cancellation.  However, if the $-1$ curve intersects other compact curves, this symmetry may be broken to a subgroup.  For instance, if a $-1$ curve carrying gauge algebra $\mathfrak{sp}_1$ touches two $-4$ curves with gauge group $\mathfrak{so}_{10}$, then the $SO(20)$ flavor symmetry will be broken and gauged to $SO(10) \times SO(10)$.  In general, a non-split fiber of type $I_k$ and a non-split fiber of type $I_{2k+1}$ will both give rise to a $\mathfrak{sp}_k$ gauge algebra.  However, if the flavor symmetry on the $-1$ curve is broken to factors $G_1 \times G_2 \times ...$ and associated representations $(r_{G_1}, r_{G_2},...)$ with any $r_{G_i}$ odd-dimensional, then the fiber type must be $I_{2k+1}^{ns}$.

As a result, we note that the configuration
$$
\overset{\mathfrak{g}_2}{3} \,\, \overset{\mathfrak{sp}_1}{1} \,\, \overset{\mathfrak{so}_{13}}{4} \,\, ...
$$
is acceptable.  The fiber type of the $-1$ curve must be $I_3^{ns}$ due to the $\mathfrak{so}_{13}$ factor, and so the mixed representation of $\mathfrak{g}_2 \oplus \mathfrak{su}_2$ is $\frac{1}{2}(\textbf{7},\textbf{2})$.  This requires 20 half-fundamentalss of $\mathfrak{sp}_{1}$ to transform in mixed representations, which is indeed the amount required by anomaly considerations.

As another example, we may consider the configuration
$$
\overset{II}{2} \,\, \overset{\mathfrak{sp}_1}{1} \,\, \overset{\mathfrak{so}_{10}}{4} \,\, ...
$$
Here, there are two possibilities for the fiber of the $-1$ curve: $I_2^{ns}$ or $I_3^{ns}$.  In the former case, there will be two full hypers of $\mathfrak{sp}_1$ localized on the intersection with the $-2$ curve, and there will be an $SO(6)$ flavor symmetry under which the 6 half-fundamentals leftover on the $-1$ curve transform.  On the other hand, if the fiber type is $I_3^{ns}$, then there will be a half-fundamental localized at the intersection with the $-2$ curve, and the leftover flavor symmetry of the $-1$ curve will be $SO(9)$.

\subsection{Some examples}

Many examples of enhancements of A-D-E graphs can be constructed by using
a variant of Schoen's construction of a fiber product of rational
elliptic surfaces with section \cite{schoen,kapustka}.  Using this construction,
one can produce F-theory examples with certain enhanced gauge symmetries
over an affine Dynkin diagram:  further details are given in
\cite{MPT}.  Since each affine Dynkin diagram contains A-D-E graphs,
restricting any of these examples to an A-D-E graph will give a contractible
configuration with the specified enhanced gauge symmetry.

The examples which can be built by the methods of
\cite{schoen,kapustka,MPT} are:

\begin{center}
\begin{tabular}{cccc}
& Kodaira type & Kodaira type & Kodaira type \\
Affine
diagram
& on mult.\ 1 & on mult.\ 2
& on mult.\ $k{\ge}3$\\
\hline
$\widehat{A}_m$ & $I_n$, $n\ge2$ && \\
$\widehat{A}_m$ & $III$ && \\
$\widehat{A}_m$ & $IV^s$ && \\
$\widehat{D}_m$ & $I_n$, $n\ge2$ & $I_{2n}$ & \\
$\widehat{D}_m$ & $II$ & $IV$ & \\
$\widehat{D}_m$ & $III$ & $I_0^*$ & \\
$\widehat{E}_6$ & $I_n$, $n\ge2$ & $I_{2n}$ & $I_{kn}$\\
$\widehat{E}_6$ & $II$ & $IV$ & $I_0^*$\\
$\widehat{E}_7$ & $I_n$, $n\ge2$ & $I_{2n}$ & $I_{kn}$\\
$\widehat{E}_8$ & $I_n$, $n\ge2$ & $I_{2n}$ & $I_{kn}$\\
\end{tabular}
\end{center}
Note:
\begin{itemize}
\item All of the $I_n$ fibers which occur here are split, and correspond to
gauge algebra $\mathfrak{su}(n)$.
\item In the case of $\widehat{D}_m$
with fibers of type $II$ and $IV$, the type $IV$ curves which meet
type $II$ curves are non-split, with gauge algebra $\mathfrak{su}(2)$,
while the type $IV$ curves in the middle of the chain (which meet no
type $II$ curves) are split, with gauge algebra $\mathfrak{su}(3)$.
\item In the case of $\widehat{E}_6$ with fibers of type $II$, $IV$, and
$I_0^*$, the type $IV$ fibers are non-split, with gauge algebra
$\mathfrak{su}(2)$, while the type $I_0^*$ fiber is split with
gauge algebra $\mathfrak{so}(8)$.  Note that the $\mathfrak{so}(8)$
 matter consists of
two vectors, two spinors of one chirality, and two spinors of
the other chirality, appearing as three different $(2,8)$ pairs,
one corresponding to each intersection point.
\end{itemize}

\newpage

\addcontentsline{toc}{section}{References}
\bibliographystyle{utphys}
\bibliography{sixDtotal}

\end{document}